\documentclass[iop]{emulateapj}
\usepackage{epsfig}
\usepackage{natbib}

\newcommand{\beq}{\begin{equation}}
\newcommand{\eeq}{\end{equation}}

\def\be{\begin{equation}}
\def\ee{\end{equation}}
\def\bea{\begin{eqnarray}}
\def\eea{\end{eqnarray}}

\def \cgssb {{\rm\,erg\,s^{-1}\,cm^{-2}\,arcsec^{-2}}}
\def \cgsflux   {{\rm\,erg\,s^{-1}\,cm^{-2}}}
\def \cgsspec   {{\rm\,erg\,s^{-1}\,cm^{-2}}\,{\rm \AA}^{-1}}
\def \cgslum   {{\rm\,erg\,s^{-1}}}
\def\mwlya{W_{\rm Ly\alpha}}
\def\wlya{$W_{\rm Ly\alpha}$}
\def\lnull{$L_{\nu_{\rm LL}}$}
\def\mlnull{L_{\nu_{\rm LL}}}
\def\mllya{L_{\nu_{\lya}}}
\def\sblya{{\rm SB}_{\lya}}
\def \hMpc      {h^{-1}{\rm\ Mpc}}



\newcommand{\nar}{New Astronomy Review}

\def\sci#1{{\; \times \; 10^{#1}}}

\def \logTd6 {\hbox{log$( T/6 \kev)$} }

\newcommand{\sdssj}{SDSSJ1204+0221}

\def\myputfigure#1#2#3#4#5%
{\vskip#5pt\makebox[0pt]{\hskip#2in
\includegraphics[width=#3\textwidth]{#1}}\vskip#4pt\hfill}

\def \arcmin     { ^{\prime} }
\def \arcsec    {^{\prime\prime}}

\def \mpc       {{\rm\ Mpc}}
\def \kpc       {{\rm\ kpc}}

\def \kms            {{\rm km~s}^{-1}}

\def \kev       {{\rm\ keV}}
\def \msol      {{\rm\ M}_\odot}

\def \hMpc      {h^{-1}{\rm\ Mpc}}

\def \cgsflux   {{\rm erg\ s^{-1}\ cm^{-2}}}

\newcommand{\mnhi}{N_{\rm HI}}

\newcommand{\mlya}{${\rm Ly\alpha}$}
\newcommand{\mlyb}{${\rm Ly\beta}$}
\newcommand{\lya}{{\rm Ly}\alpha}

\input epsf
\bibpunct{(}{)}{;}{a}{}{,}

\begin{document}

\lefthead{THE CGM in ABSORPTION \& EMISSION}\righthead{HENNAWI \& PROCHASKA}
\title{Quasars Probing Quasars IV: Joint Constraints on the 
  Circumgalactic Medium from Absorption and Emission}

\author{Joseph F. Hennawi\altaffilmark{1,2} \& J. Xavier
  Prochaska\altaffilmark{1,2,3}}

\altaffiltext{1}{Max-Planck-Institut f\"ur Astronomie, K\"onigstuhl
  17, D-69117 Heidelberg, Germany}
\altaffiltext{2}{Visiting Astronomer, W.M. Keck Telescope.
The Keck Observatory is a joint facility of the University
of California and the California Institute of Technology.}
\altaffiltext{3}{Department of Astronomy and Astrophysics, 
  UCO/Lick Observatory; University of California, 1156 High Street, Santa Cruz, 
  CA 95064; xavier@ucolick.org}

\begin{abstract}
  We have constructed a sample of 29 close projected quasar pairs
  where the background quasar spectrum reveals absorption from
  optically thick \ion{H}{1} gas associated with the foreground
  quasar. These unique sightlines allow us to study the quasar
  circumgalactic medium (CGM) in absorption and emission
  \emph{simultaneously}, because the background quasar pinpoints large
  concentrations of gas where \mlya\ emission, resulting from
  quasar-powered fluorescence, resonant \mlya\ scattering, and/or
  cooling radiation, is expected. A sensitive search ($1\sigma$
  surface-brightness limits of ${\rm SB}_{\rm \lya} \simeq
  3\sci{-18}\,\cgssb$) for diffuse \mlya\ emission in the environments
  of the foreground (predominantly radio-quiet) quasars is conducted
  using Gemini/GMOS and Keck/LRIS slit spectroscopy.  We fail to
  detect large-scale $\sim 100\,\kpc$ \mlya\ emission, neither at the
  location of the optically thick absorbers nor in the foreground
  quasar halos, in all cases except a single system. We interpret
  these non-detections as evidence that the gas detected in absorption
  is shadowed from the quasar UV radiation due to obscuration effects,
  which are frequently invoked in unified models of AGN. Small-scale
  $R_{\perp}\lesssim 50\,\kpc$ extended \mlya\ nebulosities are
  detected in $34\%$ of our sample, which are likely the high-redshift
  analogs of the extended emission-line regions (EELRs) commonly
  observed around low-redshift ($z < 0.5$) quasars.  This may be
  fluorescent recombination radiation from a population of very dense
  clouds with a low covering fraction illuminated by the quasar.
  We also detect a compact high rest-frame equivalent
  width ($W_{\lya} > 50$\AA) \mlya-emitter with luminosity $L_{\lya} =
  2.1\pm0.32\sci{41}\cgslum$ at small impact parameter $R_{\perp}=
  134\,\kpc$ from one foreground quasar, and argue that it is more
  likely to result from quasar-powered fluorescence, than simply be a
  star-forming galaxy clustered around the quasar.  Our observations
  imply that much deeper integrations with upcoming integral-field
  spectrometers such as MUSE and KCWI will be able to routinely detect
  a diffuse \mlya\ glow around bright quasars on scales $R\sim 100\,\kpc$ 
  and thus directly image the CGM.
\end{abstract}


\section{Introduction}
\label{sec:intro}

The ionizing radiation emitted by a luminous quasar can, like a
flashlight, illuminate hydrogen in its vicinity, teaching us about the
physical properties of gas in its surroundings.  This is because
photoionized hydrogen ultimately produces Ly$\alpha$ recombination
photons, which are likely to escape the system. Whether this
`fluorescent' emission arises from the highly ionized 
Ly$\alpha$
forest clouds in the intergalactic medium (IGM) which are optically
thin to ionizing photons, from the optically thick Lyman Limit Systems
(LLSs) and damped Lyman-$\alpha$ systems (DLAs) detected as the
strongest absorption lines in quasar spectra, or from the interstellar
or circumgalactic medium (CGM) of the quasar host itself, the underlying
principle is nevertheless the same -- a fraction of the energy in the
quasars' UV continuum is `focused' into line radiation and re-emitted
back into space, informing us about the physical conditions of the
emitting gas.

The idea of searching for fluorescent \mlya\ emission from the
intergalactic medium is not new. \citet{HW87} first proposed that
fluorescent emission might be detectable from the optically thin
$\mnhi \lesssim 17.2$, Ly$\alpha$ forest clouds, which emit
recombination photons because they are illuminated by the ambient
extragalactic UV background. \citet{GW96} made the crucial point that
emission from the $\mnhi \gtrsim 17.2$ LSSs should be much brighter,
because these clouds, being optically thick to Lyman continuum
photons, absorb all the ionizing radiation incident upon them. Some
$\sim 60\%$ of these ionizing photons result in \mlya\ recombinations,
and are `mirrored' back as Ly$\alpha$ photons.
Thus at the very faintest flux levels, UV background radiation
should power a ubiquitous fluorescence signal, whereby all of the
optically thick gas in the cosmic web would `glow' in the \mlya\ line
\citep{HW87,binette93,GW96,cantalupo05,juna08}.

Observationally, a number of increasingly sensitive searches for
fluorescent radiation powered by the ambient UV background have been
performed \citep{Lowenthal90,Bunker98,Bunker99,Marleau99}, the deepest
being the $\sim 100$ hour integration by \citet{rauch08} using the
Very Large Telescope (VLT). 
Notwithstanding these efforts, the fluorescence signal has yet to
be detected, and given independent estimates for the UV background
intensity \citep{meiksin04,bolton05,faucher08}, the expected surface brightness
${\rm SB}_{\lya} \simeq 1\sci{-20}\,\cgssb$ is probably out of reach of
current instrumentation.  But in the regions proximate to a quasar the
ionizing flux and resulting fluorescent surface brightness can be
significantly enhanced, provided the quasar does not photoevaporate
the nearby optically thick clouds.  Fluorescence in the vicinity of a
quasar has been searched for by a number of authors
\citep{fmw99,fbh04,francis06,ask+06,clp07,rauch08} yielding either
null or questionable detections.  Perhaps most contentious is
\citet{ask+06}, who purport to detect fluorescent emission from a DLA
serendipitously detected in a background quasar spectrum which is
$49\arcsec$ away from a luminous foreground quasar at $z=2.84$. Owing
to the proximity of the foreground quasar, the ionizing flux should be
enhanced by a factor of $\sim 5700$ over the UVB. However, the large
continuum luminosity of this source and its implied relatively modest
rest-frame \mlya\ equivalent width (EW) of 75\AA\ suggest that it may
simply be \mlya\ emission from the galaxy counterpart to the DLA,
rather than quasar powered fluorescence.

Most recently, \citet{cantalupo12} performed a deep survey for
\mlya\ emission around a very bright $z = 2.4$ quasar.  In their
narrow-band imaging, they report the detection of $\approx 100$
sources in their $3.5\,{\rm Mpc}\times 3.5\,{\rm Mpc}$ field of view,
corresponding to a $\approx 5500$ comoving Mpc$^3$ volume.  They
further note that a significant subset ($\approx 20\%$) exhibit
\mlya\ EWs far in excess of the maximum predicted value from
star-forming regions (i.e.\ $\mwlya > 240$\AA).  These sources are the
most convincing candidates to date for fluorescing gas in the extended
environment (several Mpc separation) of, and powered by, a
quasar. However it still remains unclear whether these candidate
fluorescent emitters are counterparts to typical optically thick
absorbers (LLSs and DLAs), and why more bright fluorescent emitters
were not detected very close ($\lesssim 300\,{\rm kpc}$ proper) to the
quasar. We will address both of these questions in this paper.

Besides illuminating nearby clouds in the IGM, a quasar may irradiate
gas in its own host galaxy or circumgalactic halo. \cite{Rees88}
suggested that cold CGM gas in the quasar host illuminated by the
quasar could be detectable as an extended ``fuzz'' of fluorescent
Ly$\alpha$ emission \citep[see also][]{alam02,HR01}.  A number of
groups have reported the detection of extended \mlya\ emission in the
vicinity of $z \sim 2-4$ quasars
\citep{djorgovski85,hc87,heckman91a,heckman91b,Bremer92,lehnert98,bunker03,weidinger04,fbh04,weidinger05,francis06,cjw+06,courbin08,pdla09,north12}. 
These efforts, some targeted and some serendipitous, reveal a
diversity both in emission level, detection frequency, and physical
extent, but detailed quantitative interpretation is hampered by
differing methodologies (deep longslit spectroscopy versus narrow-band
imaging), sample inhomogeneities (radio-loud versus radio-quiet), a
broad range of redshifts (significant given $(1+z)^4$ surface
brightness dimming and strong evolution in the quasar population), and
ambiguities in detection criteria and definitions of the depth of the
observations. Nevertheless, the emerging picture is that roughly
half of quasars from $z\sim 2-4$ exhibit extended emission
on scales of $10-50\,\kpc$ down to observed frame surface
brightness levels of ${\rm SB}_{\rm Ly\alpha}\sim {\rm few}\times
10^{-17}\cgssb$.  There is suggestive evidence that radio-loud quasars
have brighter emission and a higher detection frequency
\citep{heckman91a}, although high detection frequencies have also
been reported in samples which are mostly radio-quiet
\citep{cjw+06,north12}. Finally, this \mlya\ emission may be powered
by the same mechanism powering the extended emission line regions (EELRs) detected
around low redshift $z < 0.5$ type-I \citep[e.g.][]{stockton06,husemann12} and type-II \citep{greene11} quasars, traced by [\ion{O}{3}] and Balmer lines.

The brightest cases of \mlya\ nebulosities around quasars have an
average surface brightness ${\rm SB}_{\lya} \sim 10^{-16}\cgssb$
($z=2$) over a diameter of $\sim 100\,\kpc$, corresponding to a total
Ly$\alpha$ luminosity of $L_{\rm Ly\alpha}\sim
10^{44.5}~\cgslum$. These nebulae bear a striking resemblance to the
extended \mlya\ emission typically observed around high-redshift radio
galaxies \citep[HzRG;
  e.g.][]{msd+90,mccarthy93,vanojik96,nle+06,binette06,reuland07,villar07,
  mileyd08} as well as the so-called \mlya\ `blobs' \citep[LABs;
  e.g.][]{fmw99,steidel00,fwc+01,palunas04,matsuda04,dbs+05,ctf+06,
  nfm+06,sj07}.  The primary difference between the nebulae around
quasars versus those around HzRGs or \mlya\ blobs, is that a strong
source of ionizing photons (e.g. the quasar) is directly identified,
but not for the HzRGs or blobs. Obscuration and orientation effects,
as are often invoked in unified models of AGN \citep[see
  e.g.][]{Anton93,elvis00}, could be responsible for this difference.
Indeed, there is ample evidence for obscured AGN in HzRGs
\citep[e.g.][]{mileyd08}, and evidence for obscured AGN have been
uncovered in several of the LABs
\citep{csw+04,basu04,dbs+05,gsc+07,Barrio08,smith08}, although there
are several notable exceptions \citep{sj07,nfm+06,ouchi09}.

Although photoionization by an AGN is perhaps the most plausible
mechanism powering extended \mlya\ nebulae, particularly when the
quasar is directly observed, a litany of other mechanisms have also
been put forward. Other sources of ionizing photons, such as fast
radiative shocks powered by radio-jets
\citep{msd+90,heckman91a,heckman91b} or starburst driven superwinds
\citep{fwc+01,taniguchi00,taniguchi01} have been discussed. It also
been suggested that X-rays produced via inverse Compton scattering of
CMB photons (and/or local far-IR photons) by relativistic electrons in
radio jets \citep{scharf03,smail09}, could be the source of
ionization. The possibility that extended \mlya\ emission is powered
by star-formation, either via stellar ionizing photons escaping from
galaxies \citep{rauch11}, spatially extended star-formation
\citep{ouchi09,rauch12}, or by pure scattering of \mlya\ photons
\citep{dijkstra06a,dijkstra2006b,barnes09,barnes10,barnes11,zheng11,steidel11,dijkstra12},
has also been considered.  In addition to photoionization and
scattering, a large body of theoretical work has suggested that
\mlya\ emission nebulae could result from \mlya\ cooling radiation
powered by gravitational collapse
\citep{haiman00,fardal01,furlanetto05,yang06,dijkstra06,dl09,goerdt10,dayal10,bertone10,fg10,frank12,rosdahl12}.
This scenario is particularly fashionable in light of a plethora of
work on the so-called `cold mode' of cosmological accretion
\citep{fardal01,Katz03,kkw+05,keres09b,db06,birnboim07,ocvirk08,brooks09,
  dekel09}, whereby a significant fraction of the gas accreting onto
galaxies from the IGM remains cold $T\sim 10^4\,{\rm K}$, and funnels
in along large-scale filaments. In the absence of significant
metal-enrichment, collisionally excited Ly$\alpha$ is the primary
coolant of $T \sim 10^4\,{\rm K}$ gas; hence the accreting gas could
steadily radiate gravitational potential energy in the \mlya\ line as
it falls into the halo. While many studies have suggested that
gravitational cooling radiation could power the LABs
\citep{fardal01,furlanetto05,dl09,goerdt10,fg10,rosdahl12,frank12},
the predictions of these simulations are uncertain by orders of
magnitude \citep[e.g.][]{furlanetto05,fg10,rosdahl12,lyacool} because
the emissivity of collisionally excited \mlya\ is exponentially
sensitive to gas temperature. Accurate prediction of the temperature
requires solving a coupled radiative transfer and hydrodynamics
problem which is not currently computational feasible \citep[but
  see][]{rosdahl12}.

Thus despite significant observational and theoretical efforts, the
relationship between \mlya\ nebulae in quasars, HzRGs, and LABs is
rather confusing, and the physical process powering extended
\mlya\ emission remains an important unsolved problem. Questions also
remain about \mlya\ emission outside of quasar halos, on the 
larger scales relevant to quasar powered IGM fluorescence.  

In this fourth paper of the `Quasars Probing Quasars' series, we
introduce a novel technique\footnote{We note that a similar technique
  was employed for the single serendipitously discovered projected
  quasar pair by \citep{ask+06}, however our approach differs
  significantly in that we use a large sample of quasar pairs to
  statistically characterize the properties and distribution of gas in
  quasar environments, which can be used to make predictions for the
  expected emission.} to tackle the important problem of emission from
the CGM. We search for this emission from close projected quasar
pairs, which have small angular separation on the sky but large
line-of-sight separations such that the two quasars are physically
unassociated. In these unique sightlines, absorption in the background
(`b/g') quasar encodes information about the distribution of
circumgalactic and intergalactic gas in the vicinity of the foreground
(`f/g') quasar. This technique thus allows us to analyze the
absorption-line and emission-line properties of gas around quasars
\emph{simultaneously}. We perform a systematic, spectroscopic survey
for extended \mlya\ emission in the vicinity ($R_{\perp} < 600$\,kpc)
of $z \sim 2$ quasars, to typical $1\sigma$ surface-brightness limits
of ${\rm SB}_{\rm \lya} \simeq 3\sci{-18}\,\cgssb$.  The data are
drawn from our survey of projected quasar pairs \citep{QPQ1}, and we
restrict our analysis to 29 projected pair systems for which the
spectrum of the b/g quasar shows evidence for an LLS coincident with
the redshift of the f/g quasar. If the f/g quasar is illuminating this
optically thick gas, this is a prime candidate for bright
\mlya\ fluorescence and the the b/g quasar pinpoints the expected
location of the emission.  Independently, our observations provide a
sensitive search for any extended \mlya\ emission in the environment
of $z \sim 2$ quasars, such as might be powered by \mlya\ cooling
radiation or any of the other physical mechanism just discussed. Our
b/g quasar absorption-line measurements statistically map out the
covering factor of optically thick absorption in the quasar CGM
\citep{QPQ1,QPQ2,QPQ5}, and detailed absorption-line modeling places
constraints on the physical properties of this gas \citep{QPQ3}. Armed
with this knowledge about the properties of the CGM, we can make
direct predictions for the expected fluorescent emission and compare
to our observations. Similarly, by combining our emission constraints
from this study with the physical properties of the gas, we can
directly constrain the heating rate of the CGM gas we detect in
absorption, which is the subject of a companion paper \citep{lyacool}.

As we will frequently refer to other results from the `Quasars Probing
Quasars' (QPQ) series, we briefly review the basic approach and
summarize the key results of each paper. In QPQ1 \citep{QPQ1} we
introduced a novel technique to study the physical state of gas in the
CGM of $z \simeq 2-3$ quasars, which are hosted by massive galaxies,
using projected quasar pairs \citep[see
  also][]{crotts89,moller92,bhm+06,schulze12,farina12}.
Spectroscopic observations of the b/g quasar
in each pair reveals the nature of the IGM transverse to the f/g
quasar on scales of a few 10\,kpc to several Mpc. 
In QPQ1, we searched 149 b/g quasar spectra for optically thick
absorption in the vicinity of $1.8 < z_{\rm fg} < 4.0$ luminous f/g
quasars, and found a high-incidence of optically thick gas in the 
quasar hosts CGM. In QPQ2, we compared the statistics of
this optically thick absorption in b/g sightlines to that
observed along the line-of-sight, and argued that the f/g
quasars emit their ionizing radiation anisotropcially or intermittently. 
An echelle spectrum of a projected quasar pair was presented in QPQ3
\citep{QPQ3}, which resolved the velocity field and revealed the
physical characteristics (metallicity, density, etc.)  of the
absorbing CGM gas.  In QPQ5 \citep{QPQ5}, we used a significantly
enlarged sample of projected quasar pair spectra to characterize the
CGM of quasars with much improved statistics, focusing on the covering
factor and EW distributions of \ion{H}{1} and metal line absorbers.

This paper is organized as follows.  In \S\ref{sec:form} we present a
formalism which describes \mlya\ emission powered by photoionization
and scattering. In \S\ref{sec:observe} we provide details of the
quasar pair survey, our spectroscopic observations, and describe the
selection of the quasar-absorber sample that was used to search for
extended \mlya\ emission. As we will show, a key quantity for
interpreting the \mlya\ emission properties of the f/g quasars is
the covering factor of optically thick absorbers measured in
background sightlines, which we also quantify in this section.  A
custom algorithm was developed to subtract off the quasar PSFs and
search for faint diffuse extended emission which is described in
\S\ref{sec:coadd}. Our results are presented and discussed in
\S\ref{sec:discuss}, where we also compare to previous searches for
\mlya\ fluorescence. We summarize and conclude in
\S\ref{sec:summary}. In Appendix A, we quantify the attenuation of
\mlya\ by dust grains for gas clouds with properties matching our
absorbers, and conclude that dust extinction is not a significant
effect. Throughout this paper we use a cosmological model with
$\Omega_m = 0.24$, $\Omega_\Lambda =0.76$, $h=0.70$ which is
consistent with the most recent WMAP5 values \citep{Komatsu09}.
Unless otherwise specified, all distances are proper. It is helpful to
remember that in the chosen cosmology, at redshift $z=2.0$, an
angular separation of $\Delta\theta=1\arcsec$ corresponds to a proper
transverse separation of $R_\perp=8.7\,\kpc$, and a velocity
difference of $1500\,\kms$ corresponds to a radial redshift space
distance of $s=7.7\,\mpc$.  A typical quasar at $z=2.0$ with an SDSS
magnitude of $i=19$, has specific luminosity at the Lyman edge of
$\log_{10}\mlnull\simeq 30.5$ (erg\,s$^{-1}$\,Hz$^{-1}$), and at a distance of $R=100\,\kpc$
($11.5\arcsec$ projected on the sky) the flux of ionizing photons
would be a factor $g_{\rm UV}\simeq 3760$ times higher than the
ambient extragalactic UV background.  Finally, we use the terms
optically thick absorbers and LLSs interchangeably, both referring to
quasar absorption line systems with $\log N_{\rm HI}>17.2$, making
them optically thick at the \ion{H}{1} Lyman limit ($\tau_{\rm
  LL}\gtrsim 1$).

\section{Formalism and Preliminaries}
\label{sec:form}

This section provides analytical estimates for the surface brightness
of \mlya\ emission from a variety of physical processes.  We frame the
discussion in terms of a circumgalactic medium filled with cool
clouds, and we present results in terms of quantities which can be
directly measured from absorption-line observations using background
sightlines.

\subsection{Cool Cloud Model}
\label{sec:cloud}

Bright emission in the \mlya\ line typically occurs in gas at $T\sim
10^4\,{\rm K}$, thus we consider a highly idealized model of a
population of cool gas clouds in quasar halos. Several studies have
previously explored similar models \citep{mm96,mb04}, and
others have focused specifically on extended \mlya\ emission
\citep{Rees88,HR01,haiman00,dl09}. Nearly all of this work
assumes the cool gas is in pressure equilibrium with a hot tenuous
virialized plasma. Although there may be some evidence for pressure
confinement from our our absorption-line measurements \citep{QPQ3},
our conclusions about \mlya\ emission do not depend on the existence
of a hot phase.

We assume that the cool gas clouds observed in absorption are
spherical with a single uniform hydrogen volume density $n_{\rm H}$,
and that they are spatially uniformly distributed throughout a spherical halo of
radius $R$. Although extremely simple, this model is also relevant, to
within order unity geometric corrections, to more complicated
distributions which vary with radius, or the filamentary structures
expected in the cold accretion picture.  In what follows, we often
employ averages over the projected area of the spherical halo on the
sky which we denote with angle brackets, e.g. the average
line-of-sight distance through the halo is $\langle s \rangle =
1\slash A \int s dA = 4R\slash 3$. Besides the volume density $n_{\rm H}$ and the
size of the halo $R$, two other parameters completely specify the spatial distribution of
the gas, namely the hydrogen column density of the individual clouds
$N_{\rm H}$, and the cloud covering factor $f_{\rm C}$, which is
similarly defined to be an average over the sphere.  We focus on these
four parameters ($R$\,,$n_{\rm H}$\,,$N_{\rm H}$\,, $f_{\rm C}$)
because they are either directly observable or are closely related to
absorption-line observations.

All other quantities of interest can then be expressed in terms of
these four parameters. For example, of particular interest for
computing the \mlya\ surface brightness is the volume filling factor
defined as the ratio of the volume occupied by the clouds to the total
volume, $f_{\rm V} = n_{\rm c} V_{\rm c}$, where $n_{\rm c}$ is the number density 
of clouds and $V_{\rm c}$ is their volume. The definition of the covering
factor is 
\be 
f_{\rm C} \equiv \langle \int \frac{df_{\rm C}}{ds} ds\rangle \label{eqn:cov_fact}, 
\ee
where
$\frac{df_{\rm C}}{ds} = n_{\rm c}\sigma_{\rm c}$, and $\sigma_{\rm c}$ is the cloud 
cross-sectional area. This gives $f_{\rm C} = n_{\rm c}\sigma_{\rm c}4R\slash 3$ for our case of 
uniformly distributed clouds. Note that in general the covering factor can be
larger than unity. The corresponding volume filling factor can be written as
\be
f_{\rm V}=\frac{3f_{\rm C}N_{\rm H}}{4Rn_{\rm H}}, 
\ee
and the  total mass of cool gas is then
\be
M_{\rm c} = \frac{4}{3}\pi R^3 f_{\rm V} \frac{n_{\rm H} m_{\rm p}}{X}=
\pi R^2 f_{\rm C}N_{\rm H} \frac{m_{\rm p}}{X} \label{eqn:mc}, 
\ee
where $m_{\rm p}$ is the mass of the proton and $X=0.76$ is the hydrogen mass
fraction. 

As we will consider emission from objects at high redshift, for which
the size of the emitting clouds could be unresolved by our instrument,
the relevant observable is the surface brightness $(\cgssb)$ averaged
over the projected area of a sphere as measured from Earth.  The
specific intensity of \mlya\ from our spherical halo is given by
integrating the equation of radiative transfer through the gas
distribution \be I_{\lya} = \int j_{\lya} ds \label{eqn:radtrans}, \ee
where $j_{\lya}$ is the volume emissivity per steradian $({\rm
  erg\,s^{-1}\,cm^{-3}\,ster^{-1}})$.  In this and the following
sections we will ignore the attenuation of \mlya\ radiation caused by
resonant scattering of gas within the halo itself or by the IGM, as
well as extinction caused by dust.  Scattering within the halo itself
would only change the spatial distribution of the emission, and as we
are averaging over apertures on the sky, this should produce small
effects. We do however consider the resonant scattering of a point
source of \mlya\ photons from the quasar CGM in
\S\ref{sec:scatt}. Scattering by the IGM is likely negligible for the
redshifts we consider \citep{zheng11}. We estimate the impact of
extinction by dust in the Appendix, and demonstrate that it should not
result in a significant reduction of the \mlya\ emission.

We can then write the average surface brightness observed from Earth as
\be
{\rm SB}_{\lya}\equiv \frac{1}{(1+z)^4}\langle I_{\lya}\rangle = 
\frac{1}{(1+z)^4}\frac{1}{\pi R^2}\int j_{\rm \lya}dV \label{eqn:SB}
\ee
where the factor of $(1+z)^{-4}$ accounts for cosmological surface brightness 
dimming. The corresponding total \mlya\ luminosity is 
\bea
L_{\lya} &=& 4\pi^2 (1+z)^4 R^2{\rm SB}_{\lya}\label{eqn:lum}\\
&=& 1.3\sci{44} \left(\frac{1+z}{3.0}\right)^{4}\!\!
\left(\frac{R}{100\,{\rm kpc}}\right)^{2}\nonumber\\
&\times& \left(\frac{{\rm SB}_{\lya}}{10^{-17}\,\cgssb}\right)\,{\rm erg\,s^{-1}}\nonumber
\eea

\subsection{\mlya\ Fluorescence}
\label{sec:fluor}

For a low density gas, it is possible to separate contributions to the
total \mlya\ intensity into recombination and collisionally excited
components \citep{Chamberlain53}. Observationally, extended
\mlya\ emission could have contributions from both
mechanisms. Collisional excitation of neutral atoms and the subsequent
emission of \mlya\ can play an important role in the radiative cooling
of $T\sim 10^4\,{\rm K}$ gas, and we explore this cooling radiation
from the CGM of quasars in detail in a companion paper
\citep{lyacool}. In what follows we consider recombination, which we
will also refer to as \mlya\ fluorescence, from a source of ionizing
photons. A central point source with an $r^{-2}$ dependence is assumed,
but our estimates can easily be generalized to other spatial
distributions.  We consider the two limiting cases where the gas
clouds are either optically thin $N_{\rm HI} \ll 10^{17.2}\,{\rm
  cm}^{-2}$ or optically thick $N_{\rm HI} \gg 10^{17.2}\,{\rm
  cm}^{-2}$ to Lyman continuum photons.  At intermediate neutral
columns $\sim N_{\rm HI} \sim 10^{17.2}\,{\rm cm}^{-2}$ our simple
approximations break down and more detailed radiative transfer
calculations are necessary.  In each case, we consider the limit where
light from the quasar has otherwise been unattenuated by resonant
scattering or dust.

\subsubsection{Optically Thin}

The recombination emissivity from the cool clouds is
\be
j_{\lya} = f_{\rm V}\frac{h\nu_{\lya}}{4\pi}\eta_{\rm thin}n_{\rm e}n_{\rm p} 
\alpha_{\rm A}(T) \label{eqn:jthin}, 
\ee
where $n_{\rm e}$ is the electron density, $n_{\rm p}$ is the proton density, 
$\eta_{\rm thin}=0.42$ is the fraction of recombinations which result
in a \mlya\ photon in the optically thin limit
\citep{ferost06,GW96}, and the factor of $f_{\rm V}$ accounts
for the fact that the emitting clouds fill only a fraction of the volume. 
The case A recombination coefficient 
$\alpha_{\rm A}$ is a weak function of temperature $T$, and we evaluate it at 
$T=10,000\,{\rm K}$ where $\alpha_{A} = 4.18 \sci{-13}\,{\rm cm^{-3}\,s^{-1}}$ \citep{ferost06}. 

The electron and proton densities are determined by photoionization 
equilibrium,
\be
n_{\rm HI}\Gamma = n_{\rm e}n_{\rm p}\alpha_{\rm A} \label{eqn:ioneq},  
\ee
and the photoionization rate $\Gamma$ is given by 
\be
\Gamma = \frac{1}{4\pi r^2}\int_{\nu_{\rm LL}}^{\infty} 
\frac{L_{\nu}}{h\nu}\sigma_\nu d\nu. 
\ee
Here $\nu_{\rm LL}$ is the frequency at the Lyman edge, and $\sigma_{\nu}$ is 
the hydrogen photoionization cross-section. We assume that the
quasar's 
spectral energy distribution obeys the power-law form 
$L_{\nu} = L_{\nu_{\rm LL}} (\nu\slash \nu_{\rm LL})^{-\alpha_{\rm Q}}$, blueward of  
$\nu_{\rm LL}$ and adopt a slope of $\alpha_{\rm Q}= 1.57$ consistent
with the measurements of \citet{telfer02}. The quasar ionizing luminosity
is then parameterized by $L_{\nu_{\rm LL}}$, the luminosity at the Lyman edge.
Throughout this paper we estimate $L_{\nu_{\rm LL}}$
by tying the
\citet{telfer02} power-law spectrum to the composite quasar spectrum
of \cite{vanden01}, which can then be compared to the SDSS optical
photometry covering rest-frame UV wavelengths (see Appendix A of QPQ1
for a detailed description of this procedure).

For all cases of interest to us, radiation fields are sufficiently
intense that optically thin gas will always be highly-ionized. Under
this assumption, the neutral fraction $x_{\rm HI} \equiv \frac{n_{\rm
    HI}}{n_{\rm H}} \ll 1$, hence $n_{\rm p} \approx n_{\rm H}$ and
$n_{\rm e} \approx (1 + Y\slash 2X) n_{\rm H}$, where the helium and
hydrogen mass fractions are $Y$ and $X$, respectively, and we have
assumed all helium is doubly ionized.

Putting this all together, the observed \mlya\ surface brightness is
\bea
{\rm SB}_{\lya}&=&\frac{\eta_{\rm
    thin}h\nu_{\lya}}{4\pi(1+z)^4}\alpha_{\rm A}
\left(1 + \frac{Y}{2X} \right) n_{\rm H}f_{\rm C} N_{\rm H}\label{eqn:SB_thin}\\
&=& 7.7\times 10^{-19} \left(\frac{1+z}{3.0}\right)^{-4}\!\!
\left(\frac{f_{\rm C}}{1.0}\right) 
\left(\frac{n_{\rm H}}{0.1\,{\rm cm^{-3}}}\right)\nonumber\\
&\times & \left(\frac{N_{\rm H}}{10^{20}\,{\rm cm^{-2}}}\right)\cgssb\nonumber
\eea

Note that photoionization equilibrium implies that $j_{\lya} \propto
\Gamma n_{\rm HI}$, but since the neutral fraction is inversely
proportional to photoionization rate $x_{\rm HI} \approx \alpha_{\rm
  A}n_{\rm H}(1 + Y\slash 2X)\slash \Gamma$, the net effect is that
\mlya\ emission is independent of the luminosity of the ionizing
source in the highly-ionized, optically-thin regime, provided that
the ionizing source is bright enough to keep the gas highly ionized
and optically thin.  This fact can be exploited to obtain an
expression for the neutral column density 
averaged over the area of the halo, $\langle N_{\rm
  HI}\rangle$, in terms of the observed surface brightness and ionizing luminosity: 
\bea
\langle N_{\rm HI}\rangle &=& 2.2\times 10^{17}\left(\frac{1+z}{3.0}\right)^{-4}
\!\!\left(\frac{L_{\nu_{\rm LL}}}{10^{30}\,{\rm erg\,s^{-1}\,Hz^{-1}}}\right)^{-1}
\label{eqn:NHI}\\
& \times & \left(\frac{R}{100\,{\rm kpc}}\right)^{2}\!\!
\left(\frac{{\rm SB}_{\lya}}{10^{-17}\,\cgssb}\right)\,{\rm cm^{-2}}\nonumber. 
\eea
  
In order for the optically thin approximation to be valid, the
individual clouds must have $N_{\rm HI} \ll 10^{17.2}\,{\rm
  cm^{-2}}$. In optically thin photoionization equilibrium, the
neutral fraction scales as $x_{\rm HI} \propto r^2$ so that the
largest neutral columns will be attained at $R$, the edge of the
halo. It thus follows that $\langle N_{\rm HI}\rangle < {\rm
  max}(N_{\rm HI})$ and hence if $\langle N_{\rm HI}\rangle >
10^{17.2}\,{\rm cm}^{-2}$ then the clouds are definitely optically
thick at some locations in the halo and the optically thin
approximation breaks down. One can thus use eqn.~(\ref{eqn:NHI}) to
compute $\langle N_{\rm HI}\rangle$ and directly test whether the
optically thin approximation is valid, given only the observed
\mlya\ surface brightness and the size of the nebulae, provided that
\lnull\ can be estimated, as in the case of \mlya\ nebulae observed
around quasars. If $\langle N_{\rm HI}\rangle > 10^{17.2}\,{\rm
  cm}^{2}$ then the clouds cannot be optically thin.  Note however
that the converse is not true --- both optically thin and optically
thick clouds can result in $\langle N_{\rm HI}\rangle <
10^{17.2}\,{\rm cm}^{-2}$ -- a small value of $\langle N_{\rm
  HI}\rangle$ thus provides no definitive information about whether we
are in the optically thin or thick regime.

\subsubsection{Optically Thick}
\label{sec:recomb_thick}

If the clouds are optically thick to ionizing radiation, they will no
longer emit \mlya\ proportional to the volume that they occupy because
of self-shielding effects. Instead, a thin highly ionized skin will
develop around each cloud, and nearly all recombinations and resulting
\mlya\ photons will originate in this skin. The cloud will then behave
like a special `mirror', converting a fraction $\eta_{\rm thick} =
0.66$ of the ionizing photons it intercepts into \mlya\ photons
emitted at a uniform brightness \citep[i.e the specific intensity
  leaving the surface is isotropic, see][]{GW96}.  Simulations
including both ionizing radiative transfer and \mlya\ resonant
scattering \citep{cantalupo05,juna08} have shown, that for the case of
a quasar illuminating an optically thick cloud, the true surface
brightness is reduced from the `mirror' value by a factor of $\simeq
2-4$. This reduction occurs because illumination and radiative
transfer effects redistribute photons over a wide solid angle.  We
introduce a geometric reduction factor $f_{\rm gm}$ to account for
this reduction. For the case of a transversely illuminated spherical
`mirror', the resulting `half-moon' illumination pattern results in a
$f_{\rm gm}=4\slash (3\pi)=0.42$ \citep{ask+06}. We set $f_{\rm
  gm}=0.5$, which is consistent with the range seen in radiative
transfer simulations \citep{cantalupo05,juna08}.  At a distance
$R$ from a quasar, a mirror
will emit a surface brightness 
\bea 
{\rm SB}_{\lya}&=&\frac{f_{\rm gm}\eta_{\rm thick}h\nu_{\lya}}{(1+z)^4}\frac{\Phi}{\pi}\label{eqn:SB_mirror}\\
&=& 4.0\times 10^{-17}\left(\frac{1+z}{3.0}\right)^{-4}\!\!
\left(\frac{f_{\rm gm}}{0.5}\right)\!\!
\left(\frac{R}{100\,{\rm kpc}}\right)^{-2}\nonumber\\
&\times & \left(\frac{L_{\nu_{\rm LL}}}{10^{30}\,{\rm erg\,s^{-1}\,Hz^{-1}}}\right)\cgssb\nonumber
\eea
where $\Phi$ $({\rm phot\,s^{-1}\,cm^{-2}})$ is the ionizing photon
number flux 
\be 
\Phi= \int_{\nu_{\rm LL}}^{\infty}
\frac{F_{\nu}}{h\nu} d\nu = \frac{1}{4\pi r^2}\int_{\nu_{\rm
    LL}}^{\infty} \frac{L_{\nu}}{h\nu} d\nu. 
\ee
The total observed flux from the mirror is simply proportional to the area
of the mirror times this $\sblya$.

Now consider the case of a population of cool optically thick clouds, hence
also mirrors, but which are spatially unresolved by our instrument.  In this
regime we can write the \mlya\ volume emissivity as
\be
j_{\lya} = \frac{\eta_{\rm thick}h\nu_{\lya}}{4\pi}n_{\rm c}\sigma_{\rm c}\Phi,   
\ee
Note that there is no geometric reduction factor in this case. Conservation of 
photons requires that a fraction $\eta_{\rm thick}$ of the ionizing
photons emerge as \mlya\ photons, and the emissivity must be isotropic. Again 
integrating the equation of radiative transfer 
and averaging over area (eqns.~\ref{eqn:radtrans} and \ref{eqn:SB}), 
we obtain 
\bea
{\rm SB}_{\lya}&=&\frac{\eta_{\rm thick}h\nu_{\lya}}{4\pi(1+z)^4}f_{\rm C}
\Phi(R\slash \sqrt{3})\label{eqn:SB_thick}\\
&=& 6.0\times 10^{-17}\left(\frac{1+z}{3.0}\right)^{-4}\!\!
\left(\frac{f_{\rm C}}{1.0}\right)\!\!
\left(\frac{R}{100\,{\rm kpc}}\right)^{-2}\nonumber\\
& \times & 
\left(\frac{L_{\nu_{\rm LL}}}{10^{30}\,{\rm erg\,s^{-1}\,Hz^{-1}}}\right)\cgssb\nonumber. 
\eea
In the expression above it is understood that $f_{\rm C}$ is capped at $f_{\rm C} = 1$, 
corresponding to the case where all ionizing photons are absorbed. 

Thus in the optically thick limit, the emission depends only weakly on
the amount of cool gas through the covering factor of the clouds, and
scales linearly with the ionizing photon flux. This
contrasts strongly with the highly ionized optically thin limit, where
the surface brightness depends on the amount of cold gas
through the combination $f_{\rm C}n_{\rm H}N_{\rm H}$, but not
on the intensity of the ionizing source.

\subsection{\mlya\ Emission from Scattering}
\label{sec:scatt}

In this subsection we compute the surface brightness of extended
\mlya\ emission, produced via resonant scattering of a central source
of \mlya\ photons (i.e. the quasar) by the neutral gas in the CGM. By
analogy with the optically thick fluorescence, the scattering surface
brightness can be written as \be {\rm
  SB}_{\lya}=\frac{h\nu_{\lya}}{4\pi(1+z)^4}f_{\rm
  C}\Phi_{\lya}(R\slash \sqrt{3})
\label{eqn:scatt0}
\ee
where $f_{\rm C}$ now represents the covering factor of gas which 
is optically thick \emph{in the \mlya\, transition} (i.e. $N_{\rm HI} \gtrsim 10^{14}\,{\rm cm^{-2}}$). This will in general be much higher 
than the covering factor of clouds optically thick at the Lyman limit considered
in \S\ref{sec:recomb_thick}. The quantity $\Phi_{\lya}$ is the 
flux of ionizing photons emitted close
enough to the \mlya\ resonance to be scattered by gas in motion around the quasar
\be 
\Phi_{\lya} \approx  \frac{L_{\nu_{\lya}}}{4\pi r^2h}\frac{\Delta\nu}{\nu_{\lya}} = 
\frac{L_{\nu_{\lya}}}{4\pi r^2h}\frac{W_{\lya}}{\lambda_{\lya}}, 
\ee
where we have introduced the rest-frame absorption equivalent width $W_{\lya}$, and
implicitly assumed that it is a very weak function of radius, as we demonstrate
next. 

The volume averaged neutral column density through our gas distribution is 
$\langle N_{\rm HI}\rangle = f_{\rm C}N_{\rm H}\langle x_{\rm HI}\rangle$, 
where 
$\langle x_{\rm HI}\rangle$ is the volume averaged neutral fraction 
determined from photoionization equilibrium (eqn.~\ref{eqn:ioneq}). These
relations imply an average \mlya\ optical depth
\bea 
\langle \tau_{\lya}\rangle &=& 253\left(\frac{N_{\rm
    H}}{10^{20}\,{\rm cm^{-2}}}\right)\!\left(\frac{n_{\rm H}}{0.1\,{\rm
    cm^{-3}}}\right)\!\left(\frac{b}{50\,\kms}\right)^{-1}\nonumber\\ 
&\times & \left(\frac{f_{\rm C}}{1.0}\right)\!\!\left(\frac{R}{100\,{\rm
    kpc}}\right)^{2}\!\!\left(\frac{L_{\nu_{\rm LL}}}{10^{30}\,{\rm
    erg\,s^{-1}\,Hz^{-1}}}\right)^{-1}\label{eqn:taulya} 
\eea
where we have introduced the $b$-value describing turbulent and thermal motions
of the absorbing gas. On the flat portion of the curve of 
growth $1\lesssim \tau_{\lya} \lesssim 10^{4}$, the corresponding equivalent width 
is \citep[see e.g.][]{draine11}
\be W_{\lya} \approx 0.99 \left(\frac{b}{50\,\kms}\right)\left(\frac{\sqrt{\ln\left(\langle\tau_{\lya}\rangle\slash
  \ln 2\right)}}{2.1}\right){\rm \AA}\label{eqn:ewlya}, 
\ee
so that equivalent widths of $W_{\lya} \simeq 1$\AA\ are expected. We 
can  finally write 
\bea
{\rm SB}_{\lya} &=& 1.2\times 10^{-18}\left(\frac{1+z}{3.0}\right)^{-4}\!\!
\left(\frac{f_{\rm C}}{1.0}\right)\!\!\left(\frac{R}{100\,{\rm kpc}}\right)^{-2}\nonumber\\
&\times& \left(\frac{W_{\lya}}{1.0\,{\rm \AA}}\right)
\left(\frac{L_{\nu_{\lya}}}{10^{31}\,{\rm erg\,s^{-1}\,Hz^{-1}}}\right)
\label{eqn:SB_scatt}\\
& \times & 
\cgssb\nonumber, 
\eea
where it is again understood that $f_{\rm C} < 1$ in eqn.~(\ref{eqn:SB_scatt}).

Because equivalent width is such a weak function of optical depth on
the saturated part of the curve of growth (eqn.~\ref{eqn:ewlya}),
$W_{\lya}$ is approximately independent of the ionizing luminosity of
the quasar as well as the details of the gas distribution. Thus the
scattering surface brightness scales as ${\rm SB}_{\lya} \propto
f_{\rm C}\Phi_{\lya}$, completely analogous to optically thick
recombinations for which ${\rm SB}_{\lya} \propto f_{\rm C}\Phi$
(eqn.~\ref{eqn:SB_thick}).  However, scattering will typically be
about $\sim 100$ times fainter than optically thick recombinations if
the covering factors are the same because $\eta_{\rm thick}\Phi\slash
\Phi_{\lya}\sim 100$, which results from the assumption of $W_{\lya}
\simeq 1.0\,{\rm \AA}$, and from the relative amount of ionizing
photons to \mlya\ photons emitted by a typical quasar. 
One expects this scattering emission to be important, therefore, only in
regimes where the gas is optically thin at the Lyman limit.
Indeed, the surface brightness of scattering (eqn.~\ref{eqn:SB_scatt}) and
optically thin fluorescence (eqn.~\ref{eqn:SB_thin}) can be
comparable, but have very different dependencies on gas distribution
and quasar luminosity. Throughout this
work we estimate $L_{\nu_{\lya}}$ by tying the composite quasar
spectrum of \citep{vanden01} to the SDSS optical photometry (covering
rest-frame UV wavelengths) and computing the peak of the
\mlya\ emission line.

\subsection{An Illustrative Example}
\label{sec:example}

There are two important points that can be made about the surface
brightness of extended \mlya\ emission, and its implications for the
supply of cool gas.  First, optically thin and optically thick
fluorescence can result in comparable levels of emission, as can pure
resonant scattering of a central source of \mlya\ photons. To
complicate matters further, cooling radiation, that is collisional
excitation of \mlya, can also lead to comparable emission.  In
general, the only way to determine which mechanism is at work is to
use diagnostics from additional emission lines.  For instance, the
detection of strong extended Balmer line emission would be provide
compelling evidence for recombinations, and effectively rule out
scattering and cooling. But given that Balmer lines are redshifted
into the near-IR for $z > 2$, detecting even H$\alpha$ would be a
daunting observational task because of the low expected surface
brightness.  Furthermore, even if one is certain that recombination is
the physical mechanism, it is still typically impossible to ascertain
the ionization state of the gas, and hence whether it is optically
thin or thick\footnote{For extreme ratios of ${\rm SB}_{\lya}$ to 
\lnull, it is possible to rule out optically thin clouds via
eqn.~(\ref{eqn:NHI}).  However, this requires that the ionizing flux
    can be measured, which is possible for quasars but not HzRGs or
    LABs.}.

The second and related point is that measurements of the
\mlya\ surface brightness alone cannot be used to determine the cool
gas mass.  For recombinations, the surface brightness scales as ${\rm
  SB}_{\lya}\propto f_{\rm C}n_{\rm H}N_{\rm H}$ if the gas is
optically thin (eqn.~\ref{eqn:SB_thin}), or as ${\rm SB}_{\lya}
\propto \Phi f_{\rm C}$ if it is optically thick
(eqn.~\ref{eqn:SB_thick}). Whereas, for scattering the scaling is
${\rm SB}_{\lya} \propto \Phi_{\lya} f_{\rm C}$
(eqn.~\ref{eqn:SB_scatt}). The total cool gas mass scales with the
different combination $M_{\rm c} \propto f_{\rm C}N_{\rm H}$
(eqn.~\ref{eqn:mc}). Even if we have perfect knowledge of the
mechanism (recombinations versus scattering) and the ionization state
(optically thin or optically thick), the cool gas mass is not uniquely
specified by the \mlya\ emission alone. Note that the situation is
even worse for \mlya\ cooling radiation because of an exponential
dependence on an unknown gas temperature \citep{lyacool}.
Thus even in the best-case scenario where Balmer line observations point to 
recombinations, one will not know whether the gas is optically thin or
thick, and hence which dependence on physical parameters to use.

The following worked example illustrates the aforementioned
issues. Consider a relatively bright $z=2$ quasar with apparent
magnitude of $i = 17.0$. Using the procedure in Appendix A of
\cite{QPQ1}, we determine the specific luminosity 
to be, $\log_{10} \mlnull = 31.3$ at the Lyman limit, and $\log_{10} \mllya
= 32.0$ in \mlya.  Now consider three distinct cool gas
distributions, one optically thin to ionizing photons with parameters
($R,n_{\rm H},N_{\rm H}, f_{\rm C}$) = ($100\,{\rm
  kpc}$, $0.15\,{\rm cm^{-3}}$, $10^{21}\,{\rm cm^{-2}}$, 1.0), another
optically thick to ionizing photons with ($R,n_{\rm H},N_{\rm
  H}, f_{\rm C}$) = ($100\,{\rm kpc}$, $100\,{\rm
  cm^{-3}}$, $10^{20}\,{\rm cm^{-2}}$, 0.01), and a third
optically thin to ionizing photons but optically thick to
\mlya\ photons with ($R,n_{\rm H},N_{\rm H}, f_{\rm C}$) =
($100\,{\rm kpc}$, $30\,{\rm cm^{-3}}$, $10^{17}\,{\rm
  cm^{-2}}$, 2.0). These very different cool gas distributions result
in nearly identical average \mlya\ surface brightness ${\rm
  SB}_{\lya}\approx 1.1-1.2\sci{-17}\,\cgssb$ over an $R=100\,{\rm
  kpc}$ halo. For the first two models the emission is powered by
photoionization, whereas for the third it is pure
\mlya\ scattering. But note also that for the first distribution,
\mlya\ scattering contributes at very nearly the same surface
brightness level $\sim 1.1\sci{-17}\,\cgssb$ as optically thin
photoionization, implying that disentangling the two mechanisms via
say detection of Balmer lines would be challenging.  These sizes and
surface brightness levels are in line with the brightest \mlya\ blobs
surveyed by \citet{matsuda04,matsuda11} (rescaled to $z=2$) as well as
the extended \mlya\ nebulae detected around $z\sim 2$ HzRGs
\citep[e.g.][]{villar07} and radio-loud quasars 
\citep[e.g.][]{heckman91a}. For the first optically thin model the
total cool gas mass is $M_{\rm c}= 3.3\sci{11}\msol$, for the second
optically thick model it is $M_{\rm c}= 2.5\sci{8}\msol$, and for the
\mlya\ scattering case it is $M_{\rm c}= 6.6\sci{7}\msol$. Thus
variations of $\sim 5000$ in the total gas mass can all lead to
comparable levels of extended \mlya\ emission, and even if one were
sure that photoionization was powering the emission (from Balmer
lines) and/or detected the source of ionizing photons (an unobscured
quasar), there is no way to determine whether the gas is optically
thin or thick. Plugging the observed observed surface brightness ${\rm
  SB}_{\lya}$, distance $R$, and ionizing luminosity $\mlnull$ into
eqn.~(\ref{eqn:NHI}) for the area averaged neutral column gives
$\langle N_{\rm HI}\rangle = 1.3\sci{16}\,{\rm cm}^{-2}$, and thus
we cannot distinguish an optically thin from an optically thick scenario. 

\subsection{Why Quasar Pairs?}
\label{sec:cartoon}

\begin{figure*}
  \centering{\epsfig{file=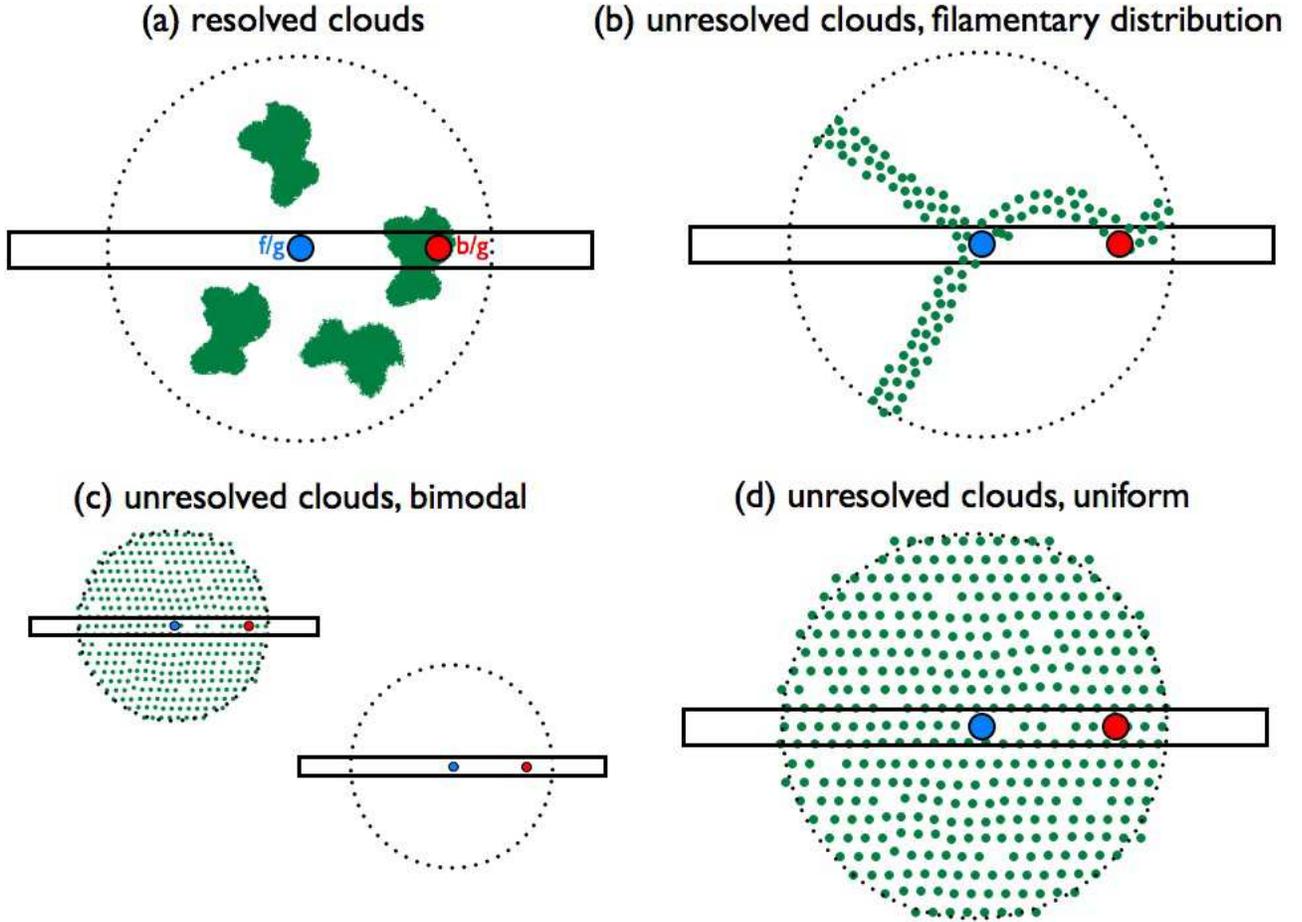,bb=0 0 1024 768,width=\textwidth,clip=}}
  \caption{Four scenarios for the spatial distribution of CGM gas in
    quasar halos which illustrate the usefulness of projected quasar
    pair observations. \emph{a)} The absorbers/emitters are large
    resolved clouds at random locations in the quasar CGM.  The
    background sightlines pinpoint the locations of optically thick
    clouds and thereby determine where to look for emission. For such
    large spatially resolved clouds, the relevant SB estimate is the
    mirror approximation (see eqn.~\ref{eqn:SB_mirror}).  \emph{b)} The
    clouds are small and hence spatially unresolved by our instrument,
    but gas resides in large resolved filamentary complexes which
    cover a fraction of the quasar halo. For longslit observations of
    quasar pairs, the b/g sightline intersects the filament by
    construction, again pinpointing where emission is expected.  The
    average SBs that we computed in \S\ref{sec:form} underestimate
    the true SB of the filament, since they include a reduction by the
    covering factor $f_{\rm C}$, due to regions devoid of
    gas. \emph{c)} The clouds are small and unresolved, but some
    quasars have cool gas complexes and others do not, and this
    bimodality in the gas supply is responsible for the variations
    that result in the observed covering factor. Quasar pairs are
    helpful because the background sightline identifies objects with
    large gas supply and hence where to search for emission.  The
    average SBs that we computed in \S\ref{sec:form} are again
    underestimated because the covering factor includes objects with
    no gas. \emph{d)} The clouds are small, unresolved, and
    distributed uniformly in the quasar halo. The background sightline 
    does not provide useful information about where to search for emission.  The
    average SB estimates in \S\ref{sec:form} appropriately account 
    for the dilution by the covering factor $f_{\rm C}$, and this is the surface brightness
    that we expect to detect at any spatial position along the slit. For
    this case, a search for diffuse emission around isolated individual quasars, i.e. 
    not residing in pairs, would be just as effective.\label{fig:cartoon}}
\end{figure*}

In the previous subsection we saw that a variety of physical processes
and a wide range of gas distributions can all lead to comparable
\mlya\ emission. Hence, the great value of quasar pairs is that
absorption-line observations of b/g quasars can provide crucial
information about the physical properties of the gas in the CGM of the
f/g quasar. Indeed, all the physical parameters ($R\,,n_{\rm
  H}\,,N_{\rm H}\,, f_{\rm C}$) can be determined from absorption line
observations, provided that the covering factor is high enough for 
the clouds to be occasionally intercepted by b/g sightlines. For
example, we map out the covering factor of optically thick absorption
as a function of impact parameter (see Figure~\ref{fig:cov_fact}) in
\S\ref{sec:cov_fact} \citep[see also][]{QPQ1,QPQ2,QPQ5}, and find a
high covering factor of cool, optically thick gas around all bright
quasars at $z \sim 2$. In \citet{QPQ3}, we conducted detailed
absorption line modeling of a single projected pair and determined
that the column densities of the absorbing clouds are typically $N_{\rm
  H}\simeq 10^{20}\,{\rm cm}^{-2}$ and volume densities $n_{\rm
  H}\simeq 0.1-1\,{\rm cm}^{-3}$, although it is unclear whether this
single system is representative. Armed with this knowledge of the gas
distribution, one can then interpret the physical implications of the
presence or absence of extended line emission.

Note however that the properties of gas with a very low covering
factor $f_{\rm C}\sim 0.01$ are much more challenging to measure,
since thousands of background sources would be required to map
it out. Thus a crucial caveat in comparing absorption line observations
to emission line observations is that the \emph{gas responsible for
  the detected emission may arise from clouds with a much lower covering factor than
  the gas dominating the absorption}. Nevertheless, one can still make
quantitative statements about the expected emission from the component
of gas which is detected in absorption, which will have the highest covering
factor.  This is the approach adopted in this work. 

But even if quasar pairs are useful for measuring the properties of
the quasar CGM, why use quasar pairs, as opposed to individual
isolated quasars, to searching for diffuse emission lines?  Consider
the following four scenarios which are illustrated schematically in
Figure~\ref{fig:cartoon}. a) The absorbers/emitters are large resolved
clouds at a `random' location in the quasar CGM
(Figure~\ref{fig:cartoon}a).  In this case, the background sightlines
pinpoint the locations of gas and thereby determine where to look for
emission. This approach is akin to that taken by \cite{ask+06}, who
purported to detect quasar fluorescence from an optically thick absorber
detected in a b/g sightline. For this case of large resolved
absorbers, the relevant SB estimate is the mirror approximation (see
eqn.~\ref{eqn:SB_mirror}).  b) Another possibility is that the clouds
are small and hence unresolved by our instrument, but that the gas
resides in large resolved filamentary complexes which cover a fraction
of the quasar halo (Figure~\ref{fig:cartoon}b). For this case, the b/g
sightline intersects a filament by construction, and identifies the
location where emission is expected via longslit observations of the
quasar pair. Note that the average SBs computed in
\S\ref{sec:form} would underestimate the true SB in this case, since they include a
reduction by the covering factor $f_{\rm C}$, due to regions devoid of gas. 
c) It could also be that the clouds are small and hence unresolved, but some quasars have cool gas 
complexes and others do not. This bimodality in gas supply could result in
the `hits' and 'misses' which give rise to the observed covering
factor (Figure~\ref{fig:cartoon}c). In this case, the quasar
pairs are helpful since the background sightline identifies which
objects to search for emission around, and the average SBs that we
computed in \S\ref{sec:form} are again underestimated because the 
covering factor includes objects with no gas. d) Finally, the clouds could be small
and hence unresolved, and distributed rather uniformly in the quasar halo
(Figure~\ref{fig:cartoon}d).  In this case the background sightline 
does not provide us with useful information about where to search for emission.  The
average SB appropriately accounts for the dilution by
the covering factor $f_{\rm C}$, and this is the SB
that we expect to detect at any spatial position along the slit. For
this case, a search for diffuse emission around isolated individual quasars, i.e. 
not residing in pairs, would be just as effective.

\section{Quasar Pair Observations}
\label{sec:observe}

Our goal is to conduct a sensitive search for extended \mlya\ emission
from the CGM and IGM around f/g quasars in quasar pairs, and to relate
these emission constraints to our knowledge of the distribution of gas
from absorption-line measurements of b/g sightlines. We concentrate on
a unique sample of projected quasar pairs useful for exploring the
CGM/IGM surrounding high-$z$ quasars and, by inference, the massive
galaxies and dark matter halos which host them. We search for extended
\mlya\ emission around 29 $z\sim 2$ quasars, which are drawn from a
parent sample of $68$ projected quasar pairs.  Before delving into
details, we briefly summarize the criteria used to select these 29
quasar pairs:
\begin{itemize}
\item the CGM/IGM around the f/g quasar exhibits strong and optically thick
  \ion{H}{1} absorption in the b/g quasar sightline. 
\item the transverse separation of the pair and magnitude of the f/g
  quasar imply a fluorescent 
  mirror surface brightness 
  ${\rm SB}_{\lya} > 5\sci{-18}\cgssb$ (see eqn.~\ref{eqn:SB_mirror}). 
\item a deep spectroscopic observation of the f/g
  quasar was obtained covering \mlya\ , allowing us to perform a sensitive search for
  \mlya\ emission on scales of tens to hundreds of kpc.
\item the two quasars are separated by velocity $> 2000\,\kms$, to
  ensure that they are indeed projected and not physically associated.
\end{itemize}
Although our search for extended \mlya\ emission is focused on
sightlines showing strong absorbers, we characterize the covering
factor of optically thick absorbers using the full unbiased parent
sample of 68 projected quasar pairs.  This analysis demonstrates that
the CGM surrounding $z \sim 2$ quasars has a high $> 50\%$ covering
fraction of optically thick gas to radius $R \approx 200$\,kpc
\citep[see also][]{QPQ1,QPQ2,QPQ5}. 

There are several reasons why we restrict our search for emission to
only pair sightlines showing strong absorbers. 
First, in the context of fluorescence from the IGM,
powered either by the UVB or boosted by a nearby quasar, emission from
optically thick clouds will always dominate over optically thin gas
because the density of gas in the optically thin \mlya\ forest is just
too low \citep[e.g.][see also eqn.~\ref{eqn:SB_thin}]{GW96}. Although
this may not hold true in the CGM, where gas densities can be higher,
note that the the mirror SB from a spatially resolved optically thick
cloud is independent of the gas properties (eqn.~\ref{eqn:SB_mirror}),
and the same holds for emission from a population of optically thick
clouds provided the covering factor has been measured
(eqn.~\ref{eqn:SB_thick}). In contrast, optically thin emission
depends on the combination $f_{\rm C}n_{\rm H}N_{\rm H}$ which is
typically unknown. The upshot is that a non-detection of emission from
gas which is optically thin could result either from $f_{\rm
  C}n_{\rm H}N_{\rm H}$ being too low, or because the gas is not
illuminated by the quasar. Whereas, interpretation of
non-detection of emission from optically thick gas is unambiguous, and implies
that the gas is not illuminated by the quasar\footnote{Another
  possibility is extinction of \mlya\ by dust, but we show in Appendix
  A that this effect is not significant.}. 

Note however, that given the high covering factor of
optically thick gas in the quasar CGM, statistically, there is a high
probability that any long-slit observation of a projected quasar pair
will intersect optically thick material, even if the b/g sightline
does not show absorption. We refer the reader to the discussion in
\S\ref{sec:cartoon} and Figure~\ref{fig:cartoon}, which illustrates,
nevertheless, why having a b/g sightline showing optically thick
absorption can be advantageous. 

In the following subsections we describe the details of the quasar pair survey, 
the spectroscopic observations, the definition of the quasar-absorber sample, 
and present our measurement of the covering factor of optically thick absorbers
from the parent sample. 

\subsection{Quasar Pair Survey}
\label{sec: sample}

Modern spectroscopic surveys select \emph{against} close pairs of
quasars because of fiber collisions.  For the Sloan Digital Sky Survey
(SDSS), the finite size of optical fibers precludes discovery of pairs
with separation $<55\arcsec$ \footnote{An exception to this rule
  exists for a fraction ($\sim 30\%$) of the area of the SDSS
  spectroscopic survey covered by overlapping plates.  Because the
  same area of sky was observed spectroscopically on more than one
  occasion, the effects of fiber collisions are reduced.} 
Thus, to find pairs with sub-arcminute separations, additional follow-up
spectroscopy is required both to spectroscopically confirm quasar pair
candidates, and to obtain spectra of sufficient quality to search for
absorption line systems.

In an ongoing survey, close quasar pair candidates are selected from a
photometric quasar catalog \citep{bovy11,bovy12}, and are confirmed
via spectroscopy on 4m class telescopes including: the 3.5m telescope
at Apache Point Observatory (APO), the Mayall 4m telescope at Kitt
Peak National Observatory (KPNO), the Multiple Mirror 6.5m Telescope,
and the Calar Alto Observatory (CAHA) 3.5m telescope. Our continuing
effort to discover quasar pairs is described in \citet{thesis},
\citet{BINARY}, and \citet{HIZBIN}. To date about 350 pairs of quasars
have been uncovered with impact parameter $R_{\perp} < 300~{\rm kpc}$
and $z_{\rm fg} > 1.6$\footnote{The lower limit on redshift is
  motivated by the ability to detect redshifted \mlya\ absorption
  above the atmospheric cutoff $\lambda > 3200$~\AA.}.  Projected pair
sightlines were then observed with Keck and Gemini to obtain `science
quality' high signal-to-noise ratio (S/N) moderate resolution
spectra. A subset of higher resolution spectra at echellette and
echelle resolution have also been obtained, but these are not used in
this work.  This observational program has several science goals:
measure the small-scale quasar clustering of quasars
\citep{BINARY,myers08,HIZBIN,Shen09BIN}, to measure small scale
transverse Ly$\alpha$ forest correlations (Rorai et al.\ in prep), to
characterize the transverse proximity effect (Hennawi et al., in
prep.), and to use the b/g sightline to characterize the
circumgalactic medium of the f/g quasar \citep{QPQ1,QPQ2,QPQ3,QPQ5},
which is relevant to the goals of this paper.

\subsection{Spectroscopic Observations}
\label{sec:spec}

High S/N ratio, moderate resolution slit spectra of $\sim 100$ 
quasars were obtained with Keck and Gemini in observing runs spanning
from 2004 until 2008.  All quasars observed have $z > 1.6$, which is
the lower limit for detecting Ly$\alpha$ set by the atmospheric
cutoff.  About two-thirds of the targeted pairs consist of projected
pairs of quasars ($\Delta v > 2500\kms$) at different redshifts; the
rest were physically associated binary quasars.

For the Keck observations, we used the Low Resolution Imaging
Spectrograph \citep[LRIS;][]{LRIS}, either in longslit mode targeting
only the two quasars in the pair, or in multi-slit mode with custom
designed slitmasks, which allowed placement of slits on other known
quasars or quasar candidates in the field.  Typically the slitmask or
slit was rotated such that both quasars in the close pair could be
observed on the same slit. LRIS is a double spectrograph with two arms
giving simultaneous coverage of the near-UV and red.  We used the D460
dichroic with the $1200$ lines mm$^{-1}$ grism blazed at
$3400$~\AA\ on the blue side, resulting in wavelength coverage of
$\approx 3300-4200$~\AA. The dispersion of this grism is
$0.50$~\AA\ per pixel and our $1''$ slits give a resolution of
FWHM$\simeq 150\kms$.  These data provide the coverage of \mlya\ at
$z_{\rm fg}$ for all of our pairs. On the red side we typically used
the R600/7500 or R600/10000 gratings with a tilt chosen to cover the
\ion{Mg}{2} emission line at the f/g quasar redshift, useful for
determining accurate systemic redshifts of the quasars.  Occasionally
the R1200/5000 grating was also used to give additional bluer
wavelength coverage.  The higher dispersion, better sensitivity, and
extended coverage in the red provided high signal-to-noise ratio
spectra of the \ion{Mg}{2} emission line and also enabled a more
sensitive search for metal-line absorption in the b/g quasar (see
Prochaska et al.\ in prep.).  Some of our older data also used the
lower-resolution 300/5000 grating on the red-side covering the
wavelength range $4700-10,000$~\AA. About half of our LRIS
observations were taken after the atmospheric dispersion corrector was installed, 
which reduced slit-losses (for point sources) in the UV.

The Gemini data were taken with the Gemini Multi-Object Spectrograph
\citep[GMOS;][]{GMOS} on the Gemini North facility.  We used the
B$1200\_G5301$ grating which has 1200 lines mm$^{-1}$ and is blazed at
5300~\AA. The detector was binned in the spectral direction resulting
in a pixel size of 0.47~\AA, and the $1\arcsec$ slit corresponds to a
FWHM~$\simeq 125\kms$.  The slit was rotated so that both quasars in a
pair could be observed simultaneously.  The wavelength center depended
on the redshift of the quasar pair being observed. But most of the the
targets considered here are at $z \sim 2.2-2.5$, so the grating was
typically centered at 4500~\AA, giving coverage from
$3750-5225$~\AA. The Gemini CCD has two gaps in the spectral
direction, corresponding to 9~\AA~ at our resolution. The wavelength
center was thus dithered by 15-50\AA\ between exposures to obtain full
wavelength coverage in the gaps. The Gemini observations were
conducted over three classical runs during UT 2004 April 21-23, UV
2004 November 16-18, and UT 2005 March 13-16 (Program IDs:
GN-2004A-C-5, GN-2004B-C-4, GN-2005A-C-9, GN-2005A-DD-4). 

Total exposure times ranged from $600-21800$s, for the Keck and Gemini
observations, depending on the magnitudes of the targets.  This was
typically broken up into several individual exposures with exposure
times of 600-1800s depending on the total planned exposure time.  The
motivation for the shorter (total) integrations was to conduct a
survey to build up a statistical sample of b/g absorption line spectra
at small impact parameter from the f/g quasars. The goal of the longer
integrations was to obtain high-quality spectra of a subset of quasar
pairs and to search for low SB \mlya\ emission near the f/g
quasar. The S/N ratio in Ly$\alpha$ forest region of the b/g quasar
varies considerably, but it is almost always S/N$>5$ per pixel for the
data we consider here.

\subsection{Defining the Quasar-Absorber Sample}
\label{sec:sample}

\begin{figure*}
  \hspace*{3.0cm} 
  \centering{\epsfig{file=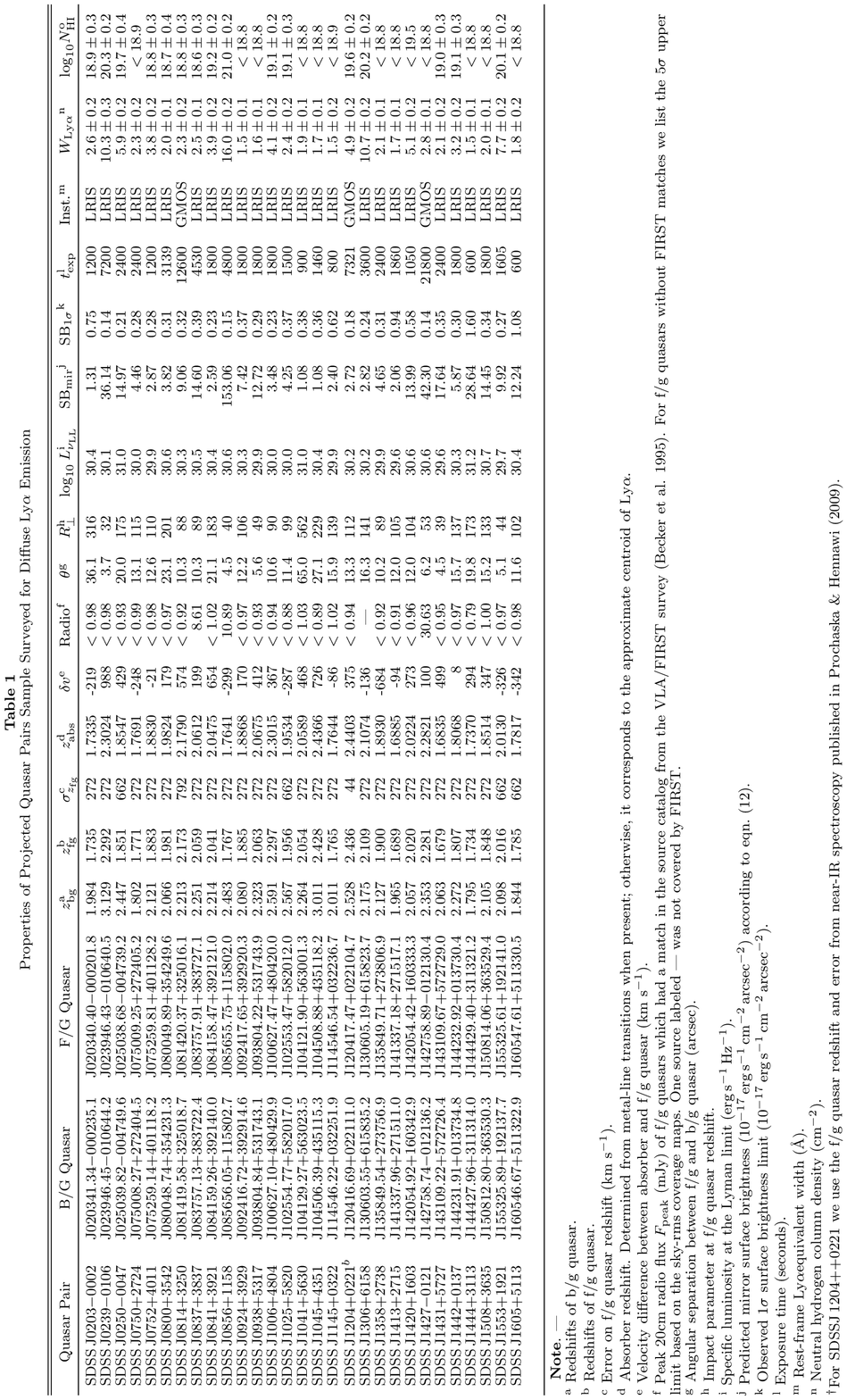,bb=70 0 612 792,width=\textwidth}}
\end{figure*}

In QPQ5 and \citet{QPQ6} (QPQ6) we analyze the statistical properties
of the CGM near quasars using our full dataset of `science quality'
projected-pair spectra, which combines spectra from the SDSS \citep{DR3} and BOSS
\citep{BOSS} spectroscopic surveys with higher quality data from
8m-class telescopes. Here we consider a subset of this full dataset
constituting 68 sightlines, and survey the spectra for optically thick
absorption coincident with the f/g quasar redshift. 
We first describe the additional criteria applied to the full dataset
to arrive at the 68 sightline parent sample, and then summarize the
procedure used in \citet{QPQ5} to identify optically thick absorbers
coincident with the f/g quasars.

The present study concerns a sensitive search for extended
\mlya\ emission, which may coincide with the location of an optically
thick absorber detected in a neighboring background sightline.
To this end, we restrict our parent sample to only the
subset of quasars which have an implied ${\rm SB}_{\lya} >
5\sci{-18}\cgssb$ given by the mirror approximation in
eqn.~(\ref{eqn:SB_mirror}), where the radius $R$ is set to be the impact
parameter $R_\perp$ of the projected quasar pair.  This surface brightness
limit is chosen because it approximately matches the median 2$\sigma$ surface
brightness limit that our longslit spectroscopic observations achieve
(see the next section for detailed discussion of surface brightness
limits). 

The only other criteria imposed are that we must have obtained a
spectroscopic observation of the quasar pair with Gemini or Keck
covering \mlya\ at the f/g quasar redshift, and we require that the
velocity difference between the f/g and b/g quasar be $> 2000\,\kms$
to ensure that the two quasar are not physically associated. This
minimum velocity difference is many times larger than both the
expected line-of-sight velocity dispersion in a $z\sim 2$ quasar
environments $\sim 300\,\kms$, and the typical redshift errors $\sim
270-660\,\kms$ (see below). We further culled the sample by excluding
pairs whose b/g quasar exhibits strong broad absorption lines (BAL),
as evidenced by large \ion{C}{4} or \mlya\ equivalent widths (EWs).
Mild BALs were also excluded if the BAL absorption clearly coincided
with the velocity window about \mlya\ at the f/g quasar redshift.

The remaining quasar pairs were visually inspected and searched for
significant Ly$\alpha$ absorption within a velocity window of $|\Delta
v| = 1500\kms$ about the f/g quasar redshift.  This velocity
window is chosen to accommodate physically associated absorbers with
extreme kinematics $\simeq 1000\kms$, as well as to bracket
uncertainties in the f/g quasar systemic redshifts. Quasar  
redshifts determined from the rest-frame ultraviolet emission lines
(redshifted into the optical at ${\rm z}\sim 2$) can differ by up to
one thousand kilometers per second from the systemic frame, because of
outflowing/inflowing material in the broad line regions of quasars
\citep{gaskell82,TF92,vanden01,Richards02,Shen08}.  Systemic redshifts
are estimated by combining the line-centering procedure used in QPQ1
with the recipe in \cite{shen07} for combining measurements from
different emission lines. The resulting typical 1$\sigma$ redshift
uncertainties using this technique are in the range $\sigma_{\rm z}
\simeq 270-770\kms$
depending on which emission lines are used. 
This analysis considered all available spectra on the f/g quasar
including the public SDSS and BOSS datasets.  We refer the reader to
QPQ6 for further details.

The Ly$\alpha$ transition saturates at $N_{\rm HI} \simeq
10^{14}\,{\rm cm}^{-2}$, and between column densities of $N_{\rm HI}
\simeq 10^{14-19}\,{\rm cm}^{-2}$ the curve of growth is flat, and
$W_{\lya}$ is a poor proxy for hydrogen column density. In principle,
additional leverage may be provided by simultaneously fitting higher
order Lyman transitions, or the detection of Lyman limit absorption,
which requires $N_{\rm HI} > 10^{16.5}\,{\rm cm}^{-2}$. However,
nearly all of our sightlines are at redshifts $z < 2.6$, for which the
Lyman limit is below the atmospheric cutoff\footnote{In fact, we
  rarely even cover \mlyb.}, and so constructing a complete and fully
certifiable sample of optically thick absorbers with $N_{\rm HI} >
10^{17.2}\,{\rm cm}^{-2}$ is extremely challenging given our
wavelength coverage.  These challenges are exacerbated by noise in the
spectra and the line blending which inevitably occurs at moderate
resolution (${\rm FWHM} \simeq 150\kms$).

We identified the strongest absorption line in the $\pm 1500\kms$
velocity window about the f/g quasar redshift, and measured the Ly$\alpha$
equivalent width $W_{\lya}$. All systems with $W_{\lya} > 1$~\AA\ were
flagged for Voigt profile fitting of the Ly$\alpha$ transition, to estimate
the \ion{H}{1} column density. For reference, a Ly$\alpha$ equivalent
threshold of $W_{\lya} > 1.5$~\AA\ ($W_{\lya} > 2$~\AA) corresponds 
to column densities of roughly $\mnhi \gtrsim 10^{17}\,{\rm cm}^{-2}$ ($\mnhi
\gtrsim 10^{19}\,{\rm cm}^{-2}$) if the absorption is dominated by a single component; but
this need not be the case.

\begin{figure}[t]
  \centerline{\epsfig{file=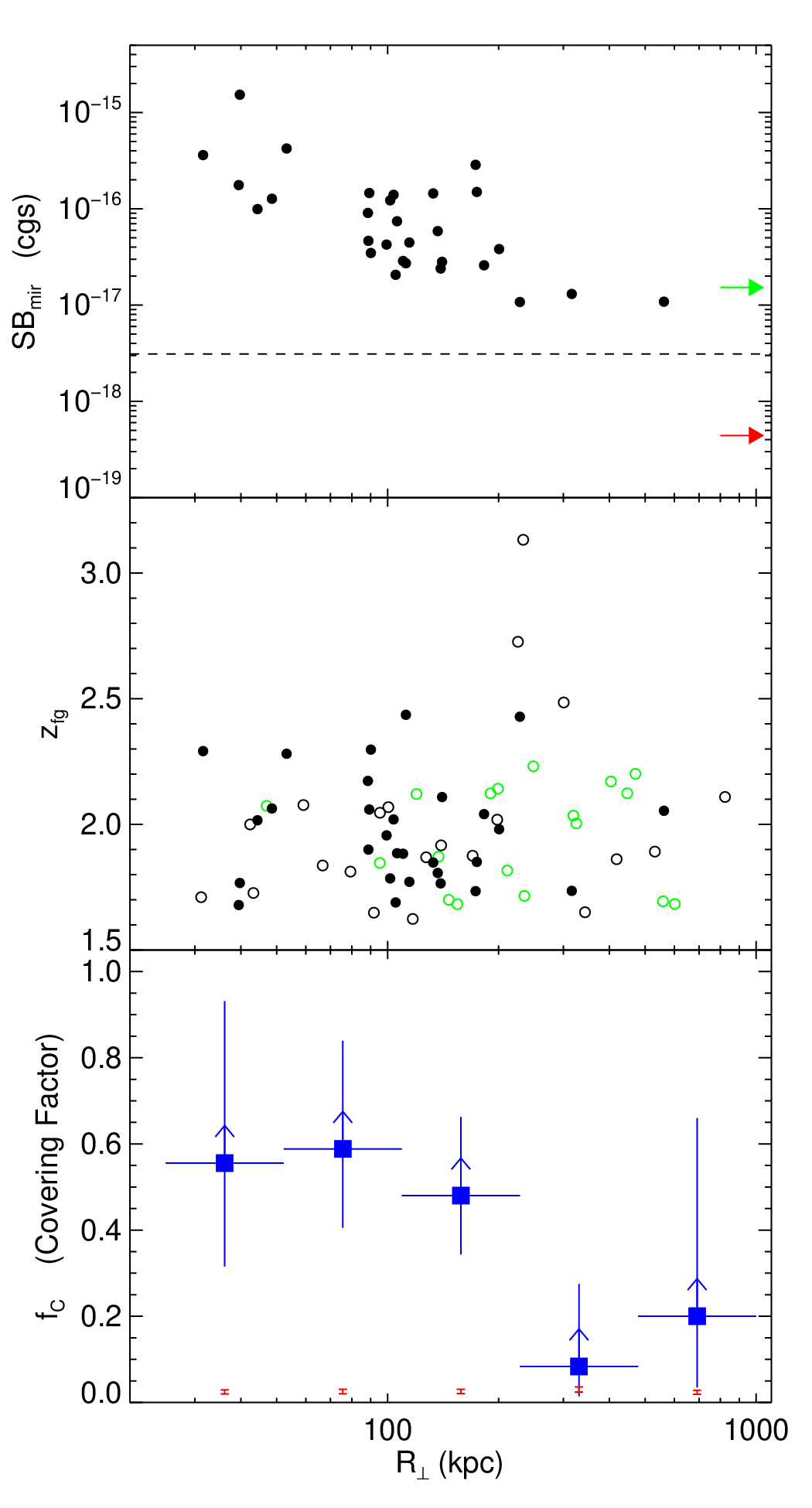,bb=72 72 576 965,width=0.46\textwidth}}
  \vskip -0.15in
  \caption{Properties of projected quasar pair sample. \emph{Middle:}
    Distribution of impact parameter and f/g quasar
    redshift for the 68 projected pairs in our
    our parent sample. Filled blue circles are 29 sightlines with an
    optically thick absorber within 
    $|\Delta v| = 1500\kms$ of the f/g quasar redshift,
    which we search for \mlya\ emission (see
    Table~1).  The black open circles are sightlines
    which are ambiguous, and the green open circles are sightlines
    classified as optically thin. \emph{Bottom:} Blue squares show the
    covering factor of optically thick absorbers in 
    logarithmically spaced bins. Vertical error bars are Poisson. 
    Because
    some of the ambiguous points (black open circles in middle panel)
    may also be optically thick, our covering factor is a conservative
    lower-limit, as denoted by arrows on the blue points. The lower
    red error bars indicate the expected covering factor within a
    $|\Delta v| = 1500\kms$ interval from the background
    abundance of LLSs \citep{ribaudo11}, with the range given
    by the current precision of these measurements. The
    covering factor of optically thick absorbers around quasars is
    high $>50\%$ for scales $R_{\perp} < 200\,\kpc$, and
    represents a significant enhancement over the 
    cosmological background. \emph{Top:} Distribution of predicted
    mirror ${\rm SB}_{\rm mir}$ (eqn.~(\ref{eqn:SB_mirror}) for the 29
    quasars searched for emission. Note the general $R_{\perp}^{-2}$
    trend.  The dashed line is the median value of the $1\sigma$ SB
    limits, ${\rm SB}_{1\sigma}$ of our sample. The red arrow
    indicates the effective $1\sigma$ SB limit of the \citet{rauch08}
    92 hour integration, where their values are rescaled from $z\simeq
    3.1$ to the median redshift of our sample $z=2$ (i.e. $(1+z)^4$
    dimming) and also rescaled to match our $700\,\kms$ velocity
    aperture. The green arrow indicates the average SB threshold given
    the \citet{matsuda04} definition of a LAB ($f_{\lya} >
    0.7\sci{-16}\cgsflux$ over a 16 arcsec$^2$ isophote) also
    rescaled to $z=2$.
    \label{fig:cov_fact}}
\end{figure}

The \ion{H}{1} analysis was complemented by a search for metal-lines
at the f/g quasar redshift in the clean continuum region redward of
the Ly$\alpha$ forest of the b/g quasar.  The narrow metal-lines
provide a more accurate redshift for the absorption line system (than
obtainable from \mlya) and, if present, they can help distinguish
optically thick absorbers from blended Ly$\alpha$ forest lines.  We
focused on the strongest low-ion transitions commonly observed in DLAs
\citep[e.g.][]{pro01}: \ion{Si}{2}~$\lambda 1260, 1304, 1526$,
\ion{O}{1}~$\lambda 1302$, \ion{C}{2}~$\lambda 1334$,
\ion{Al}{2}~$\lambda 1670$, \ion{Fe}{2}~$\lambda 1608,2382,2600$,
\ion{Mg}{2}~$\lambda 2796,2803$; and the strong high-ionization
transitions commonly seen in LLSs: \ion{C}{4}~$\lambda 1548,1550$ and
\ion{Si}{4}~$\lambda 1393,1402$. Any systems with secure metal-line
absorption were also flagged for Voigt profile fitting.  Again, this
analysis occasionally benefited from the public datasets of SDSS and
BOSS whose broad wavelength coverage complements the more limited
coverage of our large-aperture spectroscopy.

Given the equivalent width for Ly$\alpha$ absorption, our Voigt
profile fits for the $N_{\rm HI}$, and the presence/absence of low-ion
metal absorption, objects were classified into three categories: {\it
  optically thick, ambiguous, or optically thin}. Objects which show
obvious damping wings or strong (EW $> 0.3$~\AA) low-ion metal
absorption are classified as optically thick. For cases for which
metal lines are weak, not covered by our spectral coverage, or
significantly blended with the Ly$\alpha$ forest of the b/g quasar, an
object is classified as optically thick only if it has \wlya $\ge
1.7$\AA.   Table~1 lists relevant quantities for the final sample of 29
objects classified as optically thick, around which we search for extended
\mlya\ emission. 

The false positives and completeness of our optically thick sample are
sources of concern.  Line-blending, in particular, can significantly
depress the continuum near the Ly$\alpha$ profile biasing the
\wlya\ and the estimated column density high. Based on the analysis of
our current sample, and comparisons of objects observed at moderate
and echellette/echelle resolution \citep[see e.g.][]{QPQ1}, we
estimate that false-positives, that is non optically-thick objects, in
our sample is lower than $10\%$.  This low contamination follows from
our relatively conservative criteria for defining a system as
optically thick.  Specifically, there are 3 cases for which we are
less confident in the optically thick designations (see Table~1;
J1041+5630, J1045+4351, J1444+3113) but we nevertheless include them
in the following analysis.  Removing them would make no difference to
our conclusions aside from reducing the sample size by 10\%. Note that
saturated Ly$\alpha$ forest lines with $N_{\rm HI} < 10^{17.2}\,{\rm
  cm}^{-2}$ are essentially indistinguishable from optically thick
absorbers with low-metallicity.  As such, we expect the false positive
rate to be smaller than our incompleteness, and we thus consider our
estimate of the covering factor of optically thick gas (see
Figure~\ref{fig:cov_fact}) to be conservative lower limits.

Finally, given that there is suggestive evidence that radio-loud
quasars have brighter \mlya\ emission nebulae and a higher detection
frequency than radio-quiet quasars \citep{heckman91a}, and that HzRGs
at $z\sim 2$ typically exhibit bright (${\rm SB}_{\lya} \sim
10^{-16}\cgssb$) large-scale ($\sim 100\,\kpc$) \mlya\ nebulae
\citep[e.g.][]{msd+90,mccarthy93,vanojik96,nle+06,binette06,reuland07,villar07,
  mileyd08}, we consider the radio properties of our f/g quasars.  The
Faint Images of the Radio Sky at Twenty cm survey
\citep[FIRST;][]{FIRST} used the Very Large Array (VLA) to produce a
map of the 20 cm (1.4 GHz) sky with a beam size of $5.\arcsec4$ and an
rms sensitivity of about 0.15 mJy beam$^{-1}$.  The survey covers the
same 10,000 deg$^2$ sky region covered by the SDSS imaging, has a
typical detection threshold of 1 mJy, and an astrometric accuracy of
$0.05\arcsec$. Following \citep{ivezic02}, we match our f/g quasars to
the FIRST catalog with a matching radius of $1.5\arcsec$. From our
parent sample of 68 projected pairs, only 59 sources were covered by
the FIRST imaging footprint, and of these, 4/59 have matches in the
FIRST catalog ($F_{\rm peak}\gtrsim 1\,{\rm mJy}$), which is
consistent with roughly $\sim 10\%$ of quasars being radio-loud
\citep{ivezic02}. This is to be expected since our f/g quasars are
typical SDSS quasars. Visual inspection of the FIRST images indicates
that none show evidence of a complex radio morphology
\citep[i.e. core-jet, core-lobe, and
  double-lobe;][]{maglio98,mcmahon02}. Among the 29 f/g quasars with
coincident optically thick absorption, 28 sources were covered by the
FIRST footprint, and only 3/28 have matches in the FIRST catalog,
again consistent with a $\sim 10\%$ radio loud fraction. The radio
fluxes for the three sources detected in the FIRST survey are listed
in in Table~1; for the sources which were not detected we
list 5$\sigma$ upper limits based on the FIRST sky-rms coverage
maps. Given that only $\sim 10\%$ of our f/g quasars have radio
counterparts $\gtrsim 1\,{\rm mJy}$, both in our parent sample and in
the sample with coincident absorbers, we conclude that quasars which
we study here are predominantly radio-quiet.

\subsection{The Covering Factor of Optically Thick Absorbers}
\label{sec:cov_fact}

The distribution of f/g quasar redshifts and transverse separations
for all 68 projected quasar pair sightlines in our parent sample are
shown in the scatter plot in Figure~\ref{fig:cov_fact}. The 29 black
filled symbols indicate sightlines which have an optically thick
absorber within an interval $\delta v = [-1500,1500]\,\kms$ of the f/g quasar
redshift (the median absolute offset of these systems
is $|\delta v| = 256\,\kms$), 
which are objects that we search around for extended
\mlya\ emission.  
Open black circles indicate sightlines classified as
ambiguous, and green sightlines which are optically thin. Note that
the closest pairs are predominantly at redshift $z \sim 2$, because
this is where quasar selection, and hence quasar pair selection, is most efficient
\citep[see][]{qsoselect,BINARY,bovy11,bovy12}.

It is clear from Figure~\ref{fig:cov_fact} that the covering factor of
absorbers is very high for small separations: at least 28 out of 48 sightlines
with $R_{\perp} < 200$\,kpc have signatures of optically thick gas
coincident with the f/g quasar, implying a $> 58\%$ covering
fraction. This high covering factor was previously noted in
\citet{QPQ1} based on a much smaller pair sample (11 sightlines with
$R_{\rm perp} < 200\,\kpc$) which are a subset of our current parent
sample (see also QPQ5). 
Note again that because of the significant incompleteness discussed
in \S\ref{sec:sample}, this high covering factor should be
interpreted as a conservative lower limit, albeit subject to Poisson
uncertainty.

The distribution of points in
Figure~\ref{fig:cov_fact}, or equivalently the dependence of covering
factor on impact parameter, can be used to measure the quasar-absorber
correlation function, which is equivalently the spatial profile of the
absorbing clouds around the f/g quasar. In QPQ2, we analyzed the much
smaller sample of pairs from QPQ1 and determined a correlation length
of $r_0=9.2^{+1.5}_{-1.7}~\hMpc$ (comoving) assuming a power law
correlation function, $\xi\propto r^{-\gamma}$, with $\gamma=1.6$. The
high covering factor of absorbers quantified by the correlation
function, implies a comparably high incidence of proximate
absorbers along the line-of-sight; however, this absorption is not observed in our f/g
quasars, or in other isolated quasars.  This \emph{anisotropic}
clustering of absorbers around quasars indicates that the transverse
direction is less likely to be illuminated by ionizing photons than
the line-of-sight, either because of anisotropic emission which could
result if the f/g quasar is obscured from some vantage points, or
because the quasar emits radiation intermittently, on a timescale
comparable to the transverse light-crossing time.  We will return to
this important inference in the context of the search for
\mlya\ fluorescence in \S\ref{sec:discuss}.

\section{Data Reduction and PSF Subtraction}
\label{sec:coadd}

Our goal is to conduct a sensitive search for extended \mlya\ emission
in slit spectra in the CGM region surrounding the f/g quasar in quasar
pairs. Our approach is to subtract off the point spread function (PSF)
of the the quasars and other objects (galaxies or stars), which may have
serendipitously landed on the slit, and search for resolved
\mlya\ emission which is inconsistent with these spectral PSFs. In the
next sections, we describe the details of our data reduction and
spectral PSF subtraction algorithm, our procedure for flux calibration
and determining surface brightness limits, and our simulation using
synthetic sources to determine the surface brightness levels we can
visually detect.

\subsection{Data Reduction and PSF Subtraction Algorithm}

The Gemini/GMOS observations were conducted in longslit mode, whereas
the LRIS observations include a mix of longslit and multislit
exposures.  All data were reduced using the LowRedux
pipeline\footnote{http://www.ucolick.org/$\sim$xavier/LowRedux}, which
is a publicly available collection of custom codes written in the Interactive Data
Language (IDL) for reducing slit spectroscopy. Individual
exposures are processed using standard techniques, namely they are
overscan and bias subtracted and flat fielded. Flat fielding is
performed in two steps, using both a pixel flat to correct for
pixel-to-pixel sensitivity variations, as well as spectroscopic
illumination flats (taken with the sky in twilight) to flatten the
larger scale illumination pattern arising from either instrument optics or
imperfections in the slits. Cosmic rays and bad pixels are identified
and masked in multiple steps. First, a sharpness detection algorithm
is run on each image to identify and mask features smaller than the
spectral/spatial PSF. Then further downstream in the reduction procedure,
outliers from our models of the sky and the object are masked (see
below).  Wavelength solutions are determined from low order polynomial
fits to arc lamp spectra, and then a wavelength map is obtained by
tracing the spatial trajectory of arc lines across each slit. This
wavelength map allows us to model the sky counts as a function of
wavelength and slit position without needing to rectify the original
data \citep[e.g.][]{kelson03}.

Our method for spectral PSF subtraction employs a novel custom 
algorithm. We treat sky and PSF subtraction as a coupled problem to
obtain a multi-component model of the image counts. This model is
composed of a sum of 2-d `basis functions', which consists of a sky background
and a PSF model for each object on the slit. This is typically just
the two quasars (f/g and b/g), but in some cases it includes additional
objects which landed on the slit in the region of interest. By
construction, the sky-background has a flat spatial profile because
our slits are flattened by the slit illumination function. For the
object models, we first identify objects in an initial sky-subtracted
image, and trace their trajectory across the detector. We then extract
a 1-d spectrum, normalize these sky-subtracted images by the total
extracted flux, and fit a B-spline profile to the normalized spatial light
profile of each object relative to the position of its trace. This
non-parametric object profile has the flexibility to vary as a function of
wavelength, and corrections to the initial trace are simultaneously
determined and applied.  Given this set of 2-d basis functions,
i.e. the flat sky and the object model profiles, we then minimize
chi-squared for the best set of spectral B-spline coefficients (i.e. break points
are spaced in wavelength) which are
the spectral amplitudes of each basis component of the 2-d model. In
other words 
\be \chi^2 = \sum_{i}^{N_{\rm pix}} \frac{\left({\rm DATA}_i - {\rm
    MODEL}_i\right)^2}{\sigma_i^2}\label{eqn:chi} 
\ee 
where the sum is taken over all $N_{\rm pix}$ pixels in the image, `DATA' is the
image, `MODEL' is a linear combination of 2-d basis functions multiplied by
B-spline spectral amplitudes, and $\sigma$ is a model of the noise in
the spectrum, i.e. $\sigma^2 = {\rm SKY} + {\rm OBJECTS} + {\rm READ
  NOISE}$. The result of this procedure are then full 2-d models of the
sky-background (SKY), all object spectra (OBJECTS), and the noise ($\sigma^2$).  We
then use this model SKY to update the sky-subtraction, the individual
object profiles are re-fit and the basis functions updated, and 
chi-square fitting is repeated. We iterate this procedure of object
profile fitting and subsequent chi-squared modeling four times until we
arrive at our final models for the sky background, the 2-d spectrum
of each object, and the noise. 

Each exposure of a given target is modeled according to the above
procedure which then allows us to subtract the sky and the object
profiles from each individual image. These images are registered to a
common frame by applying integer pixel shifts (to avoid correlating
errors), and are then combined to form final 2-d stacked
sky-subtracted and sky-and-PSF-subtracted images.  The individual 2-d
frames are optimally weighted by the ${\rm (S\slash N)}^2$ of their
extracted 1-d spectra (using the b/g quasar), and bad pixels
identified via sharpness filtering, as outliers in the chi-square
fitting, or by sigma-clipping of the image stack, are masked.  The
final result of our data analysis are three images: 1) an optimally
weighted average of $({\rm DATA} - {\rm SKY})$, henceforth the
`stacked sky-subtracted image', 2) an optimally weighted average of 
$({\rm DATA} - {\rm SKY} - {\rm
  OBJECTS})$, henceforth the `stacked sky-and-PSF-subtracted image',
and 3) the noise model for these images $\sigma^2$.  The final noise map
is propagated from the individual noise model images taking into account
weighting and pixel masking entirely self-consistently.

Extended \mlya\ emission will be manifest as residual flux in our 2-d
sky-and-PSF-subtracted images which is inconsistent with being noise.
We thus search for \mlya\ emission by defining a $\chi$ image in
analogy with eqn.~(\ref{eqn:chi}) above, but using the stacked images
and corresponding propagated noise instead of the individual
frames. This procedure allows us to visually assess the statistical
significance of any putative emission feature. Figure~\ref{fig:maps}
presents these images for the 29 objects in our optically thick sample
(see Table~1). The middle panels show $\chi_{\rm
  sky+PSF} = ({\rm DATA} - {\rm SKY} - {\rm OBJECTS})\slash \sigma$
for the stacked sky-and-PSF-subtracted images. Recall that an
analogous $\chi^2$ has been minimized by our modeling of each
image. Thus if our model is an accurate description of the data, the
distribution of pixel values in the $\chi_{\rm sky+PSF}$ should be a
Gaussian with unit variance. The lower panel of each image shows
$\chi_{\rm sky}= ({\rm DATA} - {\rm SKY})\slash \sigma$; the numerator 
is the the stacked sky-subtracted (but not
PSF-subtracted) image. The upper panels show smoothed maps which are helpful
for identifying extended emission. Specifically, the smoothed images
are given by \be \chi_{\rm smth} = \frac{{\rm CONVOL}[{\rm DATA} -
    {\rm SKY} - {\rm OBJECTS}]}{{\rm CONVOL}[\sigma]}, \ee where the
${\rm CONVOL}$ operation denotes smoothing of the stacked images with
a symmetric Gaussian kernel (same spatial and spectral widths) with
FWHM$_{\rm smth}=235\,\kms$ (dispersion $\sigma_{\rm smth} =
100\,\kms$).  For LRIS this FWHM$_{\rm smth}$ corresponds to 5.2-6.1
pixels, or 1.4-1.5 times the spectral resolution element, and 
$1.4-1.6\arcsec$ spatially\footnote{The range of values results from
  the different f/g quasar redshifts, and hence different observed
  frame wavelength of \mlya.}. For GMOS a FWHM$_{\rm smth}=235\,\kms$
corresponds to 6.6-7.0 pixels or 1.4-1.5 times the spectral
resolution, or $1.9-2.0\arcsec$ spatially.

Note that smoothing correlates pixels in the smoothed image, and
furthermore the ratio of smoothed image to smoothed noise will no
longer obey Gaussian statistics. Nevertheless, $\chi_{\rm smth}$
proves to be a useful tool for identifying extended emission.
 
\subsection{Flux Calibration and Surface Brightness Limits}

The following describes our procedure to flux calibrate our spectra and
determine surface brightness limits. Standard star spectra were not
typically taken immediately before/after our quasar pair
observations. Instead, we construct a model sensitivity function of
the Keck LRIS-B spectrograph by first fitting an aggregate of standard
star spectra taken at different slit positions, which span the range
of wavelength coverage available with a multi-slit setup.  A
similar procedure was employed for Gemini GMOS using standard star
spectra taken at a variety of central wavelengths. We apply this
archived sensitivity function to the b/g quasar spectrum, and then
integrate the flux-calibrated 1-d spectrum against the SDSS $u$-band filter
curve. The sensitivity function is then rescaled to yield the
correct SDSS $u$-band photometry. A comparison of our 1-d
spectra flux calibrated in this way to the SDSS spectra typically show
agreement to within $\sim 10-20\%$ in the region of wavelength
overlap. Given that SDSS spectra have spectrophotometric errors of
$\sim 10\%$ \citep{DR3}, that $u$-band photometric errors (used
for our renormalization) are about $7\%$ for a typical source with $u
\simeq 21$, and that quasar variability over $\simeq 5$ year
timescales (between when the SDSS photometry and our spectra were
taken) could result in a $\sim 10\%$ fluctuation \citep[e.g.][]{schmidt10}, we
consider $\sim 20\%$ agreement between our spectra and the SDSS
spectra to be reasonable. 

This procedure of rescaling the sensitivity functions to match
photometry is effective for point source flux-calibration, however a
subtlety arises in relation to calibrating extended emission. The
point source counts will be reduced because of improper object
centering due to guiding errors or inaccuracies in slit/mask
alignment, and because a fraction of the spatial PSF lies outside the
slit area. These point-source slit-losses do not, however, reduce the
amount of spatially resolved emission, which is of interest here.
Thus, renormalizing to point-source photometry will tend to
over-estimate our surface brightness detection limit and underestimate
our sensitivity. Hence, our procedure is to apply the rescaled
sensitivity functions (based on point source photometry) to our 2-d
images, but reduce them by a geometric slit-loss factor so that we
properly treat extended emission.

To compute the slit-losses we use the measured spatial FWHM to determine
the fraction of light going through our $1.0\arcsec$ slits, but we do
not model centering errors.  A useful check on our flux calibration
procedure is to examine any residual variation in our renormalized
sensitivity functions after the effects of the atmospheric extinction
(i.e. due to airmass) and slit-losses are removed. These variations
in the spectroscopic zero-point could be due either to transparency
variations or to other systematic flux calibration errors. We find a
35\% relative variation ($1\sigma$) about the mean zero-point, which
is at a level that can be plausibly explained by transparency
variations in clear but non-photometric conditions. If these
variations in zero-point are indeed due to transparency fluctuations,
then our procedure of renormalizing to the photometry effectively
takes them out. But because we have no quantitative information about the
transparency during our observations, we conservatively assume a relative
error of $\sim 35\%$ on our surface brightness limits.

Application of the wavelength dependent sensitivity function
converts our stacked sky-and-PSF-subtracted images from units of
electrons into ${\rm erg\,s^{-1}\,cm^{-2}}\,$\AA$^{-1}$. We then tile
these calibrated images with a uniform grid of aperture centers within
$\pm 3000\,\kms$ of the f/g quasar's redshifted \mlya\ line.  Regions
within $\pm 7.0\arcsec$ of the f/g quasar trace on either side are
excluded to avoid potential contamination from \mlya\ emission. Regions
within $\pm 3000\,\kms$ of the b/g quasar's \mlya\ and also within
$\pm 7.0\arcsec$ of the b/g quasar trace are also excluded, to similarly avoid 
\mlya\ contamination (this is relevant if the
redshift difference of the pair $\delta v < 3000\,\kms$).  The spacing
between these aperture centers is 1.0$\arcsec$ spatially and $\Delta
v=400\,\kms$ spectrally. We then compute SB limits in windows of
$700\,{\rm km\,s^{-1}}\!\times 1.0\arcsec$ about each aperture center, which
corresponds to an aperture of $700\,{\rm km\,s^{-1}}\!\times 1.0\,{\rm arcsec}^2$ on the
sky because we always used a $1.0\arcsec$ slit.  The $700\,\kms$
spectral width is motivated by the observed kinematics of the
metal-line absorbers associated with the optically thick gas we detect
in absorption \citep[][Prochaska et al.\, in prep.]{QPQ3}. If this gas emits
\mlya, we expect the emission kinematics to have a comparable
velocity width (in the absence of significant resonant broadening).
Because our noise has been propagated self-consistently, we
use the noise model for our stacked sky-and-PSF-subtracted image to
determine the 1$\sigma$ surface brightness limit for the $i$th
aperture ${\rm SB}_{1\sigma,i}$, and we take the mean value of
$1\slash N \sum_i^N {\rm SB}_{1\sigma,i}$ of all apertures to be the
surface brightness limit for the quasar pair ${\rm SB}_{1\sigma}$,
which is quoted in Table~1. 

We can assess the accuracy of our noise model by examining the
distribution of $\chi_{i} = {\rm SB}_i\slash {\rm SB}_{1\sigma,i}$,
where ${\rm SB}_i$ is the extracted flux of the $i$th aperture in each
$700\,{\rm km\,s^{-1}} \times 1.0\,{\rm arcsec}^2$ window, and ${\rm
  SB}_{1\sigma,i}$ is the surface brightness limit based on our noise
model at this location.  In the absence of detectable extended
\mlya\ emission the ${\rm SB}_i$ should be pure noise, and if our
noise model is accurate, the $\chi_{i}$ should thus obey a Gaussian
distribution with unit variance. The distribution of $\chi_i$ for one
of our quasar pairs SDSSJ\,0938$+$5317 is compared to a Gaussian in
Figure~\ref{fig:chi_histo}. It is clear that our noise model provides
an accurate description of the real noise in the stacked sky-and-PSF
subtracted image.  Systematic errors in the data reduction and sky and
object modeling procedure are not accounted for in the purely
statistical error ${\rm SB}_{1\sigma,i}$, and thus we would expect to
typically underestimate the true noise level. We can quantify these
errors in our noise estimates by computing the standard deviation of
$\chi_{i}$ after clipping $> 5\sigma$ outliers (resulting from e.g.  a
small number of residual cosmic ray hits or bad pixels that remain
unmasked by our data reduction procedure).  For the distribution in
Figure~\ref{fig:chi_histo} we obtain $\sigma_{\rm chi} = 0.97$, which
indicates that for this case we actually slightly overestimated the
noise. We similarly compute $\sigma_{\rm chi}$ for each quasar pair in
our sample. The median value over all quasar pairs is 1.08, thus on
average we tend to underestimate the noise by $8\%$. Because this is
smaller than our estimated flux calibration error, we therefore quote
statistical SB limits based only on our noise model as ${\rm
  SB}_{1\sigma}$ in Table~1.

\subsection{Visually Recovering Synthetic Sources}
\label{sec:fake}

\begin{figure}[t!]
  \centering{\epsfig{file=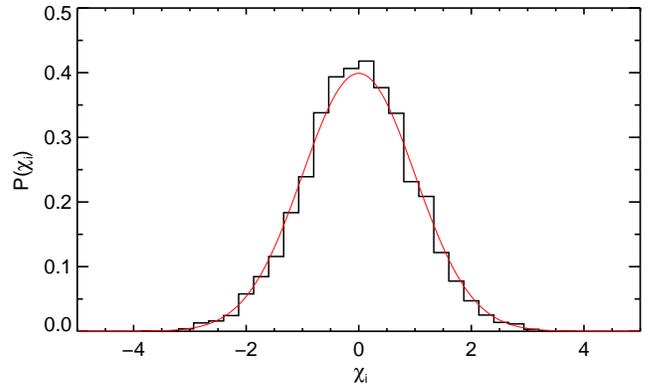,bb=0 0 648 381,width=0.50\textwidth,clip=}}
  \caption{Distribution of extracted surface brightness apertures
    relative to noise model predictions. The histogram shows the
    distribution $\chi_{i} = {\rm SB}_i\slash {\rm SB}_{1\sigma,i}$ for
    one of our sources SDSSJ\,0938$+$5317 from Figure~\ref{fig:maps}.
    In the absence of detectable extended \mlya\ emission the ${\rm
      SB_i}$ should be pure noise, and if our noise model is accurate,
    then $\chi_{i}$ should follow a Gaussian distribution with unit
    variance, which is shown by the solid red curve. The standard
    deviation of $\chi_{i}$  after clipping $> 5\sigma$ outliers is 
    $\sigma_{\rm chi} = 0.97$, which indicates that our noise model is indeed
    accurate and that fluctuations obey Gaussian statistics as expected. 
    The median value of $\sigma_{\rm chi}$,  computed similarly, 
    over all of our quasar pairs is 1.08, indicating that on average we tend to 
    underestimate the noise by $8\%$. \label{fig:chi_histo}}
\end{figure}

\begin{figure}[h!]
 \centering{\epsfig{file=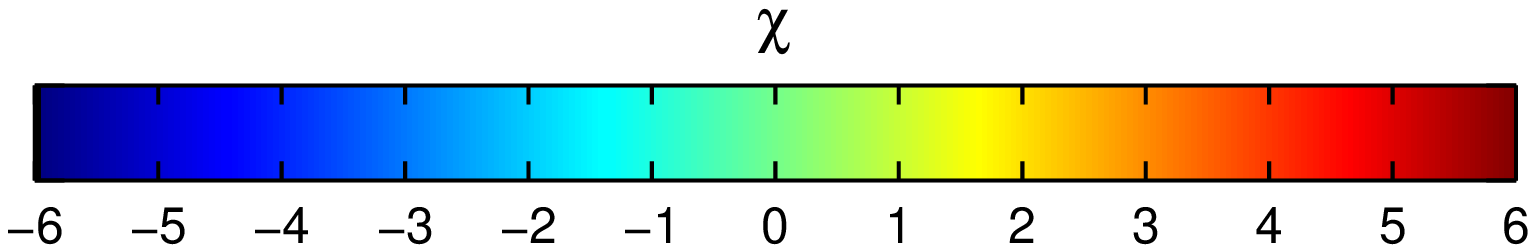,bb=10 0 540 72,width=0.50\textwidth,clip=}}
  \centering{\epsfig{file=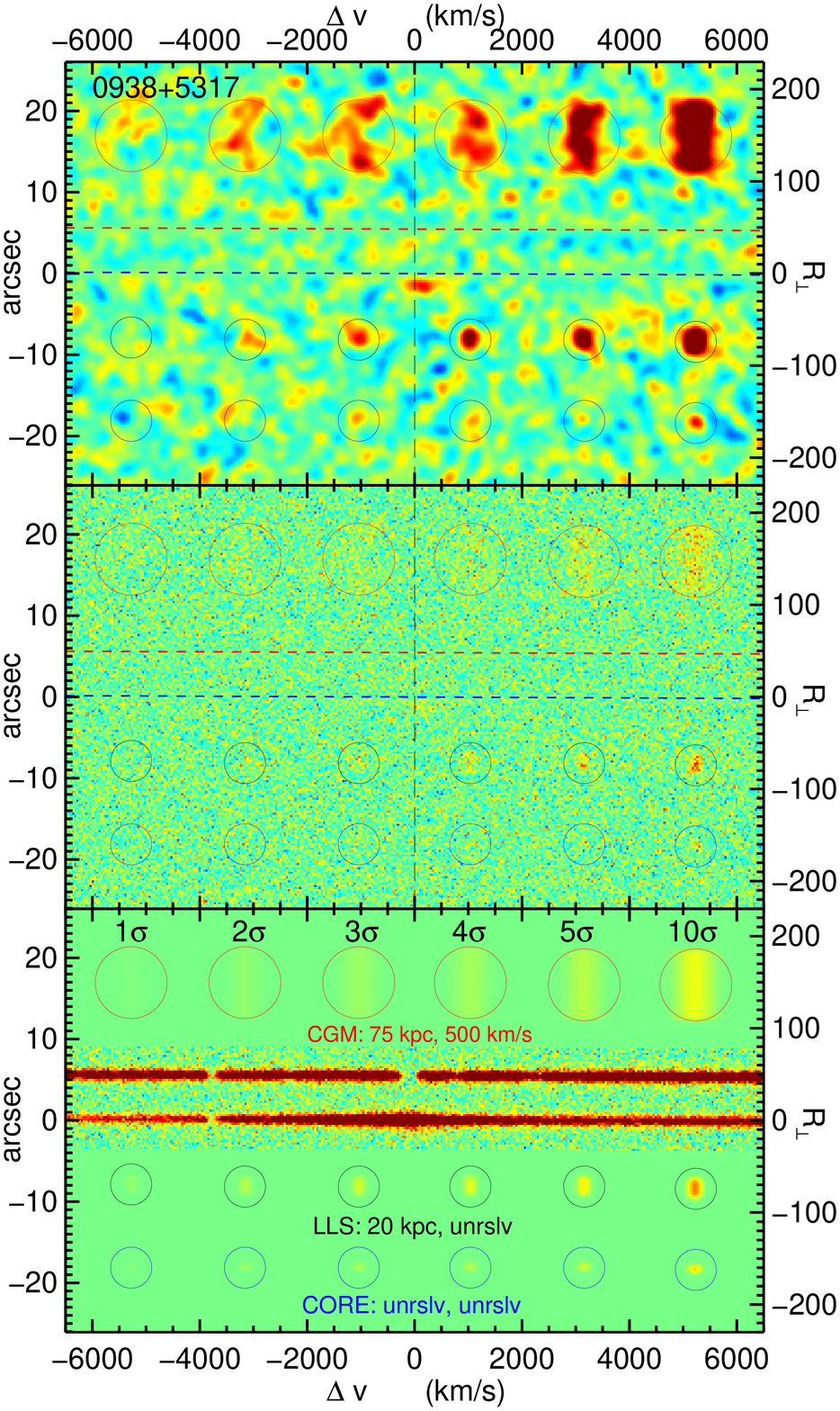,bb=10 0 540 860,width=0.50\textwidth,clip=}}
  \caption{Illustration of detection significance of synthetic sources with
    various properties.  The lower panel shows the $\chi_{\rm sky
      only}$ image for SDSSJ\,0938$+$5317 with synthetic sources added
    according to the procedure described in \S\ref{sec:fake}.  In only
    the lower panel, the synthetic sources are noiseless, and their
    brightness and relative significance can be directly compared to
    the real (noisy) spectra of the two quasars shown in the central
    inset. The middle and upper panels show $\chi_{\rm sky+object}$
    and $\chi_{\rm smth}$ images respectively, which are the same as
    the corresponding panels for SDSSJ\,0938$+$5317 in
    Figure~\ref{fig:maps}, except that noisy synthetic sources have been
    added. The stretch and colormap are identical to that used for all
    of our $\chi$ maps in Figure~\ref{fig:maps}, and is indicated by
    the upper colorbar.  The horizontal red and blue dashed lines and
    the vertical red dotted line are the same as in
    Figure~\ref{fig:maps}. The colored circles indicate the location
    of the sources for various significance levels which are labeled in
    black at the top of the bottom panel. The red circles indicate the
    large 'CGM' sources which are both spectrally and spatially
    resolved, black circles the 'LLS' sources which are spatially
    resolved but spectrally unresolved, and blue circles the 'CORE'
    sources which are both spatially and spectrally unresolved. This
    figure suggests that we should be able to marginally detect 'CGM'
    like sources that are $\gtrsim 2\times {\rm SB}_{1\sigma}$, 'LLS' like
    emission with $\gtrsim 3\times {\rm SB}_{1\sigma}$, and 'CORE like
    sources to $\gtrsim 10\times {\rm SB}_{1\sigma}$. Note that the blob of
    emission near $\Delta v=0\,\kms$
    and impact parameter $R_{\perp} \approx  -20\,\kpc$ is real small-scale 
    Ly$\alpha$ fuzz emission from the f/g quasar (see \S\ref{sec:fuzz}).\label{fig:fake}}
\end{figure}

Having quantified the formal errors in our sky-and-PSF-subtracted
images, we now determine the significance level for a convincing
detection. While we quote $1\sigma$ SB limits over $700\,{\rm km\,s^{-1}}\times 1.0\,
{\rm arcsec}^2$ apertures, our sensitivity to a given surface
brightness is of course set by the size of the source and 
its spectral characteristics (i.e.\ kinematics).  One can in principle always reach lower
SB levels by averaging over larger spatial regions, and it is easier to 
detect narrow lines over the background. To address
these issues we construct synthetic fluorescent sources and add them to
the sky-and-PSF-subtracted image for a representative pair
SDSSJ\,0938$+$5317. This pair has $z_{\rm bg} = 2.32$, $z_{\rm
  fg}=2.07$, an angular separation of $\theta=5.6\arcsec$
corresponding to $R_{\perp}=49\,{\rm kpc}$, and was observed with
LRIS, as are the bulk of our pairs. The total exposure time was 1800s,
allowing us to reach a surface brightness limit of ${\rm
  SB}_{1\sigma}=0.29\sci{-17}\cgssb$ which is close to the median value
of our sample ${\rm SB}_{1\sigma}=0.31\sci{-17}\cgssb$
(Table~1). Figure~\ref{fig:chi_histo} testifies
to the accuracy of our noise model for this object,  and illustrates
that the surface brightness noise fluctuations obey Gaussian statistics. 
Our spectral resolution for these 
observations is FWHM$=153\,\kms$ and the seeing measured from the
quasar profiles is $0.79\arcsec$ corresponding to a spatial
resolution of $7.0\,\kpc$. 

We construct synthetic sources with a uniform spatial surface brightness
distribution, where the size is specified by a single dimension along
the slit $d_{\rm fake}$. Spectrally, we assume a Gaussian velocity
profile specified by FWHM$_{\rm fake}$. The amplitude of the synthetic
sources is set such that their total surface brightness
(i.e. integrated out spectrally) is equal to ${\rm SB} =
[1,2,3,4,5,10]\times {\rm SB}_{1\sigma}$, which corresponds to ${\rm
  SB}_{\lya}=[0.29,0.57,0.86,1.1,1.4,2.9]\sci{-17}\,
{\rm\,erg\,s^{-1}\,cm^{-2}}$ ${\rm arcsec^{-2}}$ for
SDSSJ\,0938$+$5317.

We consider synthetic sources with three different spatial and kinematic
distributions which represent our expectations for the sources that
might occur near a bright quasar. The first is a large $d_{\rm fake} =
75\,\kpc$ source with kinematics FWHM$_{\rm fake}=500\,\kms$, which is
both spatially and spectrally resolved.  This source represents the
kind of emission expected if the quasar CGM is made up of a large
population of small spatially unresolved clouds which nevertheless
have sufficient covering factor, $f_{\rm C}$, to result in detectable
\mlya\ emission (see Figure~\ref{fig:cartoon}b-d); we denote this
model as `CGM'. In this scenario, the kinematics of the emission
correspond to the motions traced by the metal-line absorbers as
observed near quasars with b/g sightlines \citep[][Prochaska et al.,
in prep.]{QPQ3}.  The
second synthetic source is meant to represent a spatially resolved
\mlya\ source (see Figure~\ref{fig:cartoon}a), which is the
counterpart to the frequently observed optically thick absorption in
b/g quasar sightlines (Figure~\ref{fig:cov_fact}).  In this `LLS'
scenario, the size is taken to be $d_{\rm fake} = 20\,{\rm kpc}$,
corresponding to about three resolution elements. The kinematics are
assumed to be comparable to the narrow turbulent and/or thermally
broadened line widths $\sim 20\,\kms$ characteristic of metal
absorption in individual LLS absorption complexes, \citep[i.e., spectrally
unresolved][]{prochter10}. 
For the third synthetic source, denoted as `CORE', we assume a
size $d_{\rm fake}$ of two LRIS pixels corresponding to $4.8\,\kpc$,
and cold kinematics $\sim 20\,\kms$, such that it is both spatially
and spectrally unresolved. This source is meant to mimic the
properties of the candidate fluorescent emitters recently detected by
\citet{cantalupo12} at $\sim$\,Mpc distances from bright
quasars. These fluorescent candidates tend to be compact, are
spatially unresolved in ground-based seeing, and are believed to be
the dense cores of proto-galaxies. For spatially unresolved sources,
the total \mlya\ luminosity is the relevant quantity rather than the
surface brightness, and the corresponding luminosity of our CORE
sources are
$L_{\lya}=[0.92,1.8,2.8,3.7,4.6,9.2]\sci{41}\,{\rm\,erg\,s^{-1}}$. Note
that the median luminosity for the most compelling (i.e. those with
$W_{\lya} > 240$\AA) \citet{cantalupo12} fluorescent candidates is
$L_{\lya} = 9.1\,\sci{41}\,{\rm\,erg\,s^{-1}}$, which is at the level
of the 10$\sigma$ model source we consider.  Our model images are then
convolved with the spectral and spatial resolution of our data,
converted into counts using our sensitivity function, and Poisson
realizations of the average counts are added to the real
sky-subtracted and sky-and-PSF-subtracted images.

Figure~\ref{fig:fake} illustrates the detection significance of the
synthetic source model profiles which have been added to the real data for
SDSSJ\,0938$+$5317.  The lower panel shows the $\chi_{\rm sky}$
with synthetic sources added. In this panel only, the synthetic sources are
noiseless, and their brightness and relative significance can be
directly compared to the real (noisy) 2-d spectra of the two quasars shown
in the central inset.  Quantitatively, a uniform screen of surface
brightness at ${\rm SB}_{1\sigma}=0.29\sci{-17}\cgssb$ emitted at the
f/g quasar \mlya\ with a top-hat velocity distribution matched to our
$700\,\kms$ spectral extraction aperture would result in 
$0.44\,e^{-}$ electrons per pixel, whereas the typical value of the
$1\sigma$ noise in our stacked images (far from the quasars) is $\sigma
\simeq 4\,e^{-}$ electrons per pixel.

Inspection of Figure~\ref{fig:fake} suggests that we should be able to
marginally detect `CGM' like sources down to a level of $\gtrsim
2\times {\rm SB}_{1\sigma}$, `LLS' like sources down to $\gtrsim
3\times {\rm SB}_{1\sigma}$, and the spatially and spectrally
unresolved `CORE' like sources to $\gtrsim 10\times {\rm
  SB}_{1\sigma}$.  We emphasize that these significance limits are
rough guidelines, as the noise level varies with wavelength due to
fluctuations in the sky level. For example, the reason why the
$4\sigma$ sources in Figure~\ref{fig:fake} look scarcely more
significant than the $3\sigma$ ones, is that the $4\sigma$ sources
happen to land on a region where the sky-background is $\sim 10\%$
higher. Furthermore, in some cases systematics in the data reduction
limit our ability to make detections down to low significance levels.
These systematics manifest as large-scale correlations in the $\chi$
maps, and are particularly conspicuous in the smoothed $\chi_{\rm
  smth}$ images. In the absence of emission, the $\chi_{\rm
  sky+object}$ image should be a random Poisson noise field because
the pixel noise is completely uncorrelated, and thus the
characteristic size of structures in the $\chi_{\rm smth}$ should be
comparable to the size of the smoothing kernel $\simeq 5-6$
pixels. Some of our $\chi$ maps in Figure~\ref{fig:maps} show
correlated structure and spurious features due to systematics, the
worst case being SDSSJ\,0814$+$3250, where a large-scale gradient in
the flux level is seen across the spatial direction. This large-scale
gradient is likely due to a poorly determined slit-illumination
function. 

\section{Results and Discussion}
\label{sec:discuss}

In what follows we interpret our observational results from the
previous section in terms of the emission mechanisms discussed in
\S\ref{sec:form}.  The results further constrain the physical
properties of the CGM of the massive galaxies hosting $z \sim 2$
quasars.  Before delving into details, we briefly summarize the main findings of
our search for \mlya\ emission in the regions surrounding quasars. Careful 
inspection of the \mlya\ emission maps in Figure~\ref{fig:maps} reveals
the following:

\begin{itemize}

\item Fluorescent \mlya\ emission with properties expected in the
  `mirror' approximation, is \emph{not} detected at the locations of
  optically thick absorbers pinpointed by the b/g quasar sightlines. 
 
\item Despite the fact that the covering factor of optically thick
  absorbers on scales $R_{\perp} \sim 100\,{\kpc}$ is high ($f_{\rm C}
  > 0.5$), we do not detect diffuse \mlya\ emission on these large scales, with
  the exception of a single system. 

\item In a single system SDSSJ\,0841$+$3921, we detect a spectacular \mlya\ 
  filament around the foreground quasar with surface brightness
  ${\rm SB}\simeq 1\sci{-17}\cgssb$ and an extent of 230\,\kpc\ and possibly
  as large as 330\,\kpc (discussed further in Hennawi et al.\, in
  prep.). 

\item Extended \mlya\ fuzz is frequently detected on small-scales $R_\perp <
  50\,\kpc$. We detect definite fuzz in 10 objects, or 34\% of our sample, 
  at levels of ${\rm SB}_{\lya}\simeq 0.5-5\sci{-17}\cgssb$.

\item In SDSSJ\,0856$+$1158, a high-equivalent width ($W_{\lya}
  > 50$\AA; rest-frame) Ly$\alpha$-emitter is detected with $L_{\lya}
  = 2.1\pm 0.32\sci{41}\cgslum$, at a distance of $134\,\kpc$ and
  velocity $-540\,{\kms}$ relative to the f/g quasar.
  
\end{itemize}

As we will frequently refer to our observations in aggregate, for reference 
we list the typical properties of our sample of 29
projected quasar pair observations. The median redshift of our f/g
quasars is $z_{\rm fg} = 1.95$, which are probed by b/g sightlines at
a median impact parameter of $R_{\perp} = 106\,\kpc$. The luminosity
at the Lyman limit has a median value  $\log_{10} \mlnull \simeq 30.3$
and our f/g quasars have a
median de-reddened $i$-band apparent magnitude of 19.7. Our median
$1\sigma$ \mlya\ surface brightness is ${\rm
  SB}_{1\sigma}=0.31\sci{-17}\cgssb$, and according to the example in
\S\ref{sec:fake}, large scale emission from the quasar CGM should be
detectable for $\gtrsim 2\times{\rm SB}_{1\sigma}$ or ${\rm SB}_{\lya}
\gtrsim 0.6\sci{-17}\cgssb$.

\subsection{Fluorescent Emission from Optically Thick Gas}

In this subsection we compare our observational constraints on extended
\mlya\ emission to predictions for recombination radiation from optically
thick gas, as laid out in \S\ref{sec:recomb_thick}. We consider emission from 
spatially resolved `mirrors' and from an unresolved distribution
of clouds in turn.  

\subsubsection{Fluorescent Emission in the Mirror Approximation}

The frequent detection of optically thick gas in the CGM of quasars
quantified in \S\ref{sec:cov_fact}, and discussed in other work in
the QPQ series \citep{QPQ1,QPQ2,QPQ5,QPQ6}, provided initial optimism
that we would detect significant fluorescent \mlya\ emission from this
population of clouds. In the simple `mirror' approximation \citep[see
  eqn.~\ref{eqn:SB_mirror};][]{GW96}, $\approx 60\%$ of the ionizing
photons impinging on an optically thick cloud are converted into
\mlya\ photons, which may be `reflected' into our line-of-sight.  A
precise estimate of the surface brightness depends on the ionizing
photon flux, the cloud size, the geometrical configuration between the
source, observer, and cloud(s), as well as the velocity field
\citep{cantalupo05,juna08}.  For spatially resolved clouds (see
Figure~\ref{fig:cartoon}a), the \mlya\ emission should also be
spatially resolved and emerge from the side of the absorber that lies
closest to the quasar at a SB level given by
eqn.~(\ref{eqn:SB_mirror}). The predicted ${\rm SB}_{\rm mirror}$
values are given for each pair in Table~1, where the distance is taken
to be the impact parameter $R_{\perp}$ to the b/g sightline.

\citet{ask+06} purported to detect \mlya\ fluorescence from a
serendipitously discovered DLA ($\log \mnhi = 20.4$) situated
$49\arcsec$ away from a luminous quasar at $z=2.84$. The ionizing flux
at transverse distance $R_{\perp}=384~{\rm kpc}$ implies a very high
mirror surface brightness of ${\rm SB}_{\rm
  mirror}=1.1\sci{-16}\cgssb$ in this system, owing to the extremely
luminous foreground quasar ($r\simeq 16$). The authors detect a
marginally resolved emission-line source near the b/g sightline with
comparable surface brightness; and based on this agreement, the
presence of a double-peaked kinematic emission line profile, the small
spatial separation ($1.5\arcsec$ or $11\kpc$) of the emission from the
DLA (and b/g sightline), and importantly, the fact that the emission
is on the side of the absorber closest to the f/g quasar, they argue
that it is \mlya\ fluorescence from the same gas responsible for the
DLA absorption.

Here we have conducted a similar search for analogous `mirror'
fluorescent emission from a much enlarged sample of 29 sightlines with
the same experimental setup as \citet{ask+06}. In 24/29 of these
sightlines we could have detected emission similar to \citet{ask+06}
at $>10\sigma$ (see Table~1), but fail to detect anything comparable.
As discussed in \S\ref{sec:fake}, resolved sources with size $\simeq
20\,{\rm kpc}$, should be detectable down to the level corresponding
to $2\times{\rm SB}_{1\sigma}$. With the exception of
SDSSJ\,0841$+$3921, which we discuss further in Hennawi et al.\ (in
prep.), we do not detect any emission in the immediate vicinity of the
b/g quasar sightlines, as would be expected if the absorbing clouds
were spatially resolved \mlya\ mirrors, although we had the
sensitivity to detect 24/29 sources at $> 10\sigma$. Based on this
large number of null detections of mirror fluorescence, we conclude
that the purported detection of \citet{ask+06} is most likely the
coincidence between a DLA and a star-forming galaxy, and was
incorrectly interpreted as \mlya\ fluorescence. Indeed, the
\mlya\ emission in this source coincides with the location of a
continuum source and the corresponding rest-frame equivalent width of
only $75$\AA\ is well within the range produced by conventional
stellar populations (i.e. $< 240$\AA), and is hence entirely
consistent with being powered by star-formation. It is possible that
the \mlya\ emission in this object is boosted by a fluorescent
contribution, an effect also discussed in \citet{cantalupo12}, but
given its relatively low rest-frame equivalent width, and our large
sample of similar null detections, any interpretation of the \citet{ask+06}
object as \mlya\ fluorescence is entirely speculative.

\subsubsection{Fluorescent Emission from Unresolved Clouds}

For unresolved optically thick clouds distributed around the quasar (see
Figure~\ref{fig:cartoon}b-d), the fluorescent emission will be manifest
as an extended and spatially resolved `fuzz', similar to the large
\mlya\ nebulae observed in HzRGs and \mlya\ blobs. In this case, the
average surface brightness is given by eqn.~(\ref{eqn:SB_thick}). Given
the high $f_{\rm C} \gtrsim 0.50$ covering factor of optically thick
gas at distances $\lesssim 200\,{\rm kpc}$ \citep[see
Figure~\ref{fig:cov_fact};][]{QPQ5}, the expected surface brightness of such
nebulae is about $\simeq 30\%$ smaller than the mirror values
tabulated in Table~1 averaged over an aperture
$R_{\perp}$. However a $R_{\perp}^{-2}$ profile is expected for the
idealized case of a uniform distribution in a spherical halo (see
eqn.~\ref{eqn:SB_thick}), and hence the emission would be brighter closer
to the quasar. We emphasize that if the optically thick gas giving
rise to the high covering factor is illuminated by the f/g quasar, the
expectation is that a SB comparable to the mirror sufrace brightness
is expected at impact parameters $R_{\perp} \lesssim 200\,\kpc$
\emph{all along the slit} (see Figure~\ref{fig:cartoon}).

Such large-scale diffuse \mlya\ halos would have been very easy to
detect. Indeed our CGM synthetic source simulation with emission extending
over $70\,\kpc$ (see $\S$~\ref{sec:fake} and Figure~\ref{fig:fake})
indicates that we could have detected such resolved emission at levels
${\rm SB_{\lya}} \gtrsim 2\times {\rm SB}_{1\sigma}$. Assuming a fixed distance
$R=100\,\kpc$ and a covering factor of $f_{\rm C} = 0.50$, consistent
with our measurements in Figure~\ref{fig:cov_fact}, the optically
thick surface brightness given by eqn.~(\ref{eqn:SB_thick}) is $> 5\times
{\rm SB}_{1\sigma}$ for 28/29 targets in our sample, and $> 10\times {\rm
  SB}_{1\sigma}$ for 19/29 targets.  No diffuse \mlya\ emission is detected
along the slit at distances of $> 50\,\kpc$ from any of our f/g
quasar, with the exception of the large-scale filament in
SDSSJ\,0841$+$3921, which we discuss further in Hennawi et al.\ (in
prep.). 

\subsubsection{Why is Optically Thick Fluorescence Absent?}
\label{sec:absent}
In the preceding subsections we argued that our data shows no evidence
for fluorescent emission from optically thick clouds. The possible
explanations for these null results include:

\begin{itemize}
\item  The optically thick clouds have very small radii.
\item  The gas lies at much greater distance from the quasar than 
  implied by the observed impact parameters.
\item The emission is extincted by dust grains. 
\item The optically thick gas is not illuminated by the quasar.
\end{itemize}

The first point about the clouds sizes relates to the fact that 
in the mirror approximation
(eqn.~\ref{eqn:SB_mirror}) the \mlya\ flux at the location of the b/g
sightline is proportional to the absorber cross-sectional area
$\sigma_{\rm c}$, $f_{\lya} \propto {\rm SB}_{\lya} \times
\sigma_{\rm c}$. As noted above, we would have easily detected
emission from resolved sources with cloud sizes $\gtrsim
10\,\kpc$. But if the clouds are very small, such individual mirrors
will not be detectable.  Indeed, our detailed absorption line modeling
of one member of our sample SDSSJ\,1204$+$0221 in QPQ3, yielded
very small cloud sizes $\sim 10-100\,{\rm pc}$, relative to the
typical $\sim 5\,\kpc$ resolution implied by our ground based seeing,
although it is still unclear whether this one system represents the
typical properties of gas in the quasar CGM.  Nevertheless, even for
such unresolved clouds (cloud sizes $\ll 10\,\kpc$) the high covering
factor, $f_{\rm C}$, of optically thick gas still implies significant
resolved \mlya\ flux, resulting from the collective emission of a
population of unresolved clouds (see Figure~\ref{fig:cartoon}). Thus, while
an appeal to small clouds can explain our null detections of \mlya\ mirror
fluorescence from the individual cloud absorbing the background sightline, a
large SB is still expected from unresolved clouds. 

The second explanation is that we have substantially over-estimated
${\rm SB}_{\rm mirror}$ because we have assumed the gas is at distance $R_\perp$
(eqn.~\ref{eqn:SB_mirror}), whereas in reality it lies at a
significantly larger distance $r =\sqrt{R_{\parallel}^2 +
  R_{\perp}^2}$, where $R_{\parallel}$ is the unknown line-of-sight
distance. This would reduce our estimates for the mirror SB by $[1 +
  (R_{\parallel}\slash R_{\perp})^2]$, a large factor if
$R_{\parallel}$ is significantly larger than $R_{\perp}$. We consider
this interpretation highly unlikely for several reasons. First, note
that we have demonstrated that the covering factor of optically thick
absorbers is high at impact parameters $R_{\perp} < 200\,\kpc$ from
quasars \citep[see e.g. \S\ref{sec:cov_fact}]{QPQ1,QPQ2,QPQ5,QPQ6},
but we observe a decrease for $R_{\perp} > 200\,\kpc$, both for our
parent sample shown in Figure~\ref{fig:cov_fact} as well as in 
larger samples \citep{QPQ5,QPQ6}. The presence of this dropoff at
$R_{\perp} \simeq 200\,\kpc$ indicates that we are indeed probing gas
interior to the quasar halo on scales $r \lesssim
200\,\kpc$. Second, our absorption line analysis of one member of our
sample in \citet{QPQ3} revealed extreme kinematics and a very high
metallicity in comparison to the typical properties of LLSs, implying
that this gas is local to the very atypical f/g quasar environment. 
We also find large equivalent widths for low-ion transitions (e.g.\
\ion{C}{2}~1334) for the parent sample \citep{QPQ5}, which are
uncharacteristic of optically thick absorbers along random quasar
sightlines. 

It is conceivable however, that optically thick absorbers very close to
the quasar are photoionized away, resulting in a donut-like radial
distribution. In this case, the absorbers in our sample with the the
smallest impact parameters $R_{\perp}\lesssim 50\,\kpc$ and highest
corresponding mirror surface brightness, could actually be at much
larger physical distances of $\sim 200\,{\rm kpc}$, reducing the
anticipated SB by factors $>10$. A clustering analysis of our full
sample of optically thick absorbers around quasars will be presented
in \cite[][see also QPQ2]{QPQ6}; there we explore the degree to which
donut like distributions can be ruled out.  Note, however, that we do
observe evidence for an increased frequency of the highest column
density absorbers (sub-DLAs and DLAs) at the smallest impact
parameters $R < 50\,\kpc$, \citep[QPQ1,QPQ2][]{QPQ6}; the obvious
interpretation is that this results from increased covering factor
and/or column density at smaller radii, again implying that the true
radius is comparable to the impact parameter.  Finally, note that
even if a donut like radial distribution implies that we have
overestimated the SBs for the smallest impact parameters ($R <
50\,\kpc$), we still could have detected \mlya\ emission from
optically thick gas at $R=100\,\kpc$ in $28$ out of $29$ members of
our sample at high statistical significance. Based on all of these
arguments, we conclude that radial distances much larger than the
impact parameters are unlikely, and therefore do not explain our null
detections.

In the Appendix we consider the third explanation, that resonantly
trapped \mlya\ photons are extincted by dust grains, in detail. There
we show that, given reasonable assumptions about the column density
and ionization state of absorbers in the quasar CGM, and rather
conservative assumptions about the metallicity and hence dust content
of this gas, extinction by dust grains can result in a factor of at
most $\sim 2$ reduction in the total \mlya\ emission.  Furthermore,
this estimate applies to the case of \mlya\ photons generated
uniformly throughout the volume of an emitting cloud, as it would be
for collisionally-excited cooling radiation. It thus likely
overestimates the attenuation of fluorescent \mlya\ photons from an
externally illuminated optically thick absorber. In these
self-shielding clouds, \mlya\ photons traverse a smaller optical depth
before escaping than we assumed in the Appendix.

Having cast significant doubt on the other three interpretations, we
consider the last possibility, that the optically thick gas detected
in b/g sightlines is not illuminated by the quasar, as the most
probable explanation for our null detections. In principle this could
result from either of two physical effects, anisotropic emission of
the quasars ionizing radiation, or intermittent emission by the quasar
on timescales comparable to the light crossing time $t_{\rm cross} =
R_{\perp}\slash c \sim 3\sci{5}\,{\rm yr}$ for $R_{\perp} \sim
100\,\kpc$ (see also discussion in QPQ2). Note however that
the frequent detection of extended \mlya\ fuzz at distances
$R_{\perp}\sim 30\,\kpc$ (see \S\ref{sec:fuzz}) indicates
that quasars emit continuously for at least $t_{\rm cross} =
30\,\kpc\slash c$, hence under the intermittency interpretation the
duration of quasar emission bursts would have to follow a very narrow
distribution, which seems rather unphysical.  

We thus favor the anisotropic emission interpretation, which results
if, e.g. the f/g quasar is obscured from some vantage points, and the
optically thick gas we do detect in b/g sightlines is
\emph{shadowed}. In addition to the absence of fluorescent
\mlya\ emission, several independent pieces of evidence support this
hypothesis: (1) in the context of unified models of AGN \citep[see
  e.g.][]{Anton93,elvis00}, there is a wide and growing body of
evidence for a significant population of obscured quasars both in the
local and distant Universe
\citep[e.g.][]{willott00,treister05,gilli07,maiolshem07,treister08,reyes08,stern12},
which implies quasars emit their UV radiation anisotropically; (2) our
comparison of the incidence of optically thick gas transverse to and
along the line-of-sight to quasars indicates an anisotropic
clustering pattern (see Figure~\ref{fig:cov_fact},
\S\ref{sec:cov_fact}, and also QPQ2); (3) the physical conditions of
the absorbing gas in one of our systems SDSSJ\,1204$+$0221 determined
from detailed absorption line modeling, does not show strong evidence
that the f/g quasar is shining on it (QPQ3); (4) we find a large
ratio of low-ion to high-ion equivalent widths (\ion{C}{2}$\slash$ \ion{C}{4})
for the f/g quasars in Table~1 as well as in the parent sample (QPQ5). 
These values are comparable to typical ratios in DLAs, suggesting a 
low-ionization state; (5) at distances of
$r \sim 100$\kpc, several studies indicate that the ionizing radiation
emission from the quasar will photoionize all but the densest gas
\citep{cantalupo05,QPQ2,cmb+08,juna08,pdla09,cantalupo12}.

If the quasar emits its radiation anisotropically, we can crudely
estimate the solid angle of the emission $\Omega$, and consequently
the implied fraction of AGN which are obscured $f_{\rm obscured} = 1 -
\Omega\slash 4\pi$. In the picture we have put forth, clouds detected
as optically thick absorbers are not illuminated and hence do not emit
\mlya, whereas clouds which are illuminated are photoionized,
optically thin, and although they emit \mlya\ radiation, it is not
detectable at our current sensitivity (see discussion in
\S\ref{sec:fuzz}). Note that in this context, the covering factor that
we estimated in \S\ref{sec:cov_fact} is actually the intrinsic
covering factor of optically thick clouds, $f_{\rm C}$, reduced by the
impact of photoionization. If we assume that the quasar emits into a
solid angle $\Omega$, then in our simple model of a uniform
distribution of clouds (see \S\ref{sec:cloud}), we can rewrite
eqn.~(\ref{eqn:cov_fact}) for the observed covering factor $f_{\rm
  C,obs} = f_{\rm C}(1- \Omega\slash 4\pi)$, where $f_{\rm C}$ now
represents the intrinsic covering factor, e.g. before the quasar turns
on. This expression assumes that all illuminated clouds are
photoionized and is valid provided that the intrinsic $f_{\rm C} <
1$\footnote{For values of $f_{\rm C}$ significantly larger than unity,
  this expression breaks down because one needs to take into account
  the constraint that the quasar is always pointing towards the
  observer.}. As we do not have knowledge of the intrinsic covering
factor $f_{\rm C}$, we can only obtain an upper limit on the opening
angle, which assumes that all of the `misses' in
Figure~\ref{fig:cov_fact} occur because our b/g sightline intersects
the volume which has been hollowed out by photoionization. This limit
corresponds to setting the intrinsic covering factor $f_{\rm C}$ to
unity. Then if we take $f_{\rm C,obs}= 0.5$ consistent with our
measurements (Figure~\ref{fig:cov_fact}), we obtain $\Omega\slash 4\pi
< 0.5$. The obscured fraction of $z\sim 2$ quasars is then constrained
to be $f_{\rm obscured} > 0.5$ which is broadly consistent with
independent estimates of the obscured fraction from multi-wavelength
studies of AGN
\citep[e.g.][]{willott00,treister05,gilli07,maiol_shem07,treister08,reyes08,stern12,lusso13}.
However, we emphasize that our model for cloud photoionization is
extremely crude, and surveys for AGN at different wavelengths (X-ray,
optical, near/mid-IR, radio) often come to different conclusions about
obscured fractions \citep[see e.g.][]{lusso13}.

\subsection{Small-Scale $R_{\perp} < 50\,\kpc$ \mlya\ Fuzz}
\label{sec:fuzz}

In the previous section, we argued that the absence of significant
\mlya\ emission from optically thick gas in the f/g quasar CGM
indicates that this material is not illuminated by the quasar.
However, our longslit observations frequently reveal extended
\mlya\ fuzz at a level of ${\rm SB}\sim 0.5-5\sci{-17}\cgssb$ at
distances $R_{\perp} \lesssim 50\,{\rm kpc}$. To quantify the frequency of
this extended emission, both authors independently visually inspected
the maps in Figure~\ref{fig:maps} to search for compelling extended
\mlya\ emission. Each object in our sample (see
Table~1) was classified as having definite fuzz,
ambiguous, or having no fuzz. Classifications were merged according to
the following rules: 
1) an object is classified as definite or no fuzz only if both authors
agreed on that classification;
2) all other objects were considered ambiguous.
Out of our sample of 29 systems, this yields 10
objects with definite fuzz, 8 ambiguous objects, and 11
for which no fuzz is detected. Given that this fuzz tends to
be asymmetrically distributed, and that our single-slit orientation
covers only a fraction of the area around the quasar, the true
detection rate of such fuzz is expected to be significantly
higher. Our detection rate is hence about a factor of two lower than the
frequency of $\sim 50-70\%$ deduced from the more recent studies
of this small-scale \mlya\ fuzz \citep{cjw+06,courbin08,pdla09,north12}, although
limited statistics, sample inhomogenities, differing methodologies, and
our single position angle preclude a quantitative comparison.

We now interpret this small-scale \mlya\ fuzz in terms of the emission
mechanisms discussed in \S\ref{sec:form}. Consider our typical f/g
quasar at $z\simeq 2$, which has $\log_{10} \mlnull \simeq 30.3$.  If
we assume $R_\perp = 35\,\kpc$ as characteristic of the impact
parameters at which we detect the fuzz, the average surface brightness
for optically thick fluorescence given by eqn.~(\ref{eqn:SB_thick}) is
${\rm SB}_{\lya} = 1.0\sci{-15}(f_{\rm C}\slash 1.0)\,\cgssb$. It is
thus quite clear that small-scale fuzz cannot be arising from the same
population of clouds giving rise to the high covering factor $f_{\rm
  C} \gtrsim 0.50$ of optically thick absorbers at $R_{\perp} <
50\,\kpc$ in Figure~\ref{fig:cov_fact}, since the implied surface
brightness would be two orders of magnitude higher.  To be consistent
with optically thick emission, the covering factor of clouds
responsible for the small scale fuzz needs to instead be much lower
$f_{\rm C} \simeq 0.005-0.05$, and this low covering factor matches
the values deduced in previous work on quasar nebulae
\citep[e.g.][]{heckman91a}. 

Considering together both the detected small scale fuzz emission,
which implies a small covering factor $f_{\rm C} \simeq 0.005-0.05$,
and the lack of large-scale nebulosity from gas with a high $f_{\rm C}
\gtrsim 0.50$ covering factor established by b/g sightlines, the
picture that emerges is that clouds with a range of densities and
covering factors exist in the CGM around quasars.  In the absence of
ionizing photons, a low density $n_{\rm H}\sim 0.01-1~{\rm cm^{-3}}$
population dominates the covering factor, and it is this gas that we
typically observe in absorption with b/g sightlines.  In contrast, the
denser clouds ($n_{\rm H}\sim 10-100~{\rm cm^{-3}}$) cover only a tiny
fraction of the line-of-sight $\sim 0.01$.  When illuminated by the
f/g quasar, the lower density (high covering factor) clouds are
photoionized by the intense quasar radiation, whereas the dense clouds
(low covering factor) can self-shield and survive. It is then these
dense clouds which are responsible for the small scale \mlya\ fuzz
that we detect in our sample, and which have been previously detected
around quasars
\citep[e.g.][]{bunker03,weidinger04,cjw+06,francis06,courbin08,pdla09,north12}.
This scenario could also apply to the \mlya\ blobs and HzRGs. Indeed,
it is intriguing that detailed modeling of the spectra of extended
emission-line regions (EELRs) around quasars and FR~II radio galaxies
at low redshift $z\lesssim 0.5$, similarly show evidence for two phase
media composed of an abundant low density medium ($n_{\rm H}\sim
1~{\rm cm^{-3}}$) and much rarer high density clouds \citep[$n_{\rm
    H}\sim 400~{\rm
    cm^{-3}}$][]{Stockton02,stockton06,Fu06,Fu07,Fu09}. It thus seems
plausible that the small-scale fuzz that we detect in \mlya\ are
high-redshift analogs of the EELRs detected around low redshift $z <
0.5$ type-I \citep[e.g.][]{stockton06,husemann12} and type-II
\citep{greene11} quasars, in [\ion{O}{3}] and Balmer lines.

While the explanation of small-scale fuzz above is imminently
plausible, and has indeed been the standard interpretation for
nebulosities around both HzRGs \citep[see
  e.g.][]{msd+90,mccarthy93} and quasars
\citep[e.g.][]{heckman91a}, as emphasized in \S\ref{sec:example}, the
interpretation of \mlya\ emission alone is degenerate, and optically
thin fluorescence and \mlya\ scattering can also result in comparable
emission levels. First consider optically thin fluorescence from a
population of clouds with ($R\,,n_{\rm H}\,,N_{\rm H}\,, f_{\rm C}$) =
($35\,\kpc\,,1.0\,{\rm cm^{-3}}\,,10^{20.1}\,{\rm cm^{-2}}\,,1.0$).
This results in a ${\rm SB}_{\lya} = 10^{-17}\cgssb$ according to
eqn.~(\ref{eqn:SB_thin}), and it is rather intriguing that this column
density $N_{\rm H}$ and volume density $n_{\rm H}$ are comparable to
values deduced from our detailed absorption line modeling of
SDSSJ\,1204$+$0221 QPQ3, and the covering factor consistent
with the large values we deduced in \S\ref{sec:cov_fact}. In other
words, if such clouds are not illuminated by the quasar, they
could account for the optically thick absorption that we observe in
b/g sightlines at $R_{\perp} < 200\,\kpc$. But if illuminated at a distance $R=35\,\kpc$, such
clouds would be so highly ionized that they would appear optically
thin in absorption (e.g. along the line-of-sight). Hence, the small
scale fuzz could be arising from the illuminated (hence optically
thin) analogs of the shadowed (hence optically thick) clouds that we
detect in absorption in background sightlines. 

But if this
interpretation is correct, why don't we see fuzz at larger distances 
$R_{\perp}\simeq 100\,\kpc$?  Recall that in the optically thin
regime, fluorescence does not depend on distance provided that the
clouds remain optically thin, and at $R_{\perp}\simeq 100\,\kpc$ this
would marginally be the case ($\log N_{\rm HI} \simeq 17.2$). Thus
under this interpretation, there should be extended emission at
$R_{\perp}\simeq 100\,\kpc$ which is just as bright as the small-scale
fuzz.  Such emission is definitely not observed in our dataset. As the optically thin surface
brightness scales with the combination $f_{\rm C}N_{\rm H}n_{\rm H}$,
one solution would be to make this product substantially lower at $R\sim
100\,\kpc$ compared to $R\sim 35\,\kpc$, so that the optically thin
emission would then be too faint to be detected.

Now consider the SB levels from \mlya\ scattering.  A model with
($R\,,n_{\rm H}\,,N_{\rm H}\,, f_{\rm C}$) = ($35\,\kpc\,,0.1\,{\rm
  cm^{-3}}\,,10^{20.0}\,{\rm cm^{-2}}\,,1.0$), i.e. nearly identical
to the optically thin case considered above, but with an order of
magnitude lower volume density $n_{\rm H}$, results in scattering
${\rm SB}_{\lya} = 0.9\sci{-17}\cgssb$ (see eqn.~\ref{eqn:SB_scatt})
which is also bright enough to be the mechanism responsible for the
small-scale \mlya\ fuzz. Furthermore, the lower volume density $n_{\rm
  H}$ of this configuration lowers the optically thin fluorescence
discussed above to ${\rm SB}_{\lya} = 8\sci{-19}\cgssb$, almost an
order of magnitude fainter than our typical 2$\sigma$ detection limit ${\rm
  SB}_{\lya} \simeq 6\sci{-18}\cgssb$.  

Thus both optically thick and optically thin fluorescence, as well as
\mlya\ scattering can all plausibly explain the SB levels of the
small-scale fuzz observed at distances $R_{\perp} < 50\,\kpc$ from
quasars. Our simple estimates are crude, and more detailed numerical
treatments which properly treat ionizing and resonant radiative
transfer are required to improve their accuracy as well as make more
detailed predictions about the SB profiles and the kinematics of the
emission from these various processes.  However, we believe that the
key point, that the interpretation of the \mlya\ emission alone is
not unique, is robust. We have argued that additional information
about the gas distribution from absorption lines measurements can
ameliorate this issue. For example, if the covering factor, column
densities, and volume densities of the gas at $R< 50\,\kpc$ can be
statistically characterized, as we have undertaken in the QPQ series,
then we can make better predictions for the optically thin and
scattering surface brightness discussed above.  Moreover, observations
of additional emission line diagnostics, such as a non-resonant Balmer
line (i.e. H$\alpha$), could distinguish between fluorescence and \mlya\ resonant
scattering.  The possibility will always remain that a very low
covering factor component, which cannot be easily mapped with absorption
lines, dominates the emission, as is the case for the low covering
factor optically thick fluorescence discussed at the beginning of this
section. Nevertheless, a comparison of absorption line properties to
emission line measurements yields important constraints on the
physical properties of the gas dominating the covering factor.

To conclude our discussion of the small-scale fuzz, we note that one
can in principle also set a lower limit on the opening angle of the
quasar emission based on the fact that fuzz is frequently detected
extending to a distance $R_{\perp} \simeq 50\,\kpc$. Unfortunately,
given the unknown distance $r$ to the gas, this is not particularly
constraining. If we conservatively assume that the fuzz-emitting gas
has to lie within the quasar halo at $r < 200\,\kpc$, then simple
geometric considerations imply an opening angle $\Omega\slash 4\pi >
0.004$ ($f_{\rm obscured} < 0.996$) corresponding to emission into a
cone with half-angle $> 7^{\circ}$. If we instead assume the unknown
line-of-sight distance $R_{\parallel}$ is comparable to $R_{\perp}$,
then we obtain $\Omega\slash 4\pi > 0.04$ ($f_{\rm obscured} < 0.96$)
corresponding to a cone with half-angle $ > 23^{\circ}$.

\subsection{Comparison to Other Searches for \mlya\ Fluorescence: The
  \mlya\ Emitter in SDSSJ\,0856$+$1158} 

\subsubsection{Previous Work on \mlya\, Fluorescence}

\citet[][see also Cantalupo et al. 2007]{cantalupo12} 
conducted a survey
for fluorescent \mlya\ emission via a narrow band imaging survey of a
$3.5\,{\rm Mpc}\times 3.5\,{\rm Mpc}$ field around a very luminous quasar
with $\log_{10} \mlnull = 31.64$ \citep{dodorico08}, which is 20 times
brighter than the typical f/g quasar in our sample. They detect an
enhancement of sources with high rest-frame equivalent width ${\rm
  EW}_{\lya}> 240\,$\AA\ around the quasar, as compared to the
`field', which they interpret as the signature of
\mlya\ fluorescence.  The significance of the $240\,$\AA\ rest-frame
EW limit is that normal stellar populations are not expected to
produce EWs this large. These fluorescent candidates tend to be
compact and are spatially unresolved in ground based seeing. \citet{cantalupo12}
argue that they are the dense cores of proto-galaxies, which are either
sufficiently dense to self-shield the quasar ionizing radiation and
act as mirrors, or which are highly-ionized but sufficiently dense to
emit observable optically thin \mlya.

Given the different search strategy (narrow band imaging versus slit
spectroscopy), the fact that their observations reach a
\emph{rest-frame} SB\footnote{When comparing the \citet{cantalupo12} results at $z=2.4$ to
our results at $z\simeq 2.0$ we refer to the rest-frame
SB $({\rm erg}\,{\rm s}^{-1}\,{\rm cm}^{-2}\,{\rm kpc}^{-2})$
  which scales as ${\rm SB}_{\rm rest} \propto {\rm SB}_{\rm obs}(1 + z)^4\slash D_{\rm A}$
  where ${\rm SB}_{\rm obs}$ is the observed SB and $D_{\rm A}$ is the angular diameter
  distance.} a factor of $2.5$ deeper than our median limit,
the much brighter quasar that was observed, and the much larger
field-of-view they surveyed, a detailed comparison of our observations 
to the \citet{cantalupo12} results is not straightforward. The
biggest challenge is that the vast majority of the fluorescent candidates 
discovered by \citet{cantalupo12} are at Mpc distances corresponding
to several arcminute separations; whereas, our longslit maps in
Figure~\ref{fig:maps} typically extend only to an impact parameter of
$R_{\perp} \simeq 200\,\kpc$ or about $25\arcsec$. However, note
\citet{cantalupo12} did detect one source very near the quasar with
$\theta = 15\arcsec$ corresponding to $R_{\perp} = 127\,\kpc$,
which overlaps with the scales covered by our observations.  This
source had the highest rest-frame $W_{\lya} > 483\,$\AA\ in
their survey, and its luminosity $L_{\lya} = 9.1\pm 0.7\sci{41}\cgslum$
puts it among a handful of the brightest sources discovered, making it one
of the most compelling fluorescent candidates. 
Consider now the total effective area covered by our
aggregate slit-spectroscopic observations of 29 f/g quasars: $A_{\rm
  eff} = 29\times 150\,\kpc \times 8.7\,\kpc$. Here we compute the
total area interior to a radius of $150\,\kpc$ (comparable to the distance
of the Cantalupo et al. fluorescent emitter) and covered by the $8.7\,\kpc$
width of our $1.0\arcsec$ slits ($z=2$). This area is $1.07$ times
the total area $\pi (150\,\kpc)^2$ around a single quasar. Hence we
are motivated to search our emission maps around foreground quasars
for compact fluorescing sources near the f/g quasar, similar to that
discovered by \citet{cantalupo12}.

\subsubsection{The \mlya\ Emitter in SDSSJ\,0856$+$1158}

Careful visual inspection of the emission maps in
Figure~\ref{fig:maps} reveals statistically significant Ly$\alpha$
emission from a compact object at a distance of $-15\arcsec$ from the
f/g quasar in SDSSJ~\,0856$+$1158, which translates to $R_{\perp} =
131\,\kpc$ at the redshift of the \mlya\ emission line
$z=1.759$. Extracting a spectral aperture at the Ly$\alpha$ emitter
(LAE) location with FWHM set to the seeing of the observations, we
measure the LAE \mlya\ flux to be $f_{\lya} = 0.94\pm 0.14
\sci{-17}\cgsflux$, corresponding to $L_{\lya} = 2.1\pm
0.32\sci{41}\cgslum$.  A gaussian fit to the extracted flux gives a
velocity dispersion of $\sigma_{v} = 147\,\kms$, which corresponds to
$\simeq 2.2$ spectral resolution elements and is hence marginally
spectrally resolved. A similar analysis of the spatial profile gives a
FWHM within $20-50\%$ of the $0.97\arcsec$ seeing, which corresponds
to $8.5\,\kpc$ at the redshift of \mlya. Given the low signal-to-noise
ratio, we assume the source is spatially unresolved, although it could
be marginally resolved. The velocity difference between the LAE and
the f/g quasar redshift is $870\kms$, but this large shift may also
have a contribution from uncertainties in the quasar redshift, which
is derived from the \ion{Mg}{2} emission line and has an associated
uncertainty of $\sigma_{\rm z} = 272\,\kms$ \citep{Richards02}.  The
velocity difference between the LAE and the metal lines in the DLA
(coincident with the f/g quasar) is $-537\,\kms$, which is consistent
with the kinematics expected in the massive dark matter halos hosting
quasars (see e.g. QPQ3) as well as the typical metal-line kinematics
seen in the quasar CGM using b/g sightlines (Prochaska et al., in
prep.).

No continuum emission is detectable by eye at the location of the LAE
over the entire spectrum. To obtain quantitative 
constraints on the continuum level and hence the rest-frame EW, we
consider a 200\,\AA\ region at the red end of LRIS-B spectrum
(3668-3868\AA) where the observations are most sensitive.  The average
single pixel 1$\sigma$ error of our extracted spectrum is
$\sigma_{\lambda} = 4\sci{-19}\cgsspec,$ corresponding to an AB
magnitude $m_{3769} = 25.7$, where $3769$\AA\ is the mean wavelength
over the region considered. Formally, the error on the mean continuum
averaged down over the $n_{\rm pix} = 471$ pixels in this 200\,\AA\
stretch is $\sigma_{\rm cont} \simeq \sigma_{\lambda}\slash
\sqrt{n_{\rm pix}}$. This value is $\sigma_{\rm cont} =
2.0\sci{-20}\cgsspec$ or an AB magnitude of $m_{3769} = 29.0$, and the
corresponding $1\sigma$ lower limit on the rest-frame equivalent width
would be $W_{\lya} \approx f_{\lya}\slash\sigma_{\rm cont}
> 173$\AA. In practice, systematic errors in the spectroscopic data
reduction probably prevent us from achieving this formal sensitivity
to the mean continuum, and hence we would not actually have been able
to detect a 29th magnitude source $\sim 20$ times fainter than the $\sigma_{\lambda}$
flux level. Note however that the mean extracted flux over this
spectral region at the location of the LAE is $0.04\sigma_{\rm cont}$,
and this fluctuation is at least consistent with being drawn from a Gaussian distribution
centered on zero with standard deviation $\sigma_{\rm cont}$, i.e. the data does not
show strong evidence for systematic error in the average flux level.
We nevertheless adopt $W_{\lya} > 50$\AA\ as a conservative lower limit
on rest-frame equivalent width, which corresponds to using roughly
$3.5\sigma_{\rm cont}$ as the continuum value, or assuming that we
could have detected a source four times fainter the than 
$\sigma_{\lambda}$ pixel noise corresponding to $m_{3769}=27.6$.

How likely is the discovery of a high $W_{\lya}$ LAE within
$15\arcsec$ and at small velocity separation from a quasar at $z\sim
2$? The answer clearly depends on the physical process powering the \mlya\, 
and we consider two possibilities: 
\begin{itemize}
\item The \mlya\ is powered by star-formation just as for the general
  LAE population.  Clustering enhances the probability of finding
  such galaxies close to a quasar.
\item The \mlya\ emission is fluorescence powered by the f/g quasar's ionizing radiation. 
\end{itemize}
We consider both possibilities in turn. 

\subsubsection{Clustering of LAEs around Quasars}
\label{sec:clust}
The expected number of LAEs with rest-frame EW $> W_{\lya}$ within a
volume $V$ around a quasar can be written 
\be N_{\rm LAE}(>W_{\lya})=
{\bar n}_{\rm LAE} P(>W_{\lya})\int_{V} dV \left[1 + \xi_{\rm
    QG}(r)\right]\label{eqn:corr}, 
\ee 
where ${\bar n}_{\rm LAE}$ is
the average number density above the canonical rest-frame EW threshold
$W_{\lya}> 20$\AA\ for selecting LAEs, $P(>W_{\lya})$ is the
cumulative distribution of LAE EWs, and the integration over the
correlation function $\xi_{\rm QG}$ quantifies the increase due to
quasar-LAE clustering. 

Although there is strong evidence that the powerful HzRGs trace large
proto-cluster like overdensities at $z > 2$
\citep[e.g.][]{penter00,debreuck04,venemans05,kuiper12}, the
clustering of galaxies around more typical radio-loud quasars like
SDSSJ\,0856$+$1158 and/or the radio-quiet quasars which dominate our
sample, appears to be much more modest
\citep[e.g.][]{fynbo01,as05,trainor12}.  We thus assume that the
clustering of LAEs around quasars at $z\sim 2$ is the same as that of
the continuum selected LBG population at $z\sim
2.7$. \citet{trainor12} (see also Adelberger \& Steidel 2005) measured
the cross-correlation between LBGs and quasars at $z\sim 2.7$, and
assuming a power law form $\xi=(r\slash r_{\rm 0,QG})^{-\gamma}$ with
$\gamma=1.5$, they measured a quasar-galaxy
cross-correlation\footnote{Although the \citet{trainor12} study
  focused on hyperluminous quasars, whereas the f/g quasar we consider
  here are not hyperluminous, there is no compelling evidence that
  $z\sim 2-3$ quasar clustering strength varies with luminosity in
  either the auto-correlation \citep{Shen08} or the cross-correlation
  with galaxies \citep{trainor12,as05}.} length of $r_{\rm
  0,QG}=7.3\hMpc$ (comoving). We integrate this correlation function
over a cylindrical volume defined by an annular region $[R_{\rm
    \perp,min},R_{\rm \perp,max}]$ and a line-of-sight extent $\pm
\Delta v\slash[H(z)]$.  The minimum radius is set to $R_{\rm
  \perp,min}= 50\,\kpc$ (proper), motivated by the fact that it could
be difficult to distinguish LAE emitters from the ubiquitous
small-scale \mlya\ fuzz seen around quasars. The maximum radius is set
to $R_{\rm \perp,max}=130\,\kpc$ (proper; $15\arcsec$ at $z\sim 2$),
which is approximately the impact parameter of the LAE in
SDSSJ\,0856$+$1158 and the bright fluorescence candidate discovered
near the quasar by \citet{cantalupo12}.  Along the line-of-sight, we
take $\Delta v = 1500\,\kms$, which encompasses both gravitational
motions of galaxies in the quasar halo, quasar redshift errors, and
also agrees with the interval used in the clustering analysis of
\citet{trainor12}. Furthermore, the 4nm narrow-band filter used by
\citet{cantalupo12} also translates to a line-of-sight velocity
interval of $2900\,\kms$ or about $\simeq 2\times \Delta v$.  Based on
these assumptions, we find that over this volume, clustering alone
enhances the number density of LAEs around quasars by a factor of $19$
above the cosmic average.

For the background number density of LAEs ${\bar n}_{\rm LAE}$, we
first need to decide which \mlya\ luminosity threshold to
consider. The LAE in SDSSJ\,0856$+$1158 has $L_{\lya} =
2.1\sci{41}\cgslum$, but owing to the long-exposure (4800s) and relatively low
redshift $z=1.76$, this represents the most sensitive observation in
our sample. As we wish to consider our sample in aggregate, we scale
this value up by a factor of three, which represents the ratio of
the luminosity limit in SDSSJ\,0856$+$1158 to the median value of our
sample. 
This value of $L_{\lya} = 6.4\sci{41}\cgslum$ is comparable to
the luminosity of the fluorescent candidate detected closest to the
quasar $L_{\lya} = 9.1\sci{41}\cgslum$ from \cite{cantalupo12}, and
also matches the median luminosity of their best fluorescence
candidates. We use the \citet{ciardullo12} Schechter luminosity
function fits to the $z=2.1$ LAE survey data of \citet{guaita11}, to
determine that the comoving number density of LAEs with $L_{\lya} >
6.4\sci{41}\cgslum$ is ${\bar n}_{\rm LAE} = 5.1\sci{-3}\,h^3{\rm 
  Mpc^{-3}}$. \citet{ciardullo12} also estimated the rest-frame
equivalent width distribution, and found that $z=2.1$ LAEs selected to
have rest-frame $W_{\lya} > 20$\AA, follow an exponential distribution
with rest-frame scale length $W_0=50$\AA. Hence the cumulative
probability distribution of rest-frame EW is $P(> W_{\lya}) =
\exp{[-(W_{\lya} - 20 {\rm \AA})\slash W_0]}$. Given our conservative
lower limit for the rest-frame $W_{\lya} > 50$\AA\ of the LAE in
SDSSJ\,0856$+$1158 we obtain a probability $P(> 50\,{\rm \AA}) =
0.55$, whereas our formal limit of $W_{\lya} > 173$\AA\ gives $P(>
173\,{\rm \AA}) = 0.047$. Putting all the pieces together (see
eqn.~\ref{eqn:corr}), we find that the expected number of LAEs at
impact parameter $50\,\kpc < R_{\perp} < 130\,\kpc$ per quasar is
$N_{\rm LAE} = 0.34$ for our conservative lower limit $W_{\lya} >
50$\AA\, and $N_{\rm LAE} = 0.029$ for our formal lower limit
$W_{\lya} > 173$\AA. Note that this calculation is conservative and
would tend to over-predict the number of LAEs near quasars for two
reasons.  First, we used the quasar-LBG correlation function, but LBGs
are known to cluster slightly stronger than LAEs.  Second, quasar
clustering varies rather strongly with redshift \citep{Shen08} and so
the \citet{trainor12} quasar-galaxy clustering measurements at $z\sim
2.7$ probably overestimate the clustering around $z\sim 2$ quasars.

But given that we expect $N_{\rm LAE}$ sources in the annular region
between $R_{\rm \perp,min}$ and $R_{\rm \perp,max}$ around a given
quasar, what is the probability that one such source should land on
our slit given our observed sample? As we used our median luminosity
limit above (i.e. the limit in SDSSJ\,0856$+$1158 scaled up by three)
to compute the number density of LAEs ${\bar n}_{\rm LAE}$, we assume
that we could have detected a $L_{\lya} > 6.4\sci{41}\cgslum$ in 15/29
sources in our sample. Each longslit observation covers 3\% of the
annular region around a given quasar, hence the total number of LAEs
expected in our sample is $15\times 0.03\times N_{\rm LAE}$, which
equals $0.15$ LAEs for our conservative EW limit, and $0.013$ for our
formal equivalent width limit. Although our calculation is crude, and
does not fully take into account the sensitivity variations in our
sample, it seems rather unlikely that the clustering of LAEs around
the quasar can trivially explain our discovery of a relatively high-EW
LAE at $R_{\perp} = 131\,\kpc$ from the f/g quasar in
SDSSJ\,0856$+$1158.

\subsubsection{Is SDSSJ\,0856$+$1158 Powered by Fluorescence?}

It is intriguing that the properties of the LAE in SDSSJ\,0856$+$1158
resemble those of the candidate fluorescing source closest to the
quasar in \citet{cantalupo12}. Namely the two sources are at nearly
identical impact parameter and our LAE is only a a factor of four
fainter then the \citet{cantalupo12} candidate.  Is the luminosity of
the SDSSJ\,0856$+$1158 LAE consistent with \mlya\ mirror fluorescence?
Taking the distance from the quasar to be the impact parameter
$R_{\perp}$, we compare the mirror surface brightness (see
eqn.~\ref{eqn:SB_mirror}) to the observed $L_{\lya}$, to deduce the
area, and hence size of the emitting `mirror'. For a spherical source
the implied diameter is $d_{\rm LAE} = 2.5\,\kpc$ which subtends an
angle $\theta_{\rm LAE} = 0.29\arcsec$. This small size is consistent
with our assessment that the LAE in SDSSJ\,0856$+$1158 is likely
spatially unresolved, although we reiterate that the data are 
noisy. Note also that our inferred size depends on two unknown
quantities, namely the geometric reduction factor $f_{\rm gm}$ (we
assumed $f_{\rm gm} =0.5$) and the line-of-sight distance to the
quasar $R_{\parallel}$ (we assumed $R_{\parallel} = 0$), and scales as
$d_{\rm LAE}\propto f_{\rm gm}^{-1\slash 2}[1 + (R_{\parallel}\slash
  R_{\perp})^2]^{1\slash 2}$.  It is thus quite plausible that the
source could be a factor of $\sim 3$ larger and hence marginally
resolved.  We conclude that, within these uncertainties, the
luminosity of the LAE in SDSSJ\,0856$+$1158 is consistent with the
mirror fluorescence interpretation, given our observational
constraints on its distance, size, and the ionizing luminosity of the
foreground quasar.

Again, how likely is it that one such fluorescing source should land
on our slit given our observed sample? Here we can only speculate, but
the single \citet{cantalupo12} fluorescing source provides
circumstantial evidence that the number of such sources
expected within the annular region considered is approximately unity
$N_{\rm LAE}\sim 1$. Analogous to our argument at the end of
the previous section, assuming we had the sensitivity to detect such a
source in roughly half of our sample, and that our slit covers 3\% of
the annular region, we get $15\times 0.03\times N_{\rm LAE}$ or 0.46. This
is a factor of $\sim 3$ larger than the number of LAEs
expected from clustering around the quasar, and we hence favor the fluorescence interpretation, 
although we emphasize that our estimates, particularly for the
fluorescence, are very crude. 

The two physical scenarios, clustering and fluorescence, can be
distinguished by obtaining sensitive constraints on the rest-frame EW.
For instance, if we adopt our formal limit on the EW in
SDSSJ\,0856$+$1158 $W_{\lya} > 173\,{\rm \AA}$, then the probability
of finding one such LAE in our sample due to quasar-LAE clustering is
$\sim 1\%$, which would make a stronger case for
fluorescence. Furthermore, if it were established that $W_{\lya} >
240$\AA, then this would be very compelling evidence for fluorescence
because normal stellar populations are not expected to produce EWs above
this value.  Given our current
relatively weak constraints on the EW, it is not possible to make a
convincing statement about whether the LAE in SDSSJ\,0856$+$1158
results from clustering or fluorescence. Deep broad band imaging of
the LAE in SDSSJ\,0856$+$1158 could be used to obtain a much more
sensitive constraint on the continuum, distinguishing between these
two scenarios.

\section{Summary and Concluding Remarks}
\label{sec:summary}

In this paper we introduced a novel technique to simultaneously
analyze the absorption line and emission line properties of CGM and
IGM gas around a quasar using close projected quasar pairs. We analyze
the absorption spectra of the b/g quasars in a sample of 68 projected
quasar pairs, and identify 29 systems for which the spectrum of the
b/g quasar shows evidence for an LLS coincident with the redshift of
the f/g quasar. The presence of self-shielding circumgalactic gas,
pinpointed by the b/g sightline, makes these 29 objects prime
candidates for the detection of \mlya\ emission from the CGM, either
from fluorescent recombinations powered by the quasar, pure resonant
scattering of \mlya\ photons emitted by the quasar, or \mlya\ cooling
radiation from $T\sim 10^4\,{\rm K}$ circumgalactic gas. We perform a
systematic, slit-spectroscopic survey for extended \mlya\ emission in
the vicinity ($R_{\perp} < 600$\,kpc) of these 29 $z \sim 2$ quasars,
achieving typical $1\sigma$ surface-brightness limits of ${\rm
  SB}_{\rm \lya} \simeq 3\sci{-18}\,\cgssb$. The primary results of
this survey and our analysis are:

\begin{itemize}

\item Fluorescent \mlya\ emission with properties expected in the
  `mirror' approximation, is \emph{not} detected at the locations of
  any of the 29 optically thick absorbers pinpointed by the b/g quasar
  sightlines, although we had the sensitivity to detect this signal in
  24/29 sources at $> 10\sigma$ significance. We thus seriously
  question the reality of the purported detection of this phenomenon
  by \citet{ask+06}.

\item We measure the covering factor of optically thick absorption in
  the quasar CGM to be $f_{\rm C}\gtrsim 0.5$ on $R_{\perp}\lesssim
  200\,\kpc$ scales from our sample of 29 absrobers detected in a
  parent sample of 68 pair sightlines. This high covering factor
  implies a level of large scale $\sim 100\,\kpc$ diffuse
  \mlya\ emission that would have been easily detected for 28/29
  sightlines we searched for emission.  Despite this fact, we only
  detected large scale \mlya\ emission in a single system.

\item Both the lack of mirror fluorescence at the discrete location of
  absorption in the b/g sightline, and the lack of diffuse fluorescent
  emission extended along the slit at $R_{\perp} \sim 100\,\kpc$ can
  be explained if the gas absorbing b/g sightlines is shadowed from
  the f/g quasar ionizing radiation, due to the obscuration which is
  expected from unified models of AGN. This same obscuration argument
  (previously put forth in QPQ1 and QPQ2), also explains the anisotropic
  distribution of absorbers around f/g quasars (QPQ2). That is, 
  the very high incidence of optically thick absorption observed in b/g 
  sightlines $f_{\rm C}\gtrsim 0.5$, compared to the much lower
  incidence of strong absorption along the line-of-sight.
  In this context, we crudely estimate the average opening angle of
  our f/g quasar UV emission to be $\Omega < 2 \pi$ which implies an
  obscured fraction $f_{\rm obscured} > 0.5$, consistent with values
  deduced from multi-wavelength studies of AGN.

\item Extended \mlya\ fuzz is detected on small-scales $R_\perp <
  50\,\kpc$ at levels of ${\rm SB}_{\lya}\simeq 0.5-5\sci{-17}\cgssb$
  in 10/29 objects, or 34\% of our sample. Given that our single slit
  orientation only covers a fraction of the area around the quasar,
  the true detection rate is expected to be significantly higher $\sim
  50-70\%$, as deduced by other recent studies. 
  This \mlya\ fuzz is far too faint to be arising from the high
  covering factor $f_{\rm C}\simeq 0.5$ optically thick gas we
  frequently detect in b/g sightlines.  If the emission arises from
  optically thick gas, the covering factor has to be very low $f_{\rm
    C}\sim 0.01$, which could arise from a population of rare clouds
  with very high density sufficient to self-shield against the intense
  quasar radiation; however this interpretation is not unique. This
  emission is likely the high-redshift analog of the extended
  emission-line regions (EELRs) commonly observed around low-redshift
  ($z < 0.5$) quasars.

\item A Ly$\alpha$-emitter with high-equivalent width $W_{\lya} >
  50$\AA\ (rest-frame) and a luminosity $L_{\lya} = 2.1\pm
  0.32\sci{41}\cgslum$ is detected at a distance of $134\,\kpc$ from
  the f/g quasar in SDSSJ\,0856$+$1158. Assuming the \mlya\ is powered
  by star-formation, we estimate the probability of discovering such
  an LAE in our survey, including the enhancement due to clustering
  around the quasar, to be small $\lesssim 15\%$. This source has properties
  comparable to the fluorescent emission candidates recently uncovered by 
  \citet{cantalupo12}, and their discovery of fluorescent LAE at about the same
  impact parameter from the quasar, provides circumstantial evidence that the
  LAE in SDSSJ\,0856$+$1158 is also powered by fluorescence. 

\end{itemize} 

In the next few years, new sensitive optical integral-field
spectrometers will be coming online on large-aperture telescopes, such
as the \emph{Multi Unit Spectroscopic Explorer} \citep[MUSE;][]{MUSE}
and the \emph{Keck Cosmic Web Imager} \citep[KCWI;][]{KCWI}.  These image
slicing integral-field units (IFUs) will enable 3-d spectroscopy over
$\sim 1\arcmin \times 1\arcmin$ fields-of-view, comparable to the
linear angular dimension we mapped out around quasars with our
$1.0\arcsec$-wide slit observations.  The multiplexing gain in speed
of these instruments is thus a factor $\sim 60$ for mapping out
emission from the quasar CGM. It is anticipated that these instruments
will achieve SB limits ${\rm SB}_{\rm \lya}\simeq 3\sci{-19}\cgssb$ in $\sim
50$ hour integrations. 

In light of the the results in this paper, it is interesting to
speculate about what these instruments might detect at these
unprecedented sensitivity levels.  Consider a quasar with $\log_{10}
L_{\nu_{\lya}} = 31$, which is comparable to the brightest sources
among the 29 we considered, corresponding to a typical $i$-band
magnitude $i=17.7$, or about $1.6$ magnitudes brighter than our median
source. Based on the statistics of absorption in background sightlines
(e.g. see Figure~\ref{fig:cov_fact} and QPQ1,QPQ2,QPQ5) and our
detailed analysis of the absorption in a single system (QPQ3), our
favored model for the properties of the quasar CGM is ($R\,,n_{\rm
  H}\,,N_{\rm H}\,, f_{\rm C}$) = ($100\,\kpc\,,0.1\,{\rm
  cm^{-3}}\,,10^{20}\,{\rm cm^{-2}}\,,0.5$). If illuminated by the
quasar, such clouds will be highly ionized and optically thin to
ionizing radiation, but optically thick to \mlya\ photons.  The
predicted level of optically thin fluorescence is ${\rm SB}_{\lya} =
3.8\sci{-19}\cgssb$ and scales as the product $n_{\rm H}N_{\rm
  H}$. Pure \mlya\ scattering gives ${\rm SB}_{\lya} =
2.7\sci{-18}\cgssb$. The \mlya\ scattering would be easily detectable
by these future instruments for long integrations, and even the
fainter optically thin emission would also likely be detectable if one
averages spatially. We thus conclude that our observations and
calculations imply that future deep integrations by MUSE and KCWI are
expected to uncover glowing \mlya\ emission on $R\sim 100\,\kpc$
scales around nearly every bright quasar, allowing one to directly
image the CGM.  The level and kinematics of this emission, when
combined with absorption line information, will provide new
information about properties of the quasar CGM, and a new window in
studying the physics of galaxy formation in the massive halos hosting
quasars.

\acknowledgments 

We acknowledge helpful discussions with S. Cantalupo about
\mlya\ fluorescence, B. Draine about the extinction of \mlya\ by dust
grains, and J. Hodge about the FIRST catalog. We also thank the members of
the ENIGMA group\footnote{http://www.mpia-hd.mpg.de/ENIGMA/} at the
Max Planck Institute for Astronomy (MPIA) for helpful discussions.
JFH acknowledges generous support from the Alexander von Humboldt
foundation in the context of the Sofja Kovalevskaja Award. The
Humboldt foundation is funded by the German Federal Ministry for
Education and Research.  JXP acknowledges support from the National
Science Foundation (NSF) grant AST-1010004, and thanks the Alexander
von Humboldt foundation for a visitor fellowship to the MPIA where
part of this work was performed, as well as the MPIA for hospitality
during his visits.

Much of the data presented herein were obtained at the W.M. Keck
Observatory, which is operated as a scientific partnership among the
California Institute of Technology, the University of California, and
the National Aeronautics and Space Administration. The Observatory was
made possible by the generous financial support of the W.M. Keck
Foundation.  Some of the Keck data were obtained through the NSF
Telescope System Instrumentation Program (TSIP), supported by AURA
through the NSF under AURA Cooperative Agreement AST 01-32798 as
amended.

Some of the data herein were obtained at the Gemini Observatory, which
is operated by the Association of Universities for Research in
Astronomy, Inc., under a cooperative agreement with the NSF on behalf
of the Gemini partnership: the NSF (United
States), the Science and Technology Facilities Council (United
Kingdom), the National Research Council (Canada), CONICYT (Chile), the
Australian Research Council (Australia), Minist\'{e}rio da
Ci\^{e}ncia, Tecnologia e Inova\c{c}\~{a}o (Brazil) and Ministerio de
Ciencia, Tecnolog\'{i}a e Innovaci\'{o}n Productiva (Argentina). 

The authors wish to recognize and acknowledge the very
significant cultural role and reverence that the summit of Mauna Kea
has always had within the indigenous Hawaiian community. We are most
fortunate to have the opportunity to conduct observations from this
mountain.

\begin{figure*}
  \begin{minipage}{0.41\textwidth}
    \centering{\epsfig{file=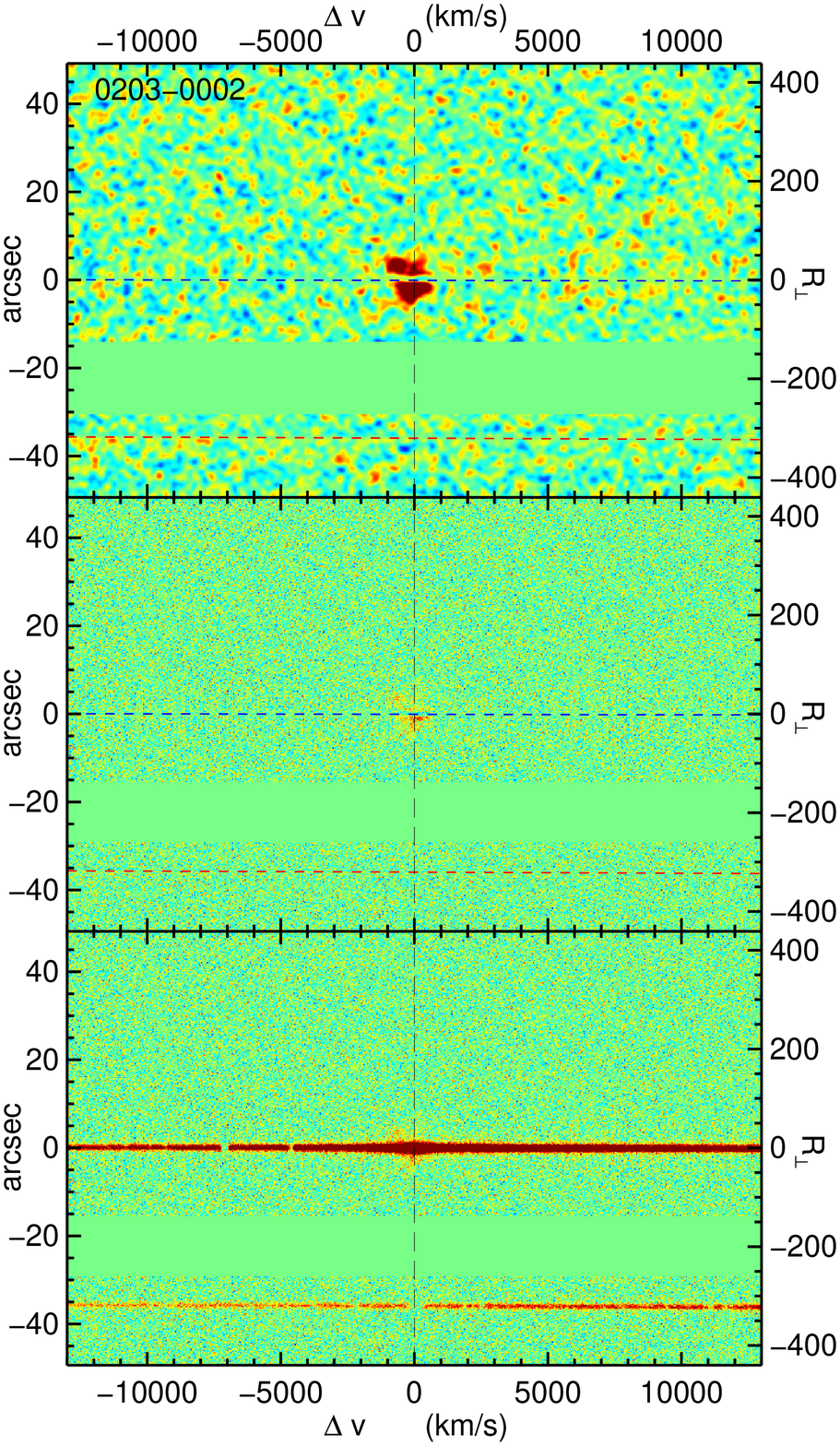,bb=10 0 540 860,width=\textwidth,clip=}}
    \centering{\epsfig{file=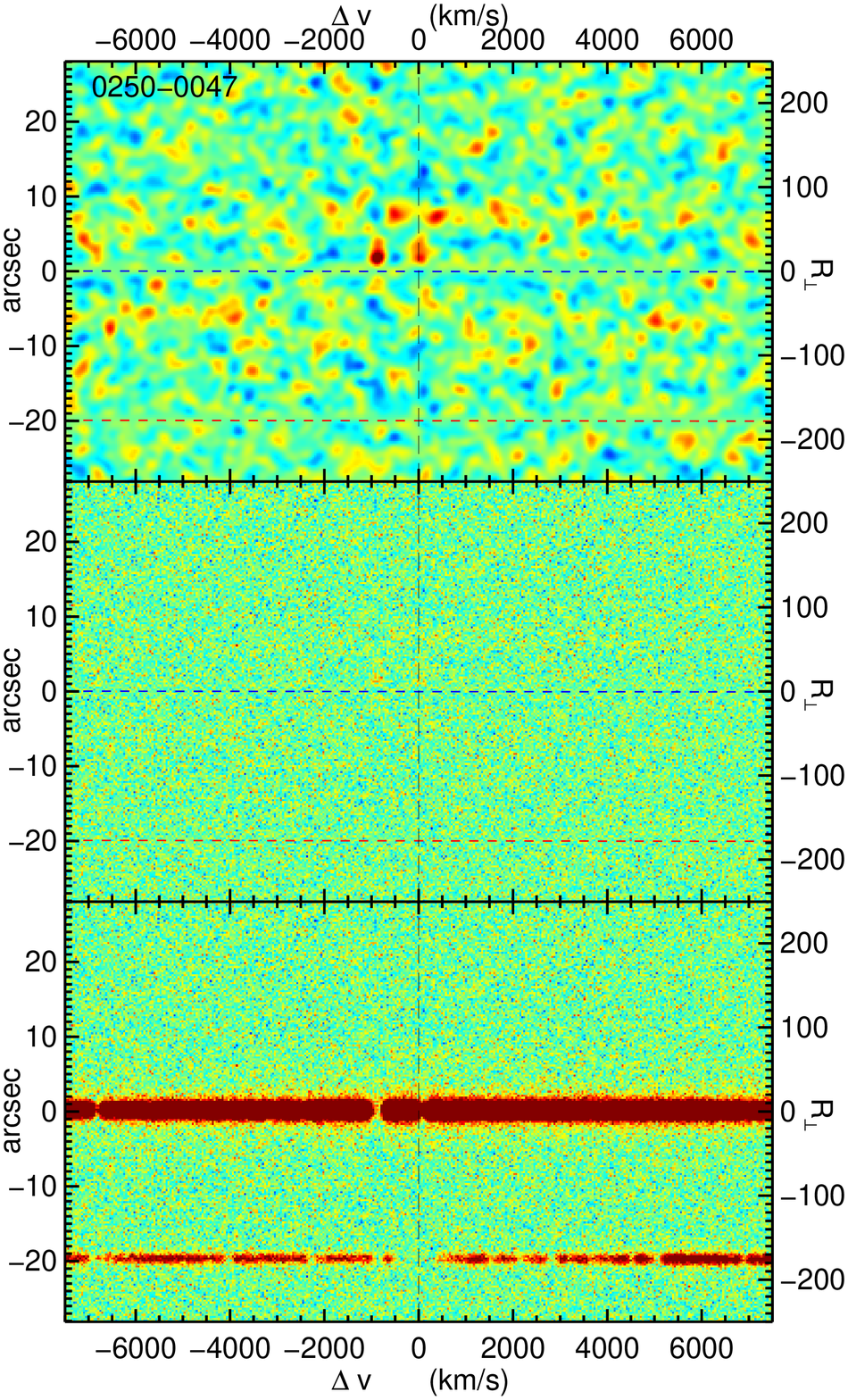,bb=10 0 540 860,width=\textwidth,clip=}}
  \end{minipage}
  \begin{minipage}{0.06\textwidth}
    \vskip 0.55cm
    \hskip -0.2cm
    \epsfig{file=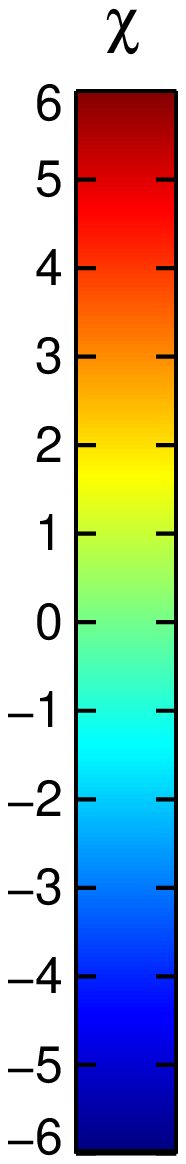,bb=5 0 55 360,width=\textwidth,clip=}
  \end{minipage}
  \begin{minipage}{0.41\textwidth}
   \centering{\epsfig{file=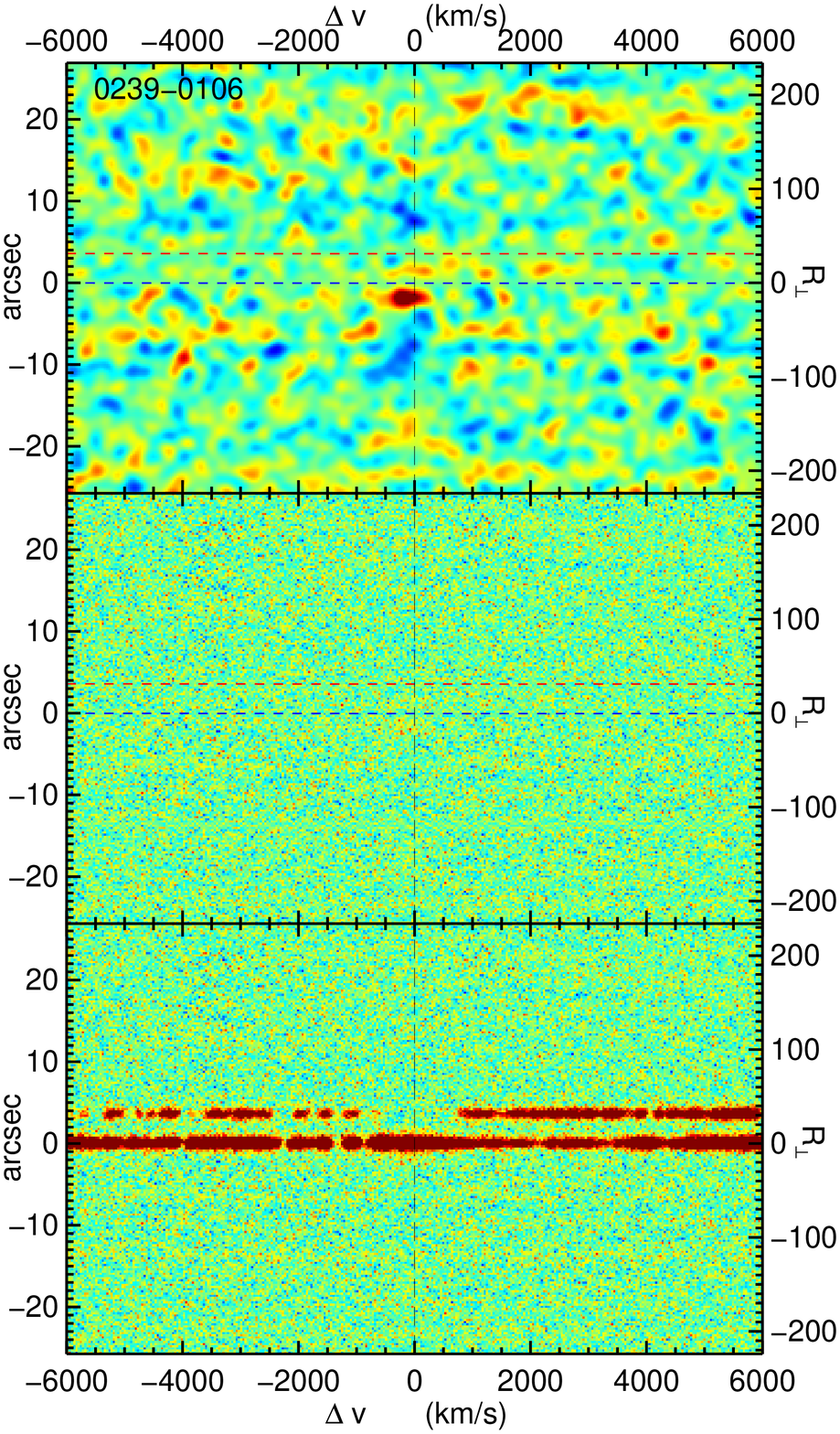,bb=10 0 540 860,width=\textwidth,clip=}}
   \centering{\epsfig{file=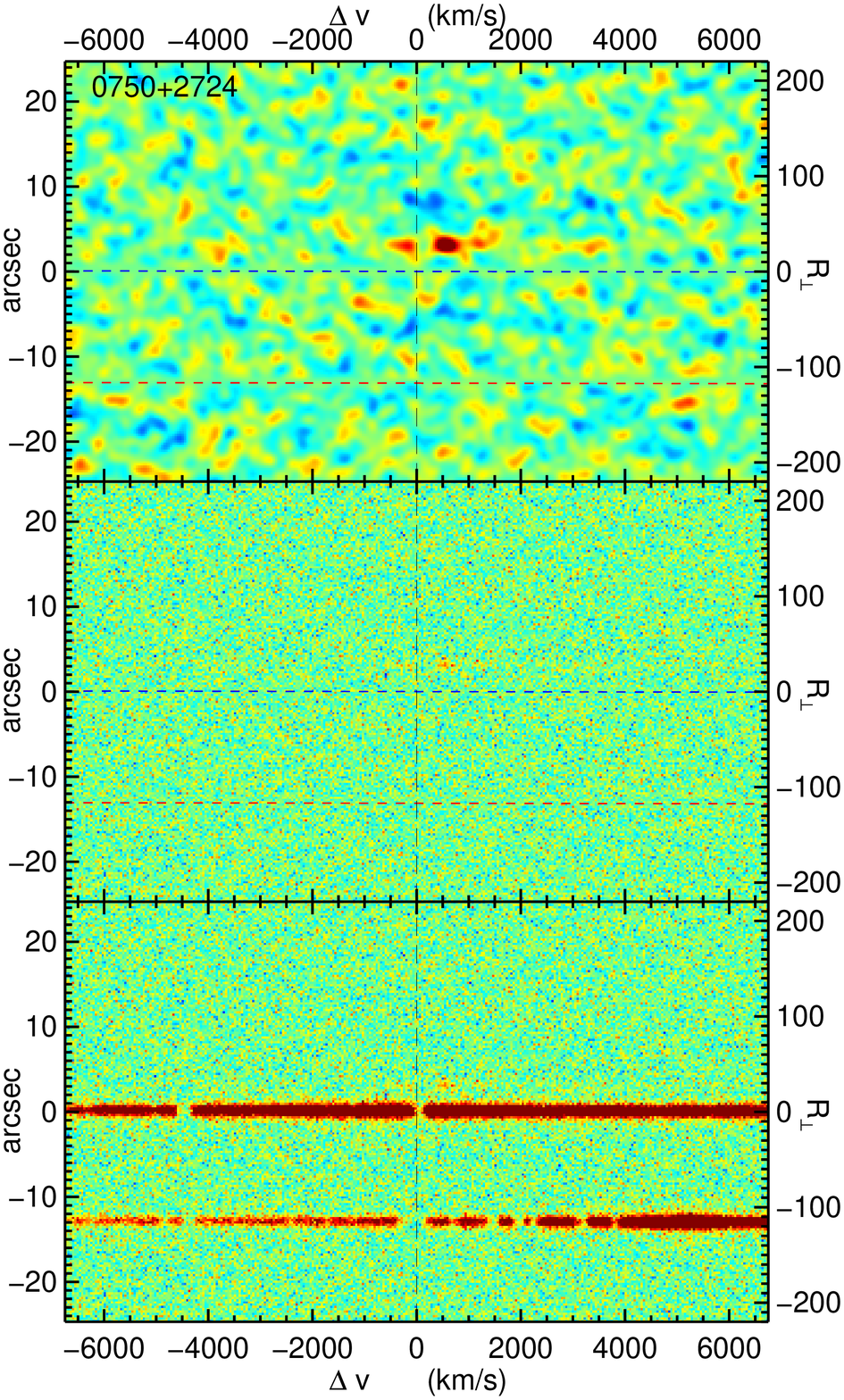,bb=10 0 540 860,width=\textwidth,clip=}}
  \end{minipage}
  \caption{ Two dimensional spectra plotted as $\chi$-maps for our
    entire projected quasar pair sample. The bottom and middle panels
    show $\chi_{\rm sky}$ (sky-subtracted only) and $\chi_{\rm
      sky+PSF}$ (sky and PSF subtracted), respectively, as
    defined in \S\ref{sec:coadd}. In the absence of extended
    emission, the distribution of pixel values in the the $\chi_{\rm
      sky+PSF}$ (middle panels) images should be Gaussian with unit
    variance (see Figure~\ref{fig:chi_histo}). All images are displayed
    with a linear stretch ranging from $-6\sigma$ to $6\sigma$ as
    indicated by the central color bar. Upper panels show smoothed
    maps $\chi_{\rm smth}$ (see \S\ref{sec:coadd}), which are
    helpful for identifying extended emission. The stacked images were smoothed
    with a symmetric Gaussian kernel (same spatial and spectral
    widths) with dispersion $\sigma_{\rm smth} = 100\,\kms$ (FWHM
    $=235\,\kms$).  For LRIS this corresponds to spatially to $0.60-0.70\arcsec$, 
    whereas for GMOS it is $0.80-0.86\arcsec$. Note that $\chi_{\rm smth}$ 
    is the ratio of smoothed image to smoothed noise, and will hence 
    no longer obey Gaussian statistics. Nevertheless, the same stretch and
    color-map are used. The horizontal red and blue dashed lines show the
    spectral traces for the f/g and b/g quasar, respectively, and the
    vertical red dotted line indicates $\Delta v=0$, corresponding to
    the location of the optically thick absorber coincident with the
    f/g quasar redshift. The name of the target is displayed at the upper
    left.\label{fig:maps}
}
\end{figure*}

\begin{figure*}
  \setcounter {figure}{4}
  \begin{minipage}{0.41\textwidth}
    \centering{\epsfig{file=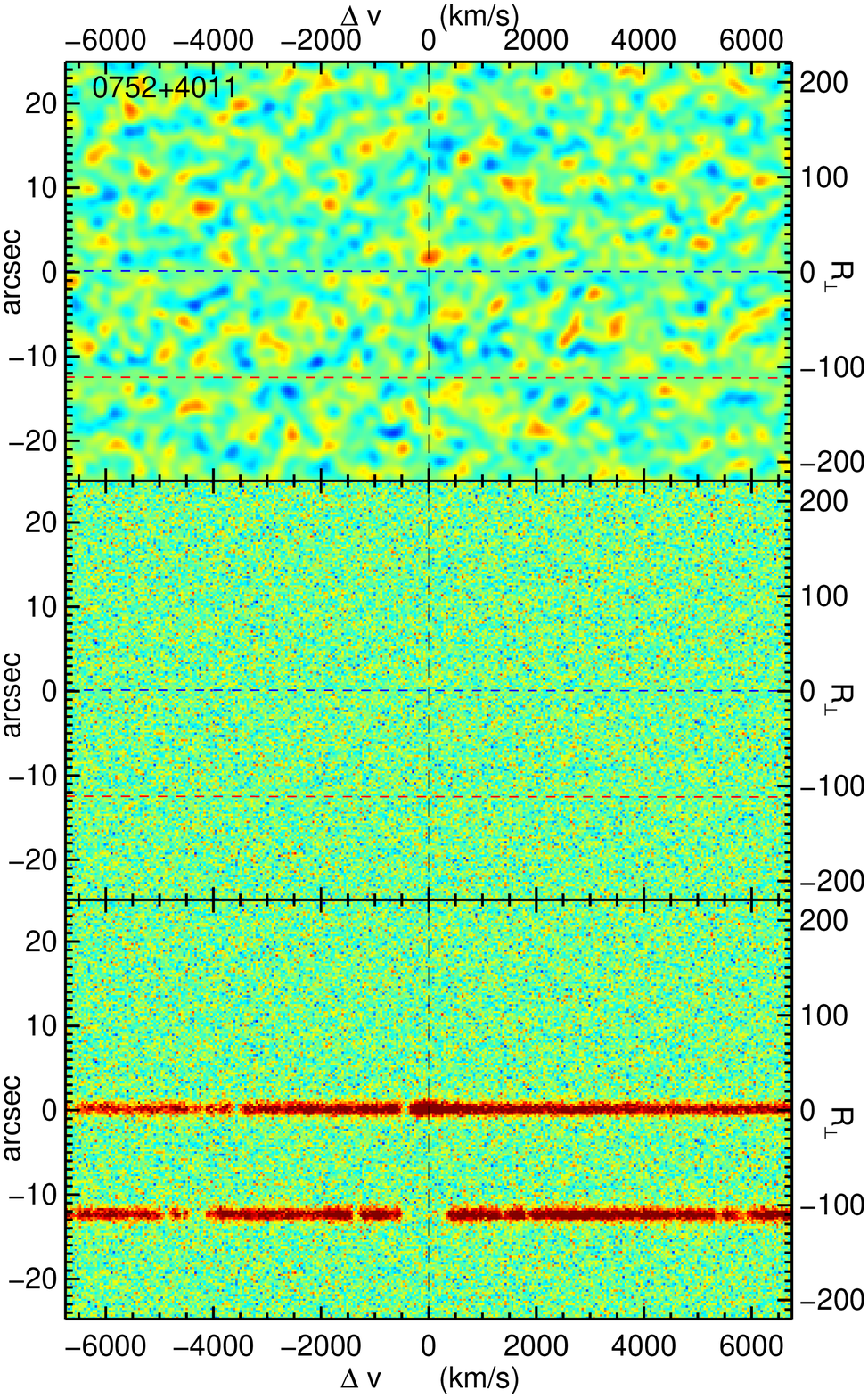,bb=10 0 540 860,width=\textwidth,clip=}}
    \centering{\epsfig{file=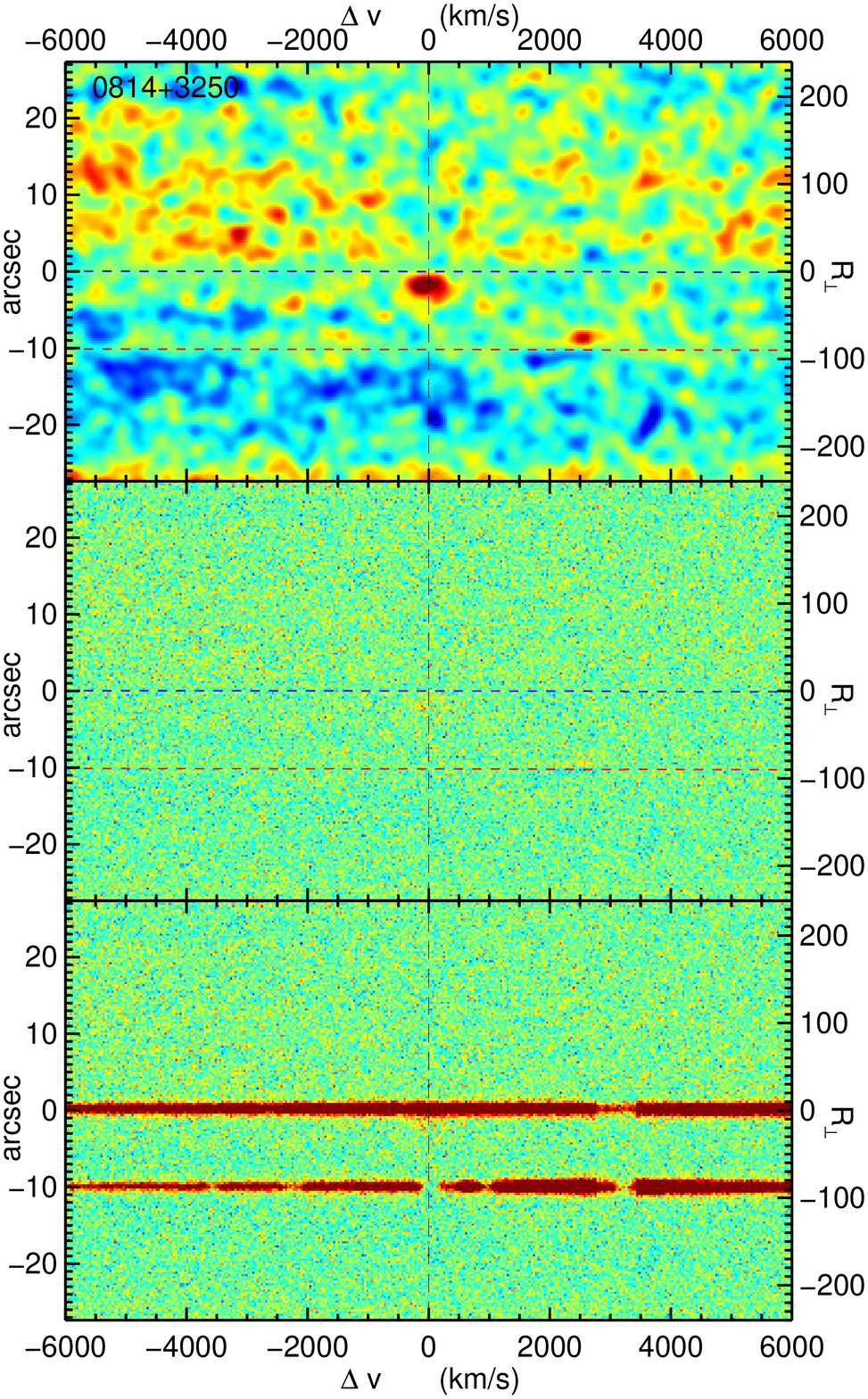,bb=10 0 540 860,width=\textwidth,clip=}}
 \end{minipage}
  \begin{minipage}{0.06\textwidth}
    \vskip 0.55cm
    \hskip -0.1cm
    \epsfig{file=f5a.eps,bb=5 0 55 360,width=\textwidth,clip=}
 \end{minipage}
  \begin{minipage}{0.41\textwidth}
    \centering{\epsfig{file=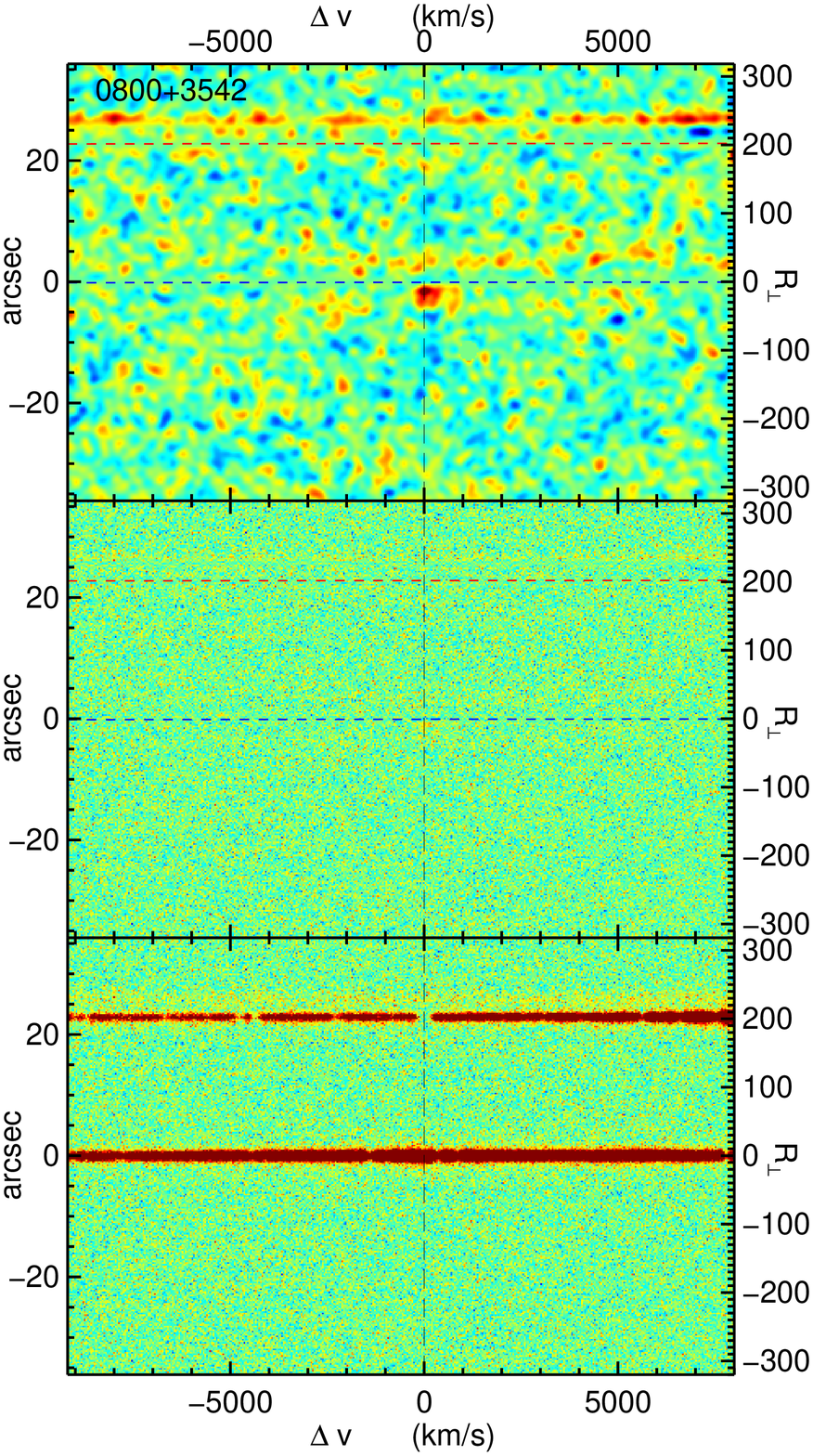,bb=10 0 540 860,width=\textwidth,clip=}}
    \centering{\epsfig{file=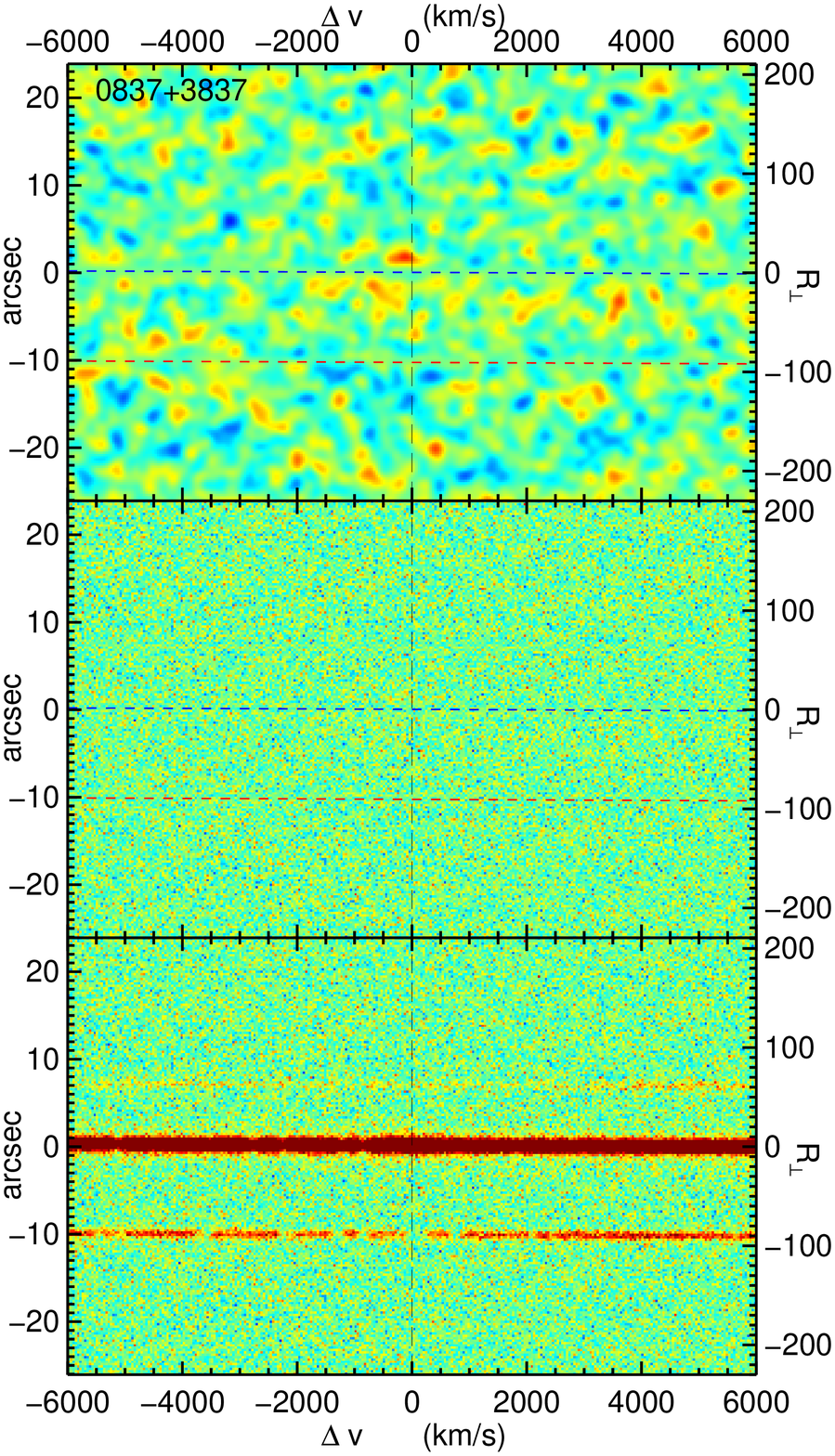,bb=10 0 540 860,width=\textwidth,clip=}}
  \end{minipage}
  \caption{Continued\label{fig:maps2}.}
\end{figure*}

\begin{figure*}
  \setcounter {figure}{4}
  \begin{minipage}{0.41\textwidth}
    \centering{\epsfig{file=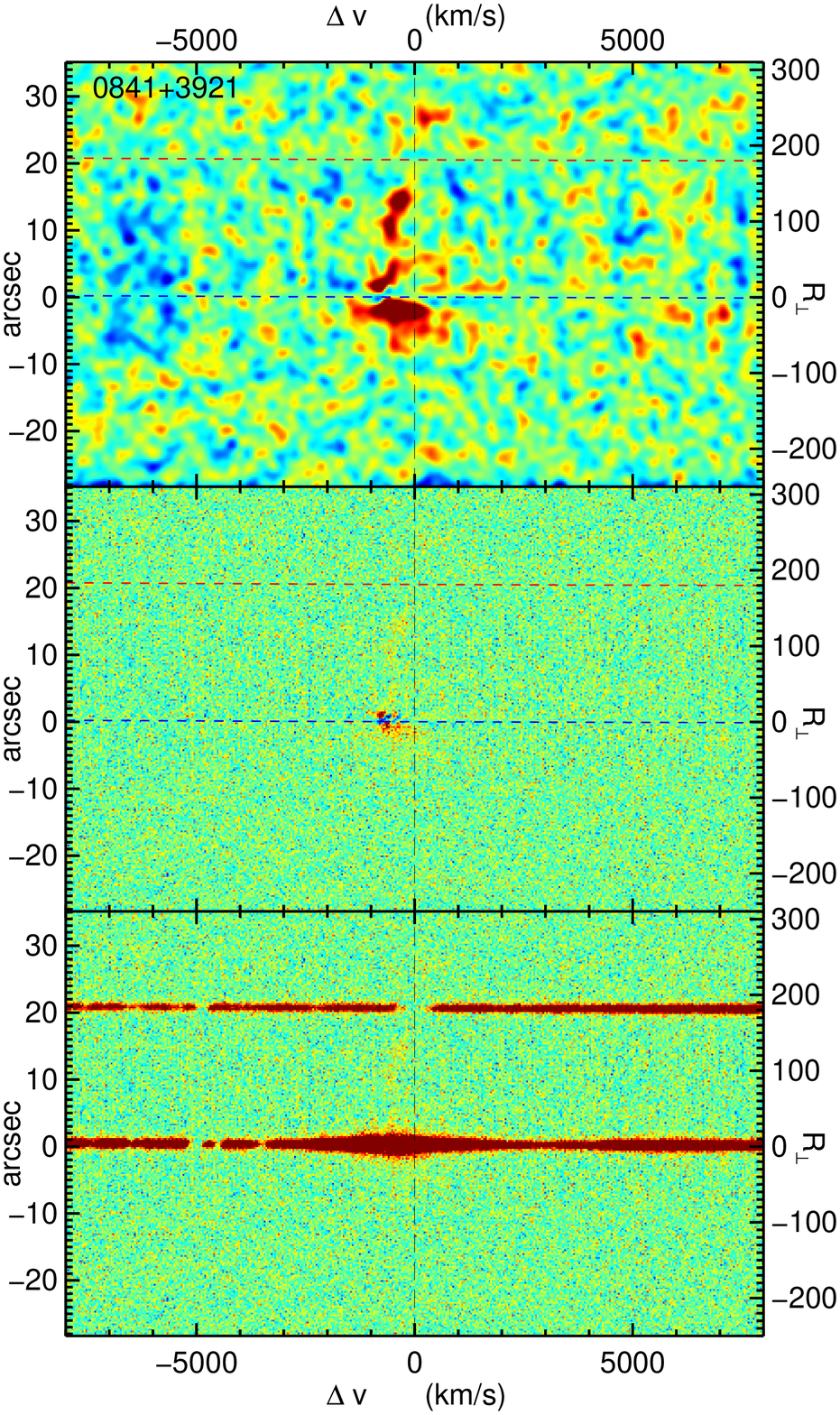,bb=10 0 540 860,width=\textwidth,clip=}}
    \centering{\epsfig{file=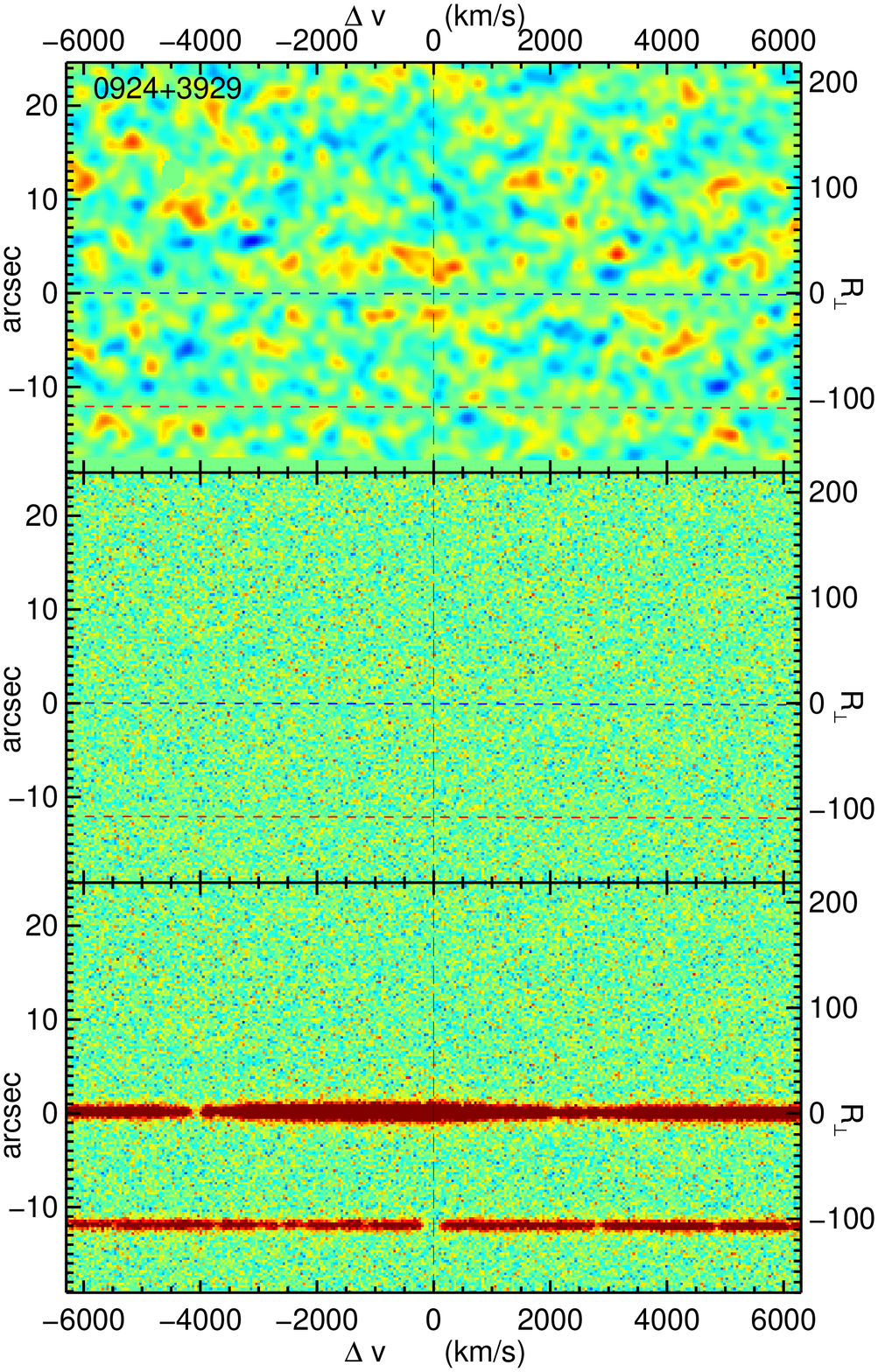,bb=10 0 540 860,width=\textwidth,clip=}}
  \end{minipage}
  \begin{minipage}{0.06\textwidth}
    \vskip 0.55cm
    \hskip -0.2cm
    \epsfig{file=f5a.eps,bb=5 0 55 360,width=\textwidth,clip=}
  \end{minipage}
  \begin{minipage}{0.41\textwidth}
    \centering{\epsfig{file=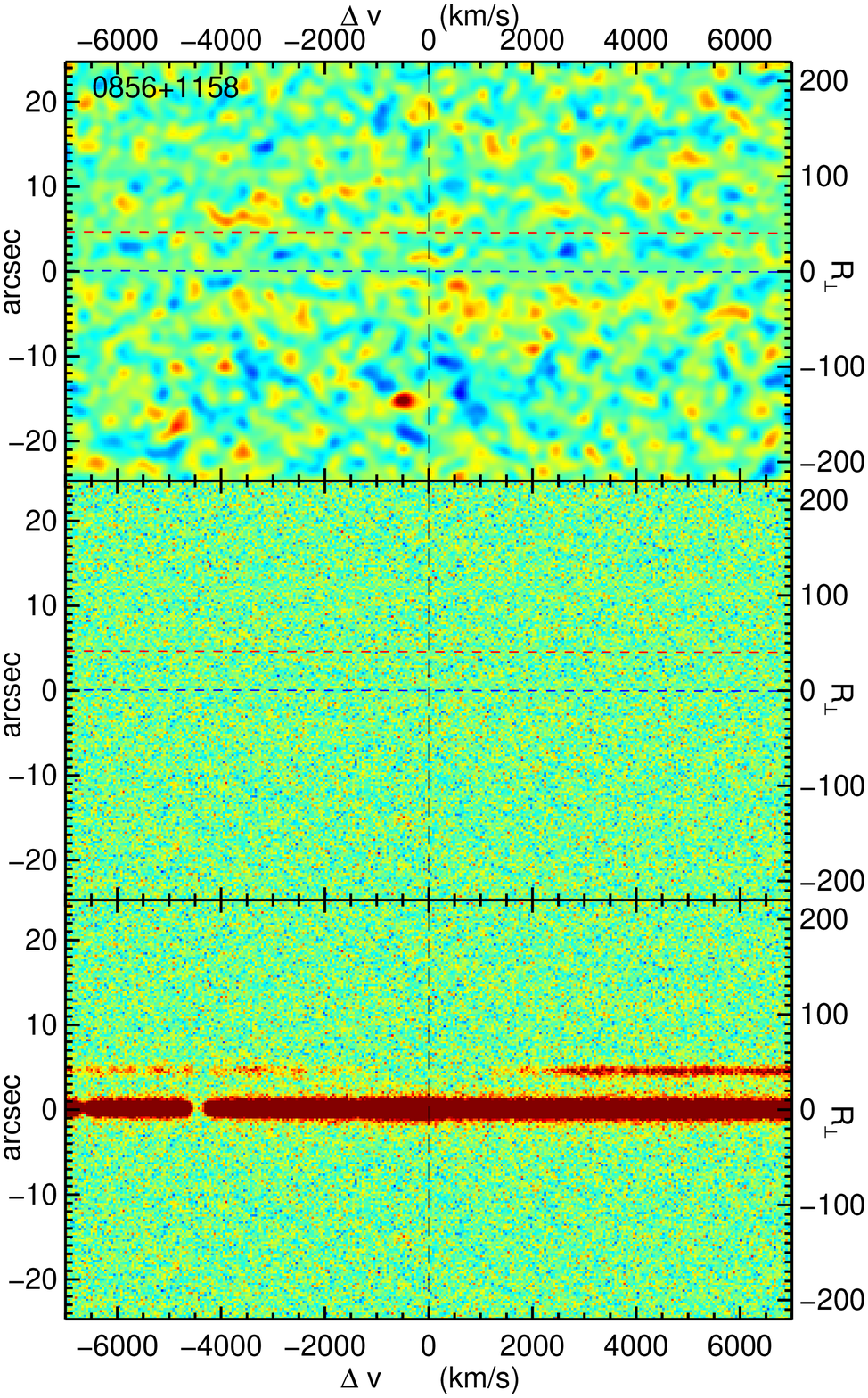,bb=10 0 540 860,width=\textwidth,clip=}}
    \centering{\epsfig{file=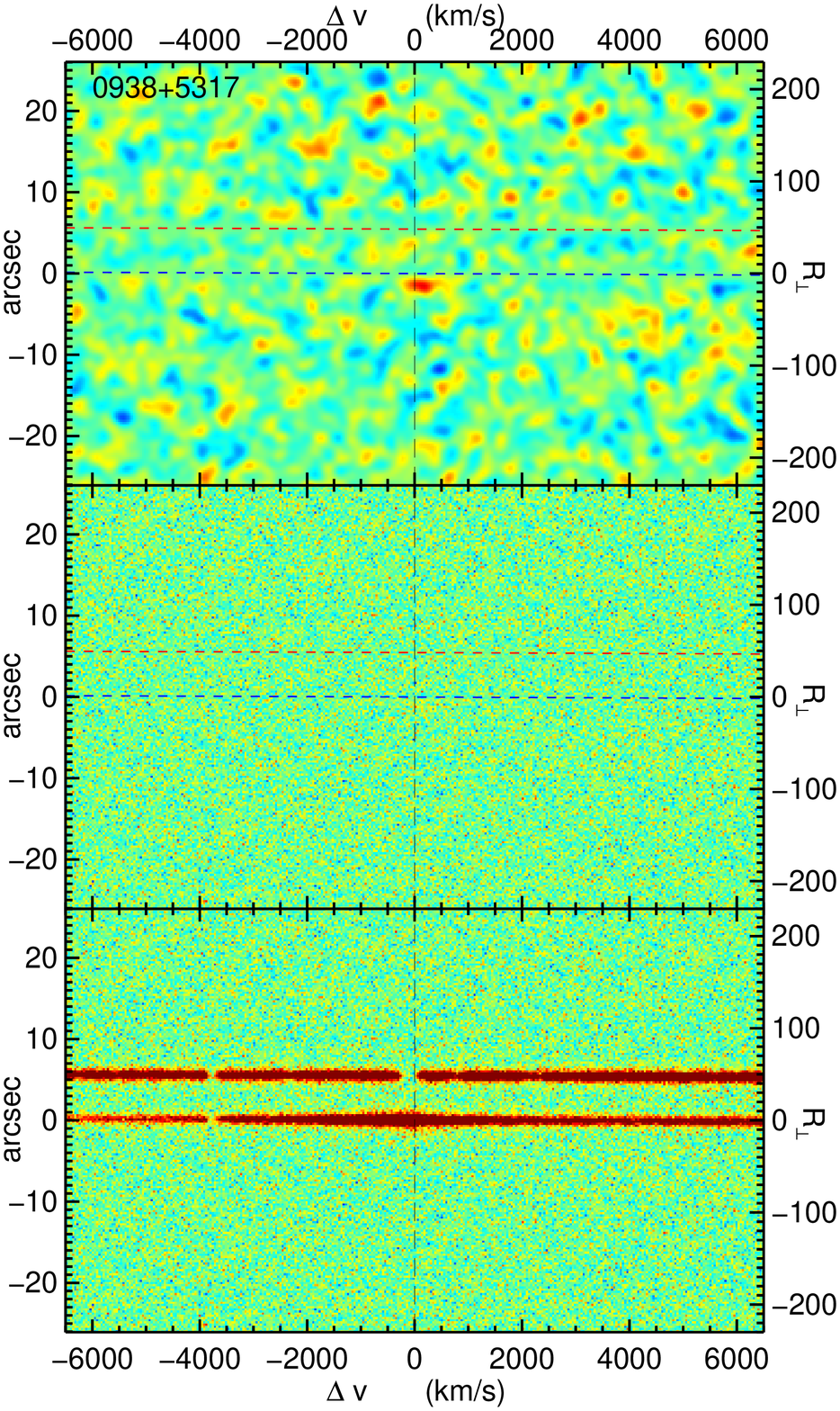,bb=10 0 540 860,width=\textwidth,clip=}}
  \end{minipage}
  \caption{Continued\label{fig:maps3}.}
\end{figure*}

\begin{figure*}
  \setcounter {figure}{4}
  \begin{minipage}{0.41\textwidth}
    \centering{\epsfig{file=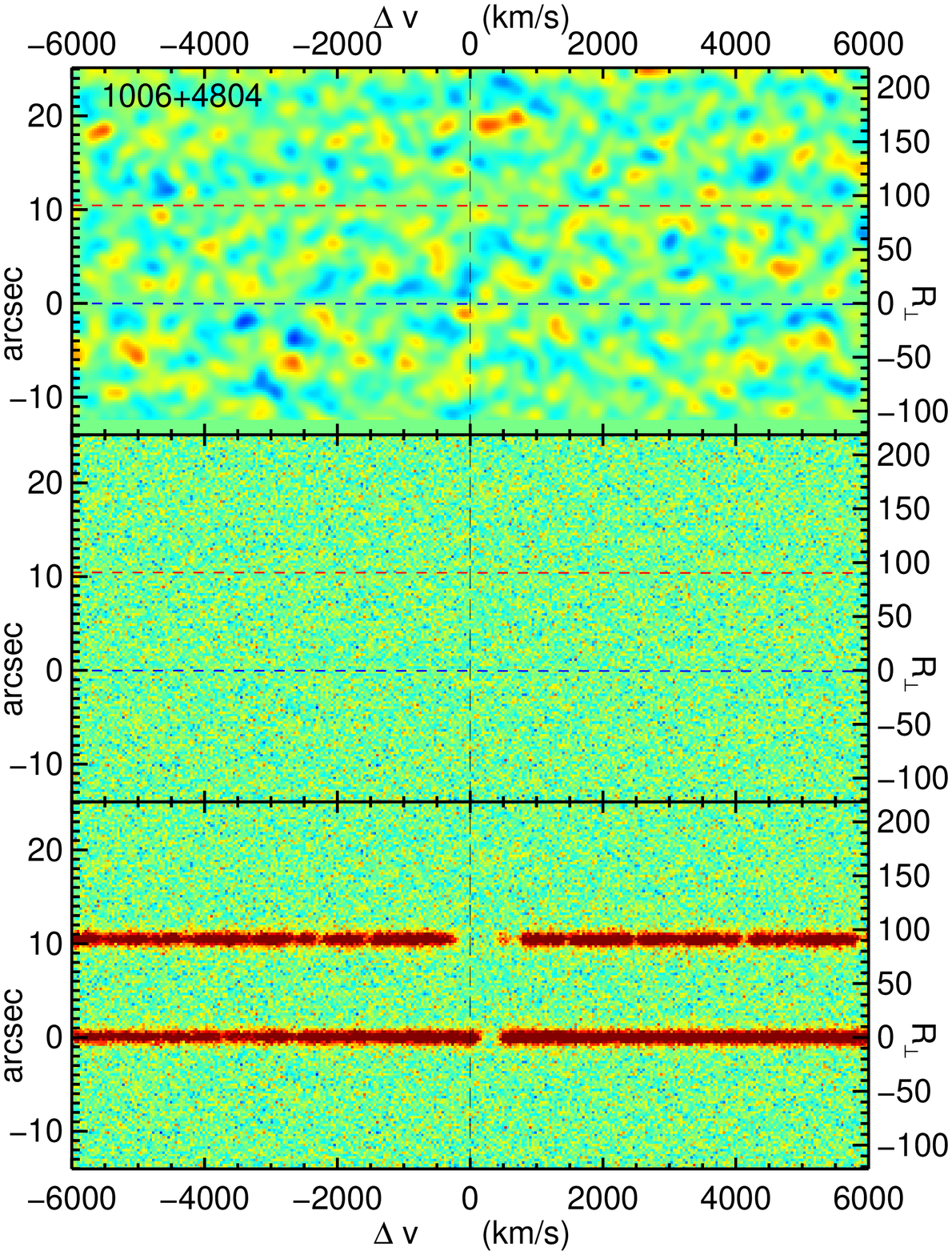,bb=10 0 540 860,width=\textwidth,clip=}}
   \vskip -2.4cm
    \centering{\epsfig{file=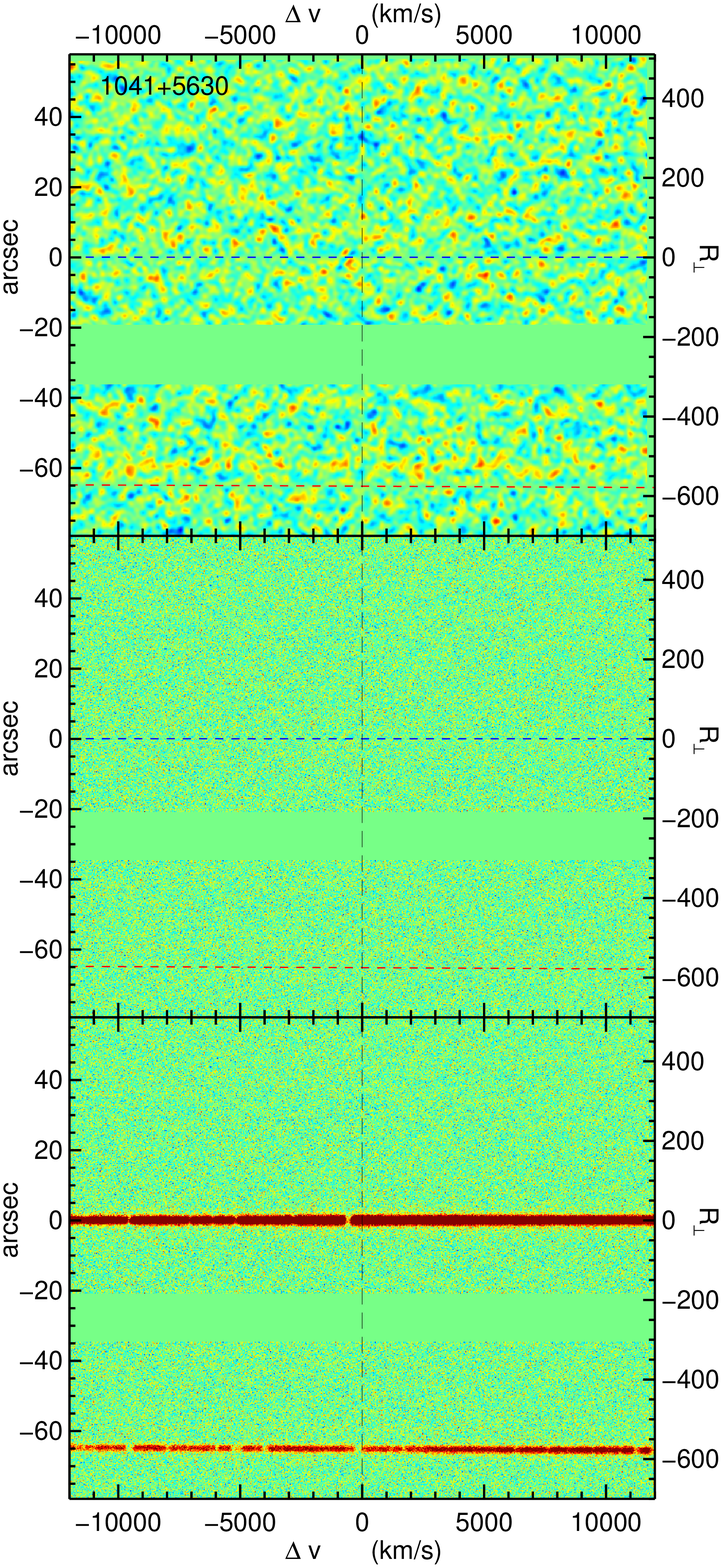,bb=10 0 540 1123,width=\textwidth,clip=}}
  \end{minipage}
  \begin{minipage}{0.06\textwidth}
    \vskip 0.55cm
    \hskip -0.25cm
    \epsfig{file=f5a.eps,bb=5 0 55 360,width=\textwidth,clip=}
  \end{minipage}
  \begin{minipage}{0.41\textwidth}
    \centering{\epsfig{file=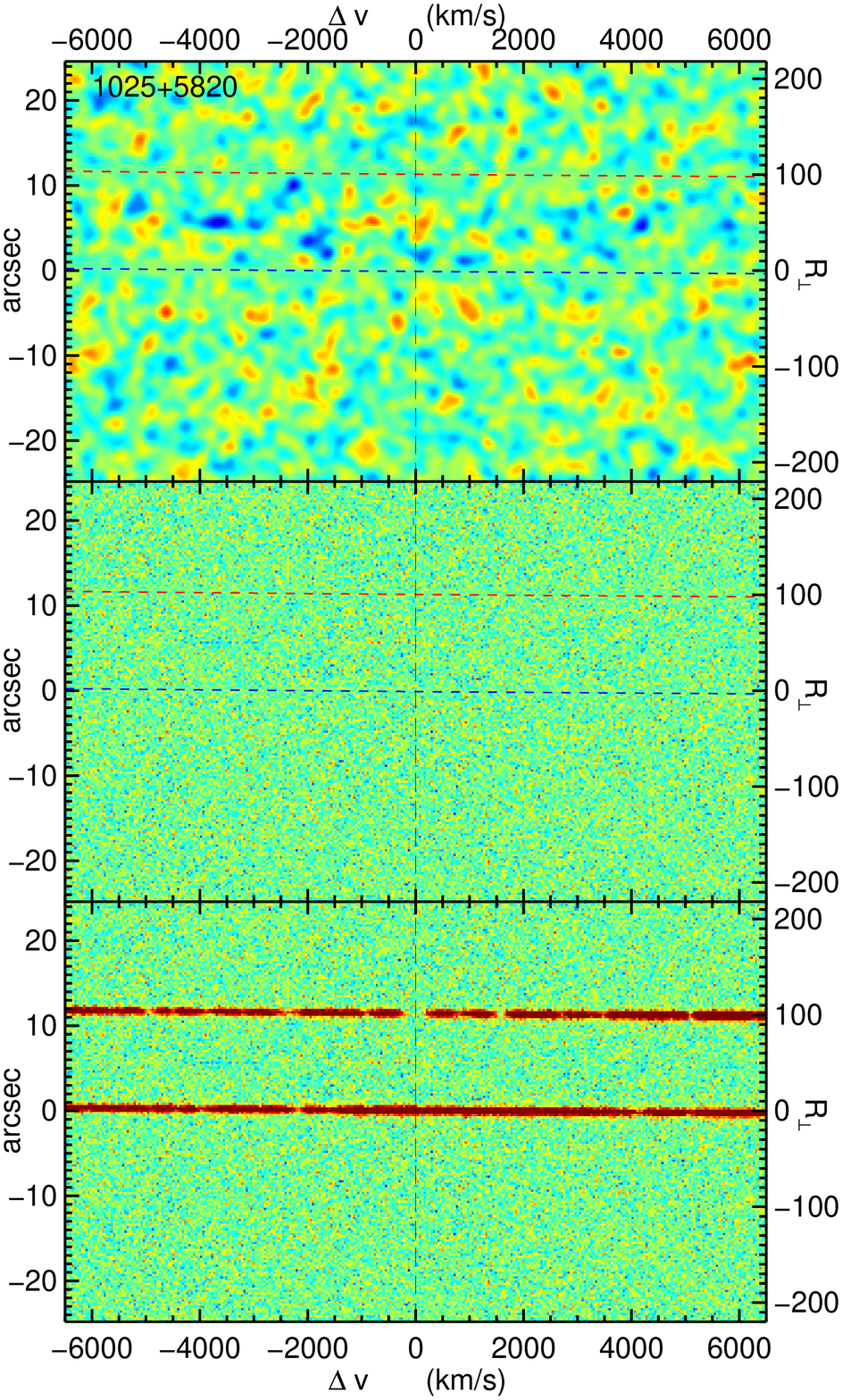,bb=10 0 540 860,width=\textwidth,clip=}}
    \centering{\epsfig{file=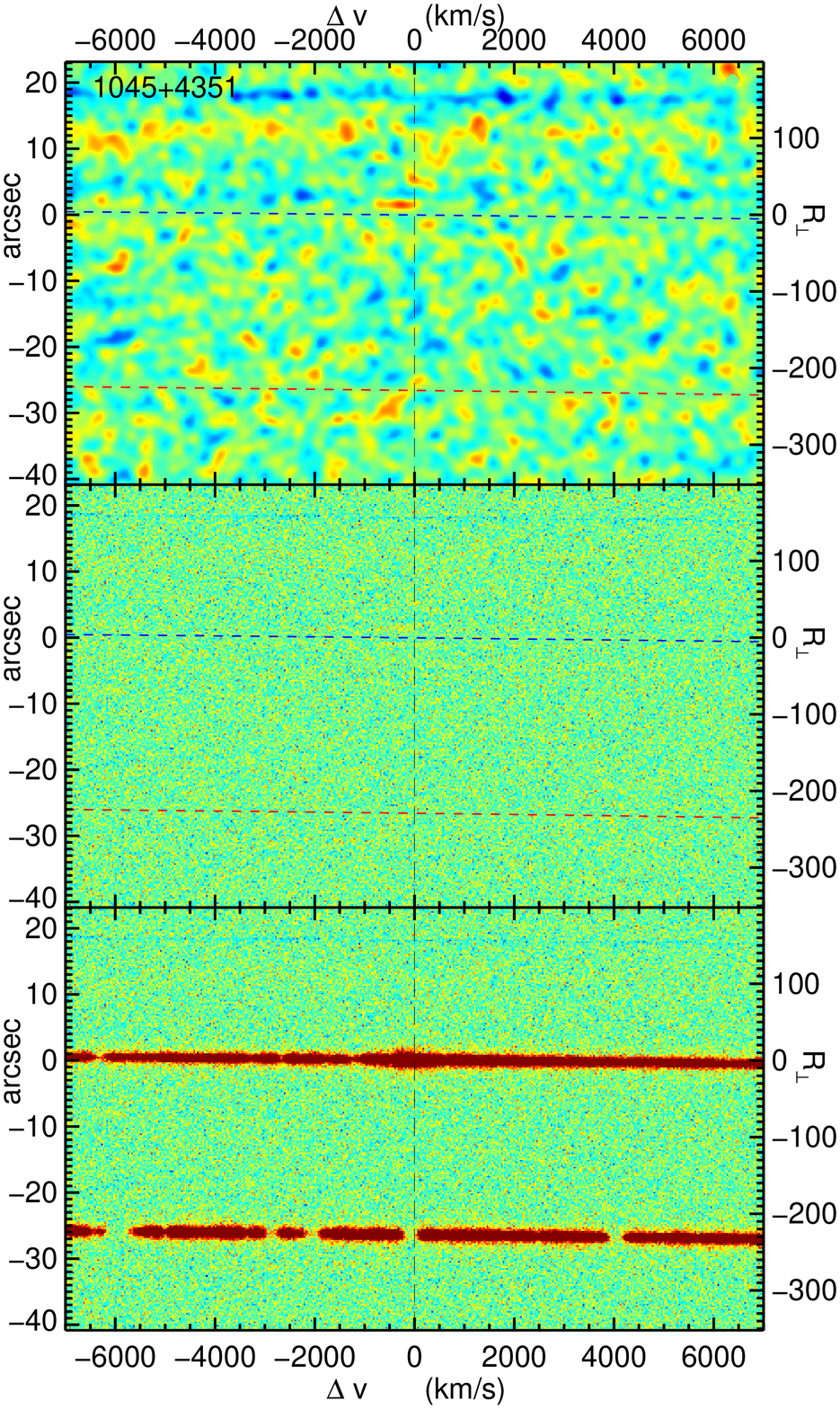,bb=10 0 540 860,width=\textwidth,clip=}}
  \end{minipage}
  \caption{Continued\label{fig:maps4}.}
\end{figure*}

\begin{figure*}
  \setcounter {figure}{4}
  \begin{minipage}{0.41\textwidth}
    \centering{\epsfig{file=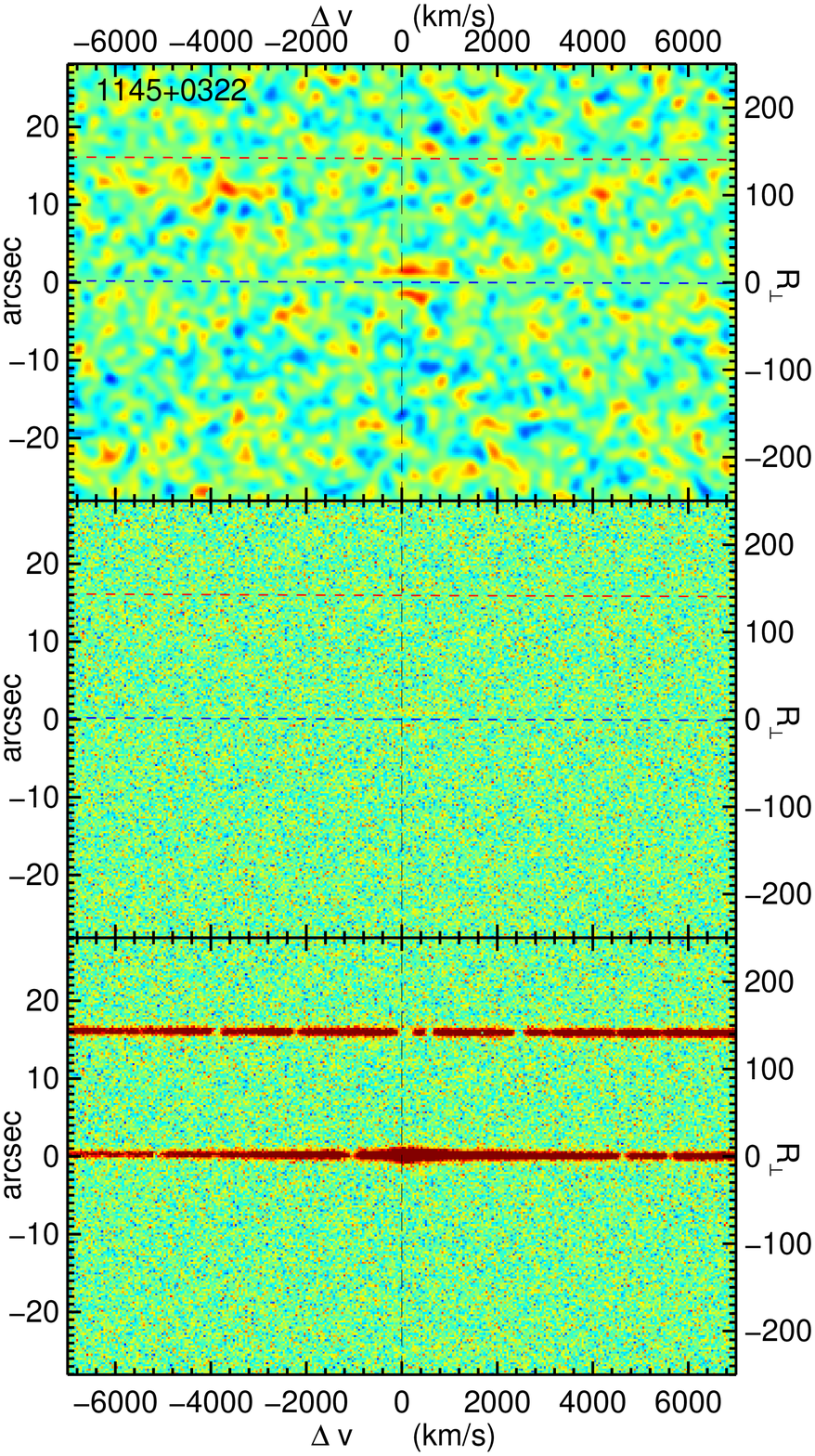,bb=10 0 540 860,width=\textwidth,clip=}}
    \centering{\epsfig{file=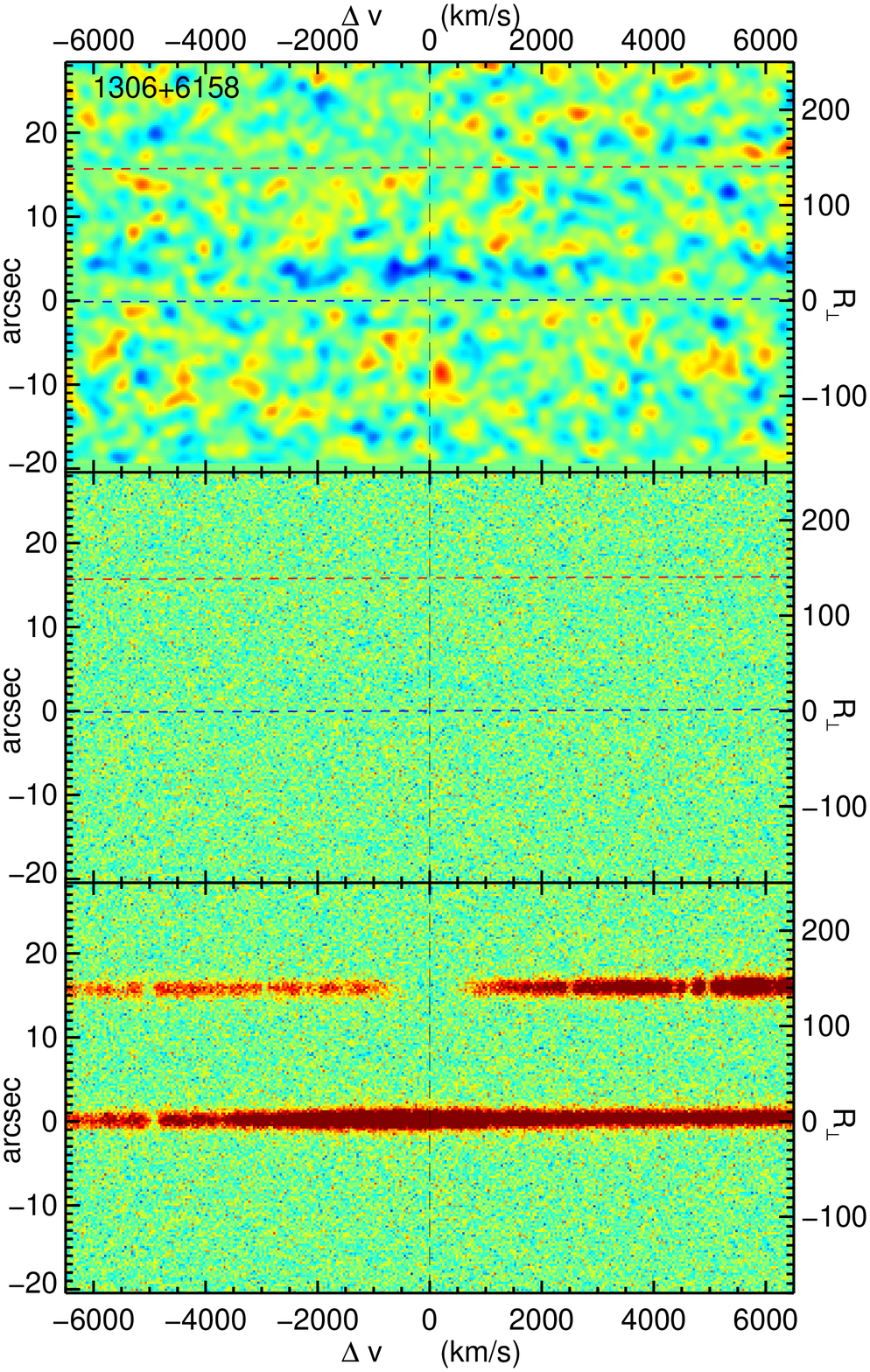,bb=10 0 540 860,width=\textwidth,clip=}}
  \end{minipage}
  \begin{minipage}{0.06\textwidth}
    \vskip 0.55cm
    \hskip -0.1cm
    \epsfig{file=f5a.eps,bb=5 0 55 360,width=\textwidth,clip=}
  \end{minipage}
  \begin{minipage}{0.41\textwidth}
    \centering{\epsfig{file=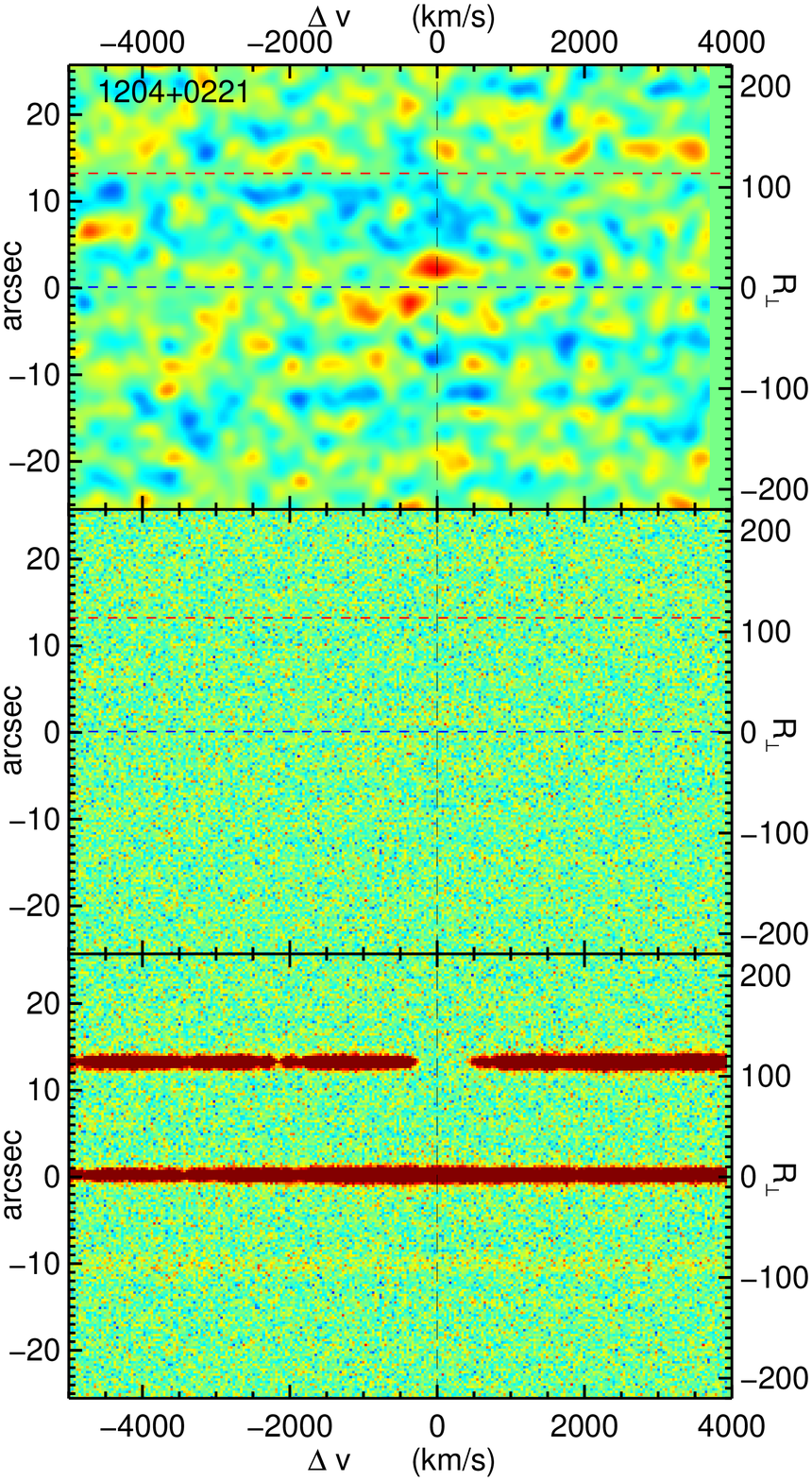,bb=10 0 540 860,width=\textwidth,clip=}}
    \centering{\epsfig{file=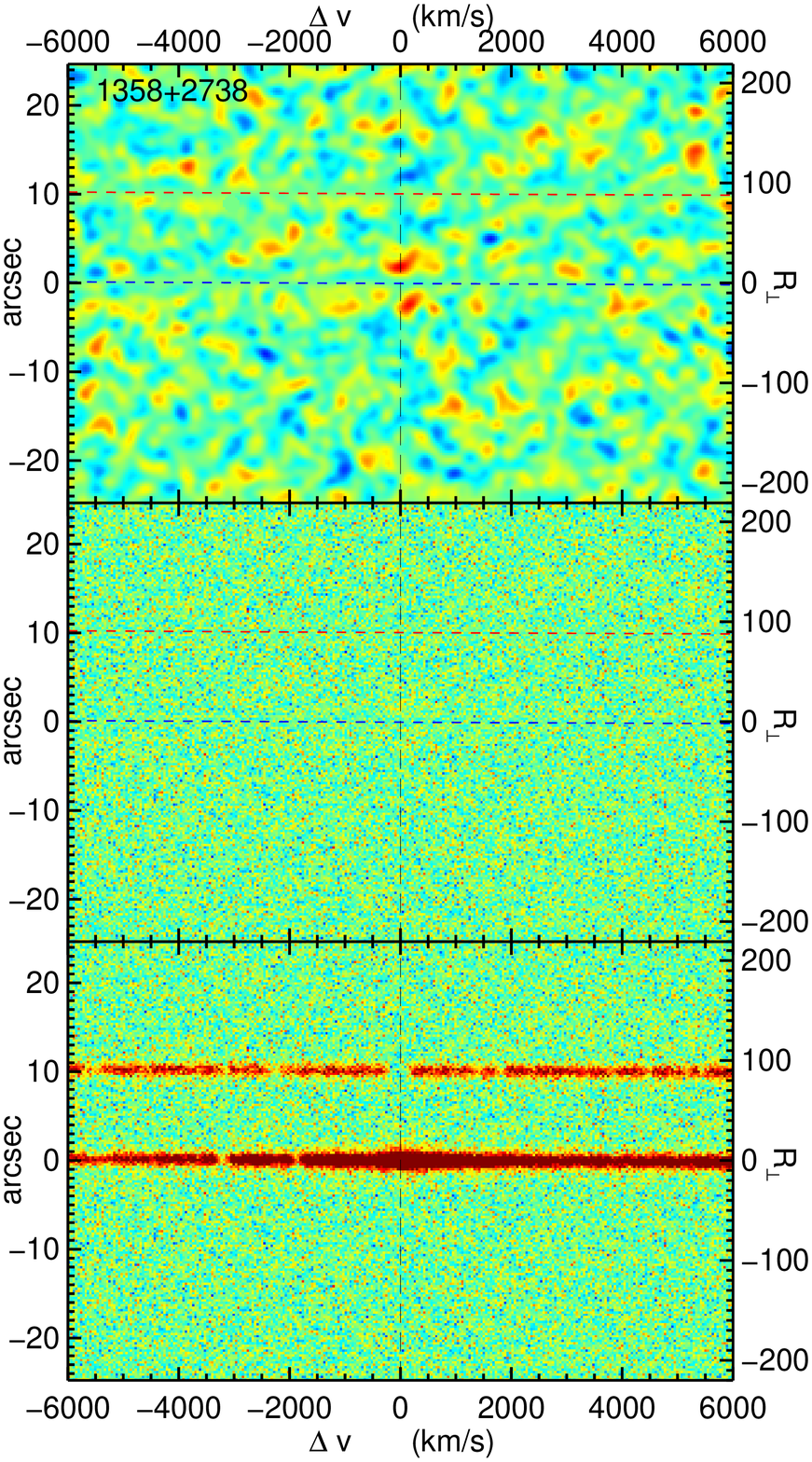,bb=10 0 540 860,width=\textwidth,clip=}}
  \end{minipage}
  \caption{Continued\label{fig:maps5}.}
\end{figure*}

\begin{figure*}
  \setcounter {figure}{4}
  \begin{minipage}{0.41\textwidth}
    \centering{\epsfig{file=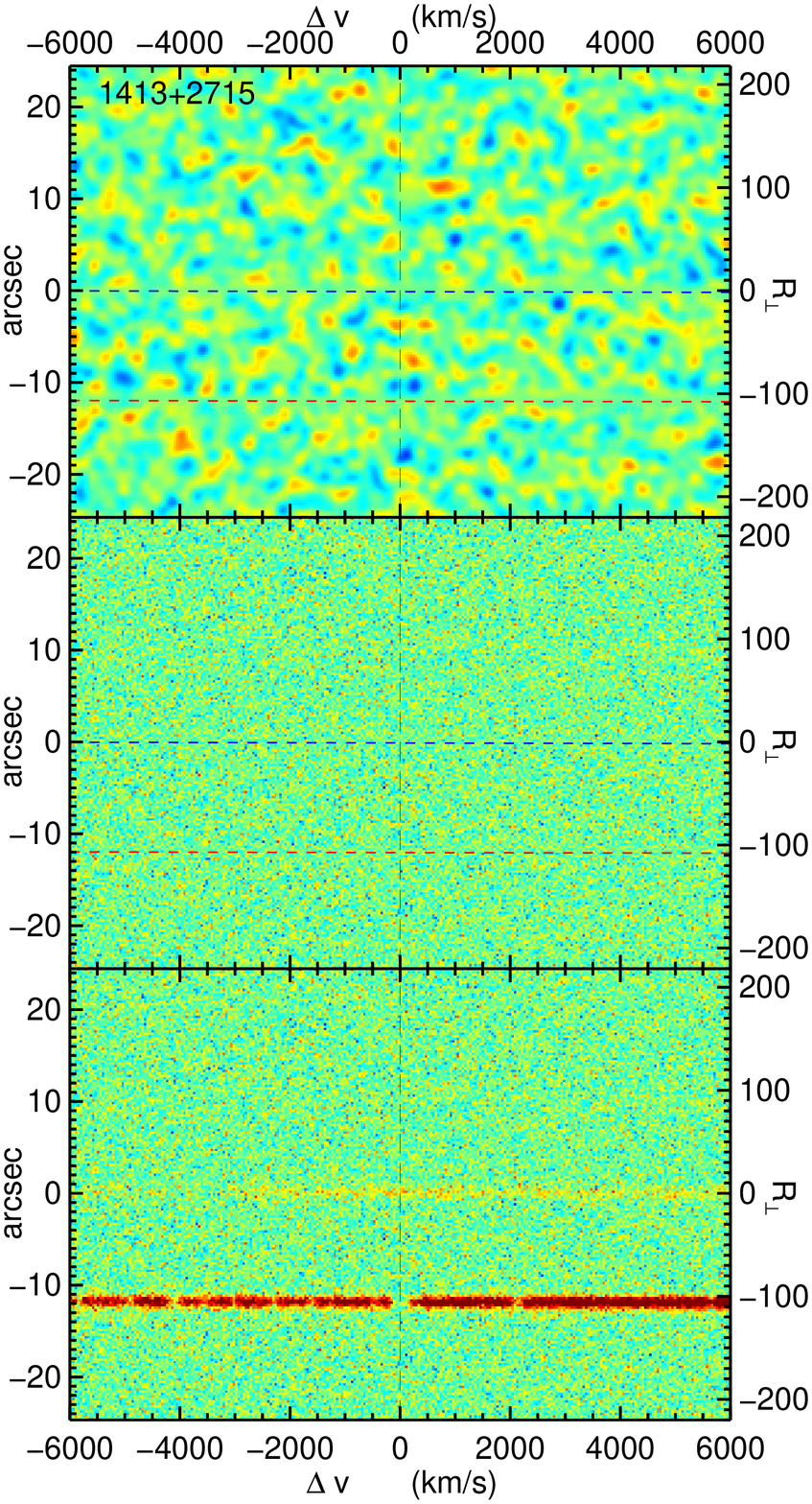,bb=10 0 540 860,width=\textwidth,clip=}}
    \centering{\epsfig{file=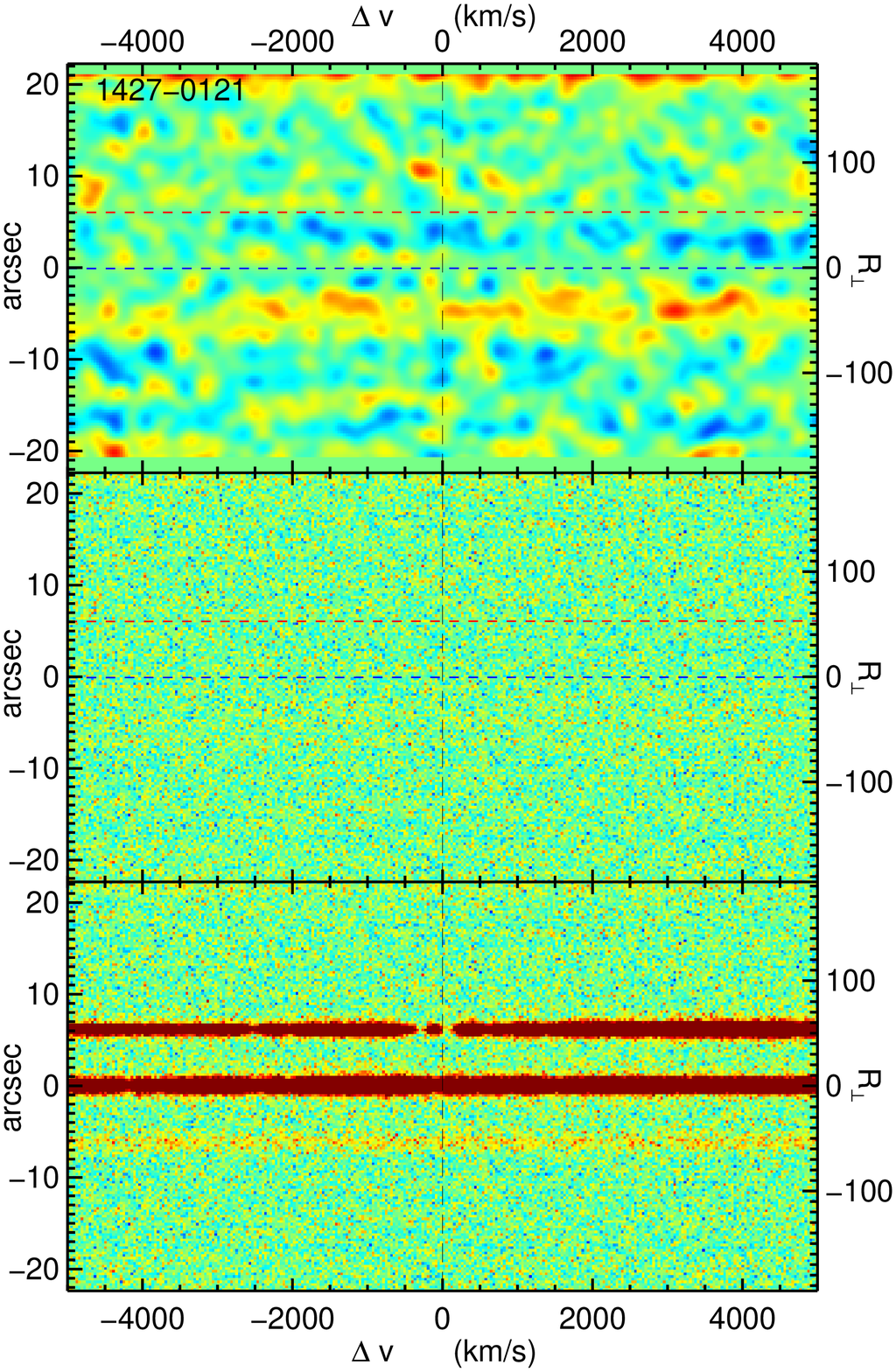,bb=10 0 540 860,width=\textwidth,clip=}}
  \end{minipage}
  \begin{minipage}{0.06\textwidth}
    \vskip 0.55cm
    \hskip -0.1cm
    \epsfig{file=f5a.eps,bb=5 0 55 360,width=\textwidth,clip=}
  \end{minipage}
  \begin{minipage}{0.41\textwidth}
    \centering{\epsfig{file=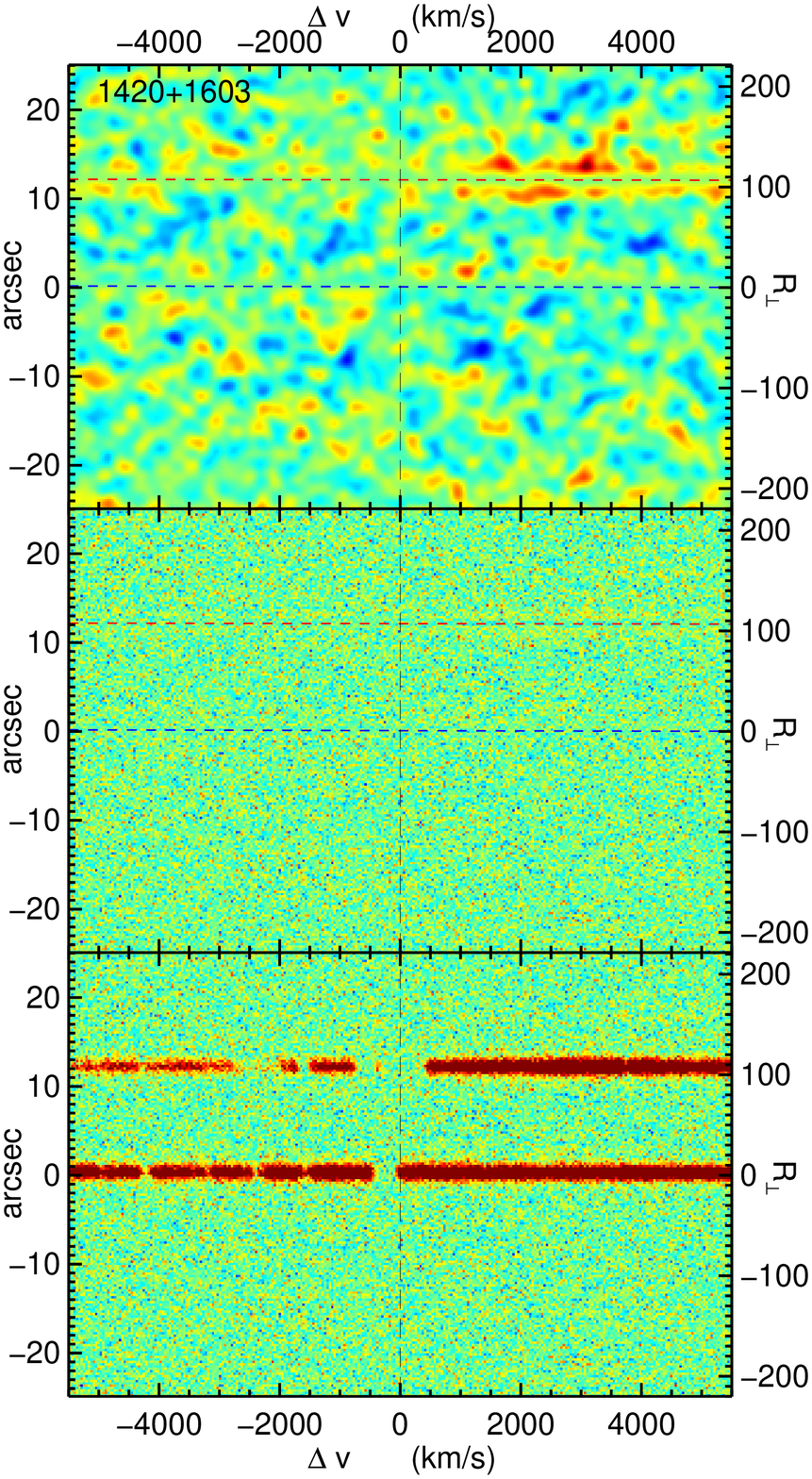,bb=10 0 540 860,width=\textwidth,clip=}}
    \centering{\epsfig{file=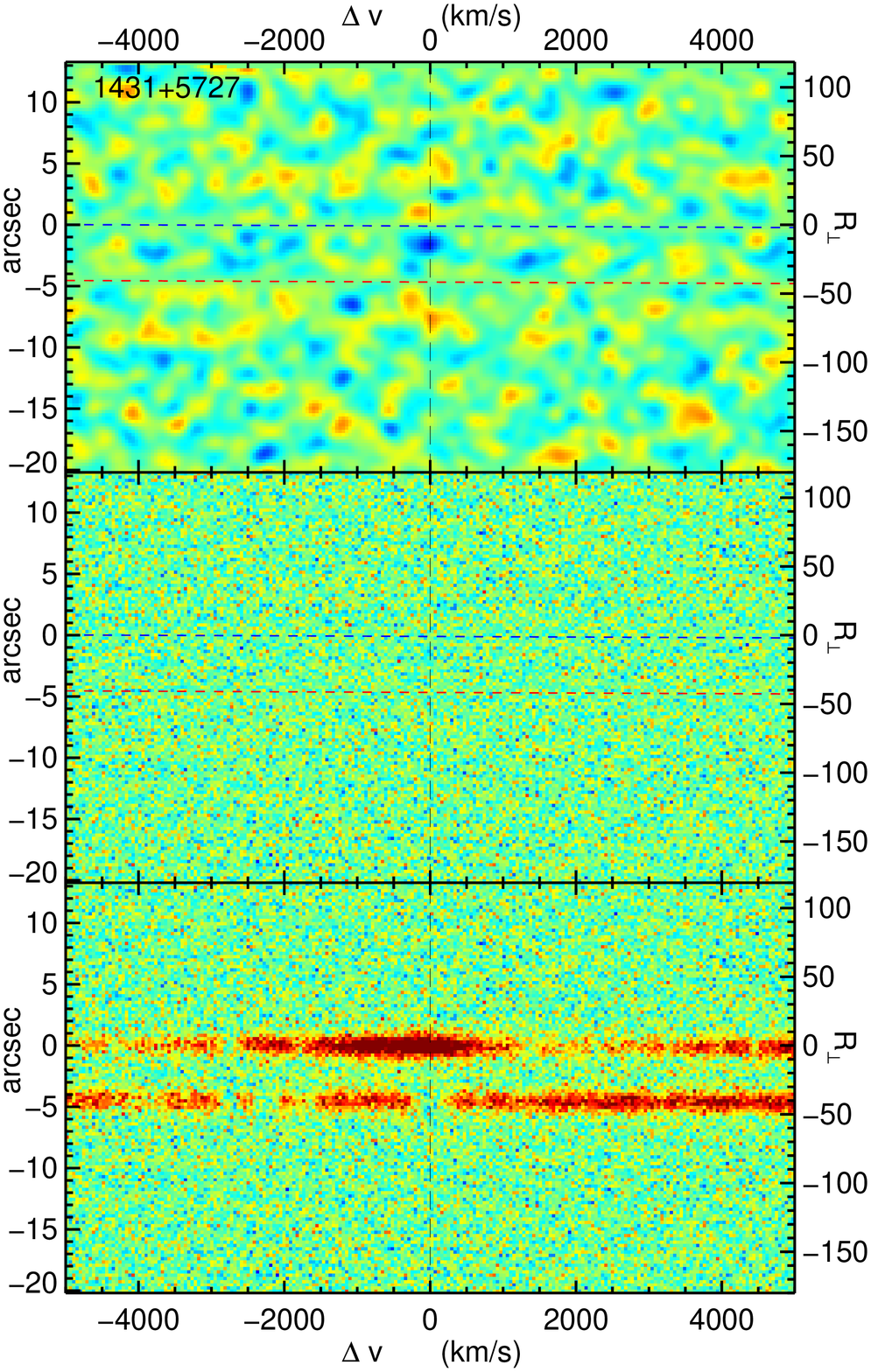,bb=10 0 540 860,width=\textwidth,clip=}}
  \end{minipage}
  \caption{Continued\label{fig:maps6}.}
\end{figure*}

\begin{figure*}
  \setcounter {figure}{4}
  \begin{minipage}{0.41\textwidth}
    \centering{\epsfig{file=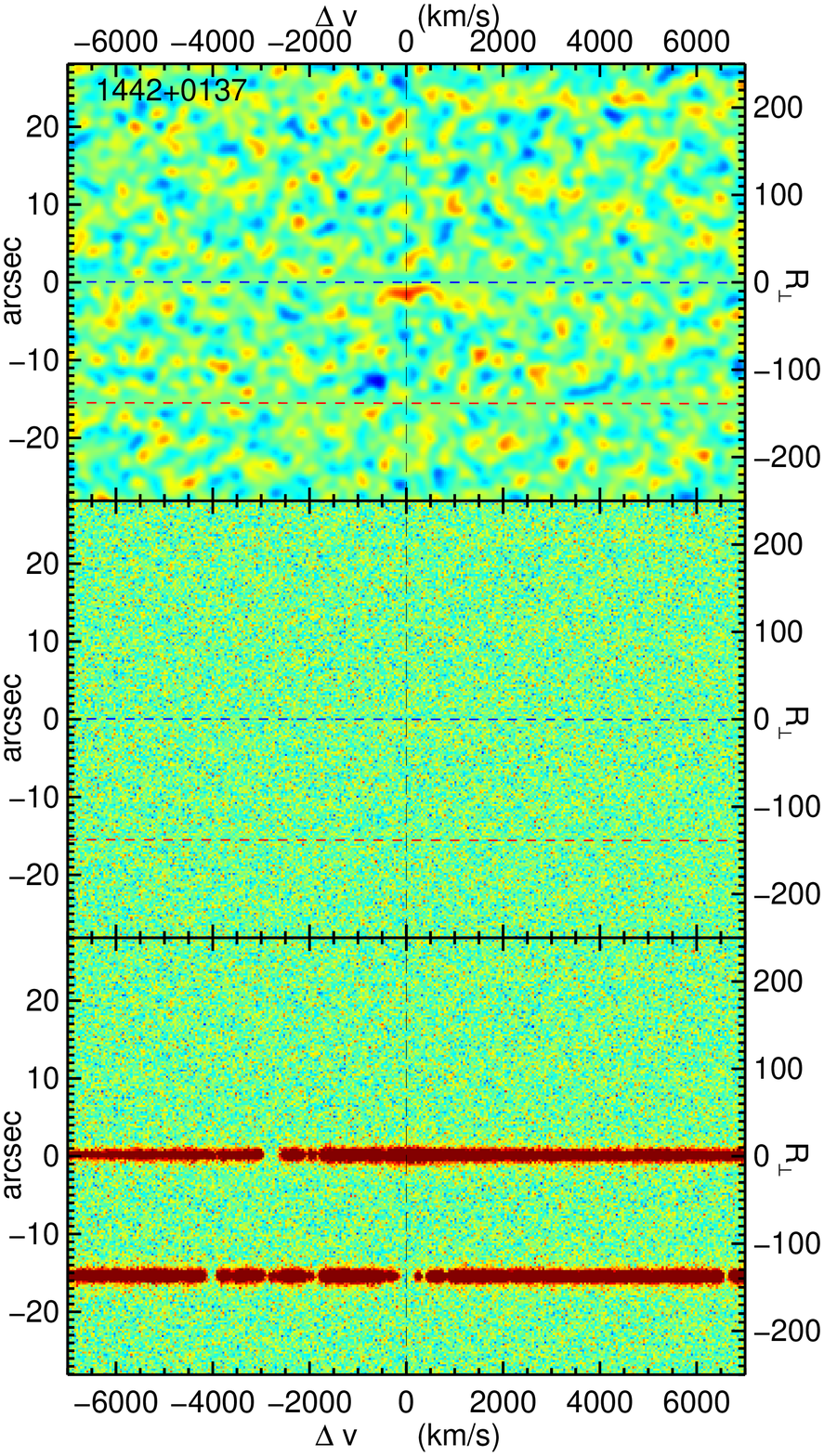,bb=10 0 540 860,width=\textwidth,clip=}}
    \centering{\epsfig{file=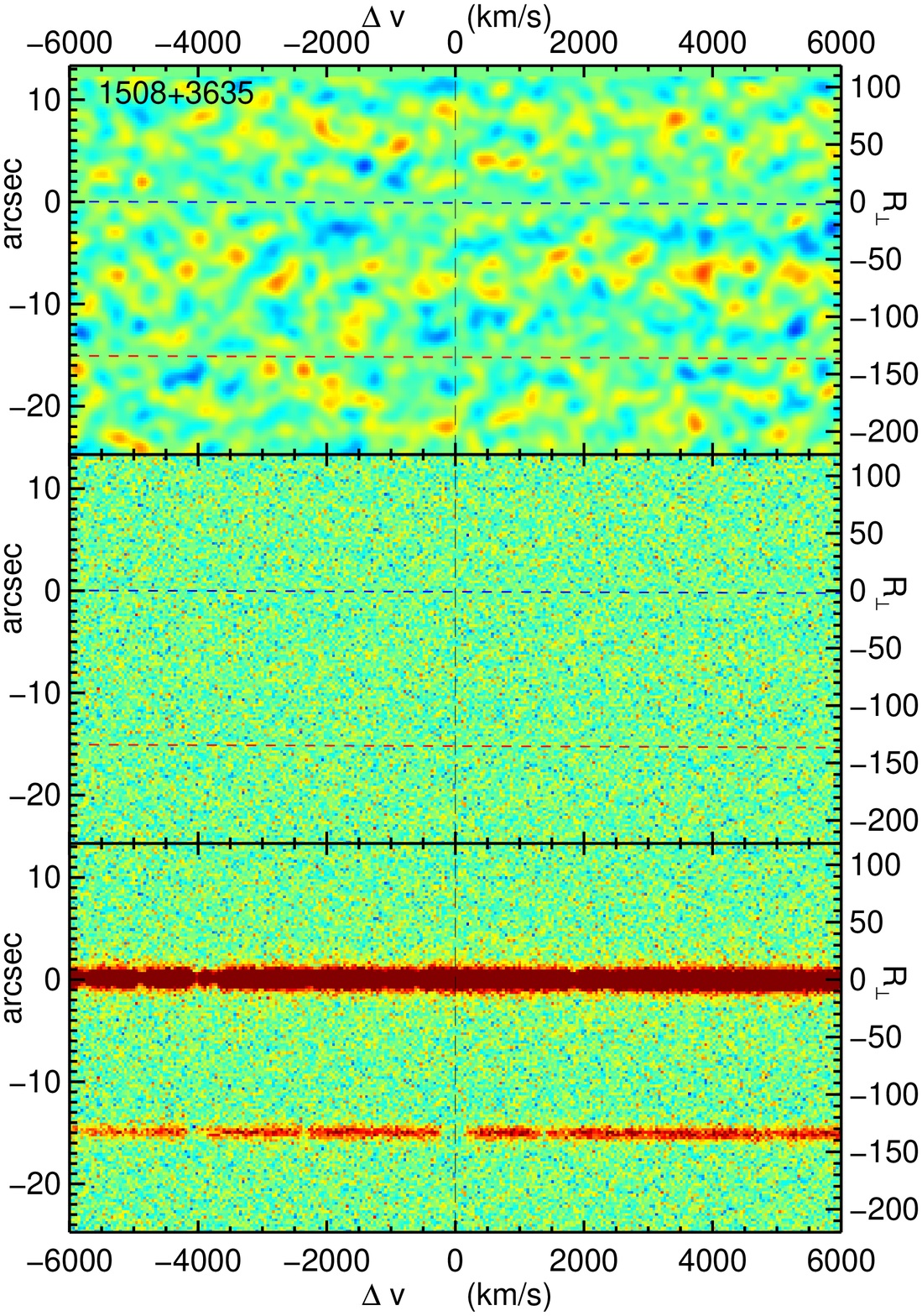,bb=10 0 540 860,width=\textwidth,clip=}}
  \end{minipage}
  \begin{minipage}{0.06\textwidth}
    \vskip 0.55cm
    \hskip 0.0cm
    \epsfig{file=f5a.eps,bb=5 0 55 360,width=\textwidth,clip=}
  \end{minipage}
  \begin{minipage}{0.41\textwidth}
    \centering{\epsfig{file=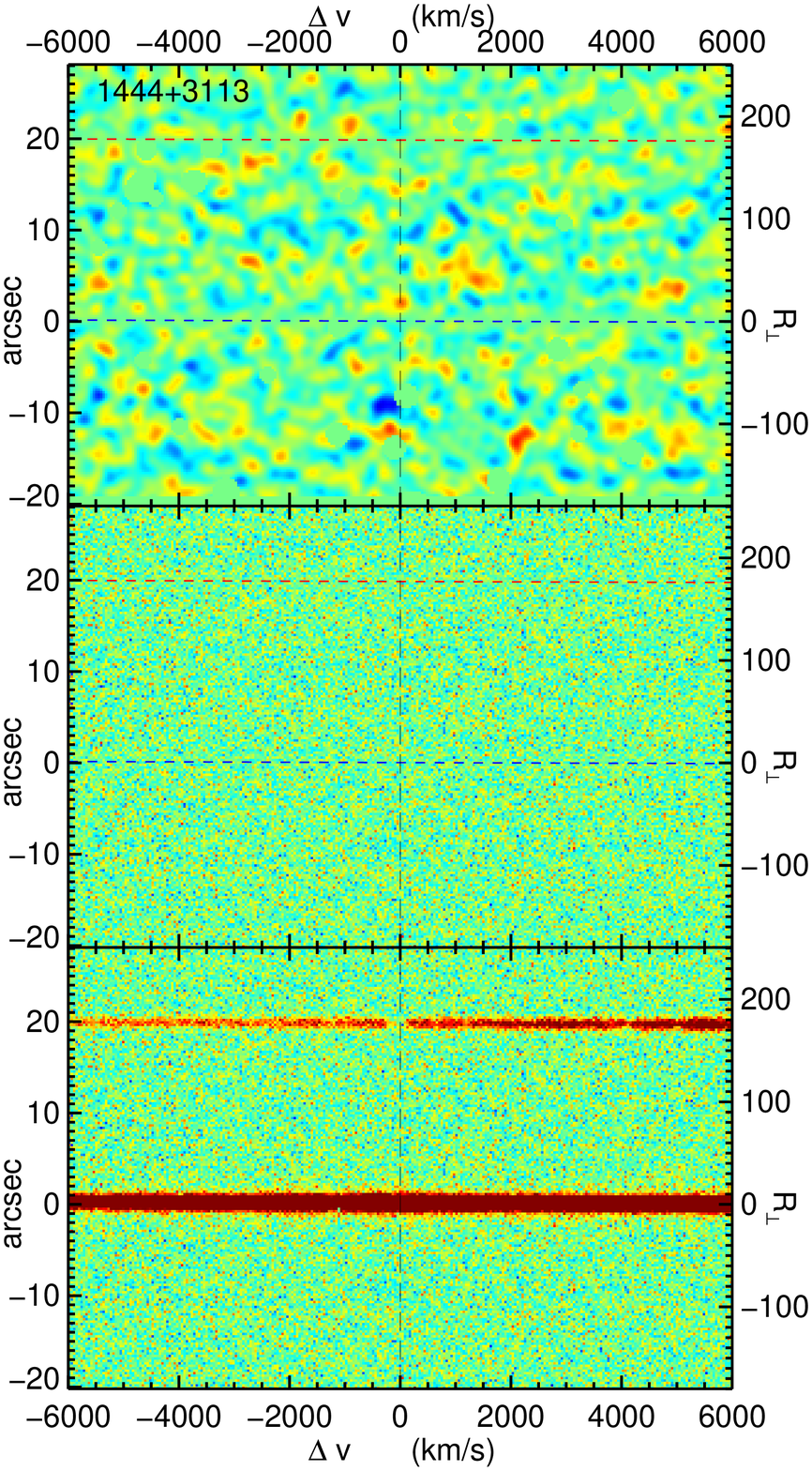,bb=10 0 540 860,width=\textwidth,clip=}}
    \centering{\epsfig{file=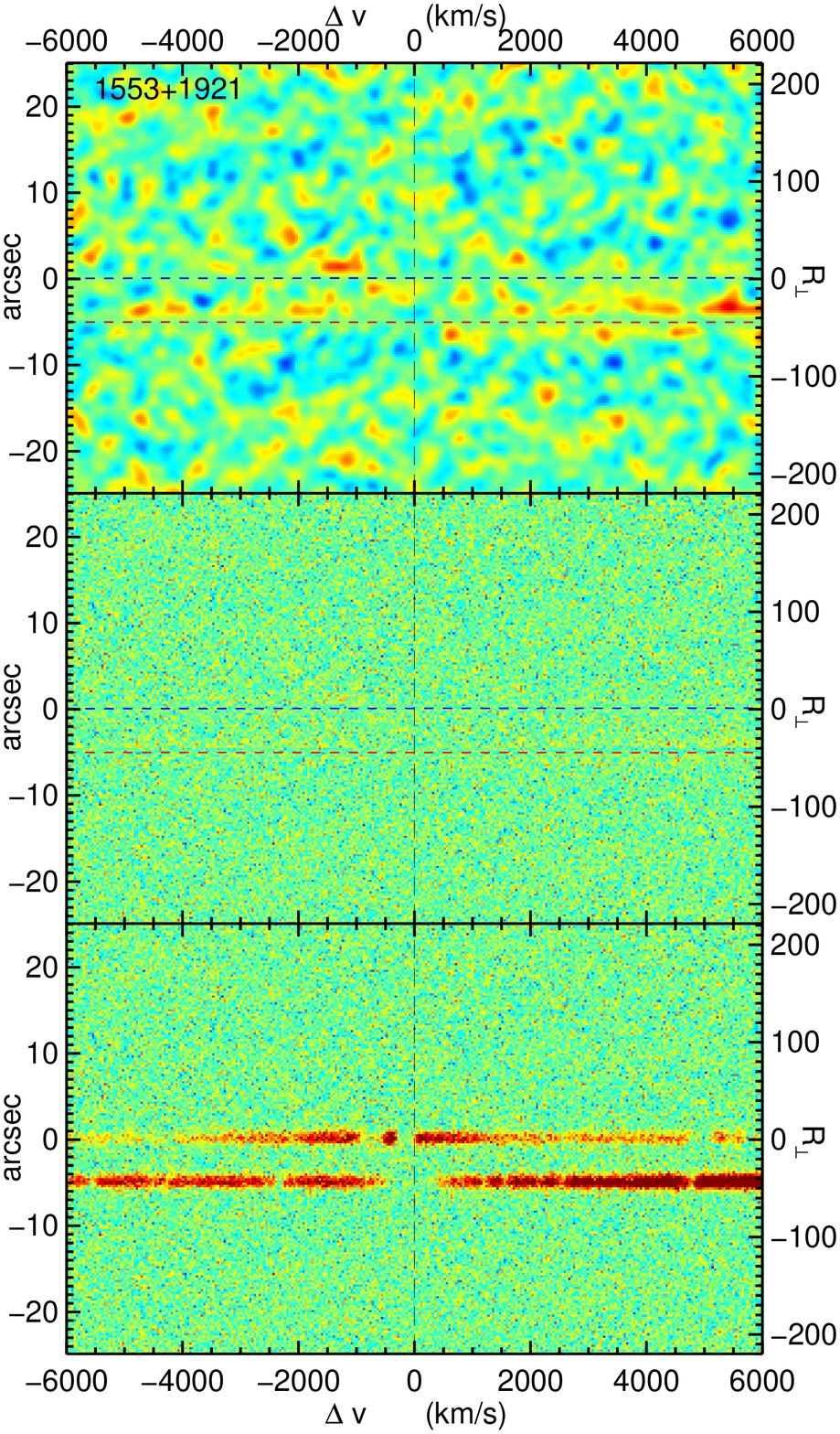,bb=10 0 540 860,width=\textwidth,clip=}}
  \end{minipage}
  \caption{Continued\label{fig:maps7}.}
\end{figure*}

\clearpage

\begin{figure}
  \setcounter {figure}{4}
  \centering{\epsfig{file=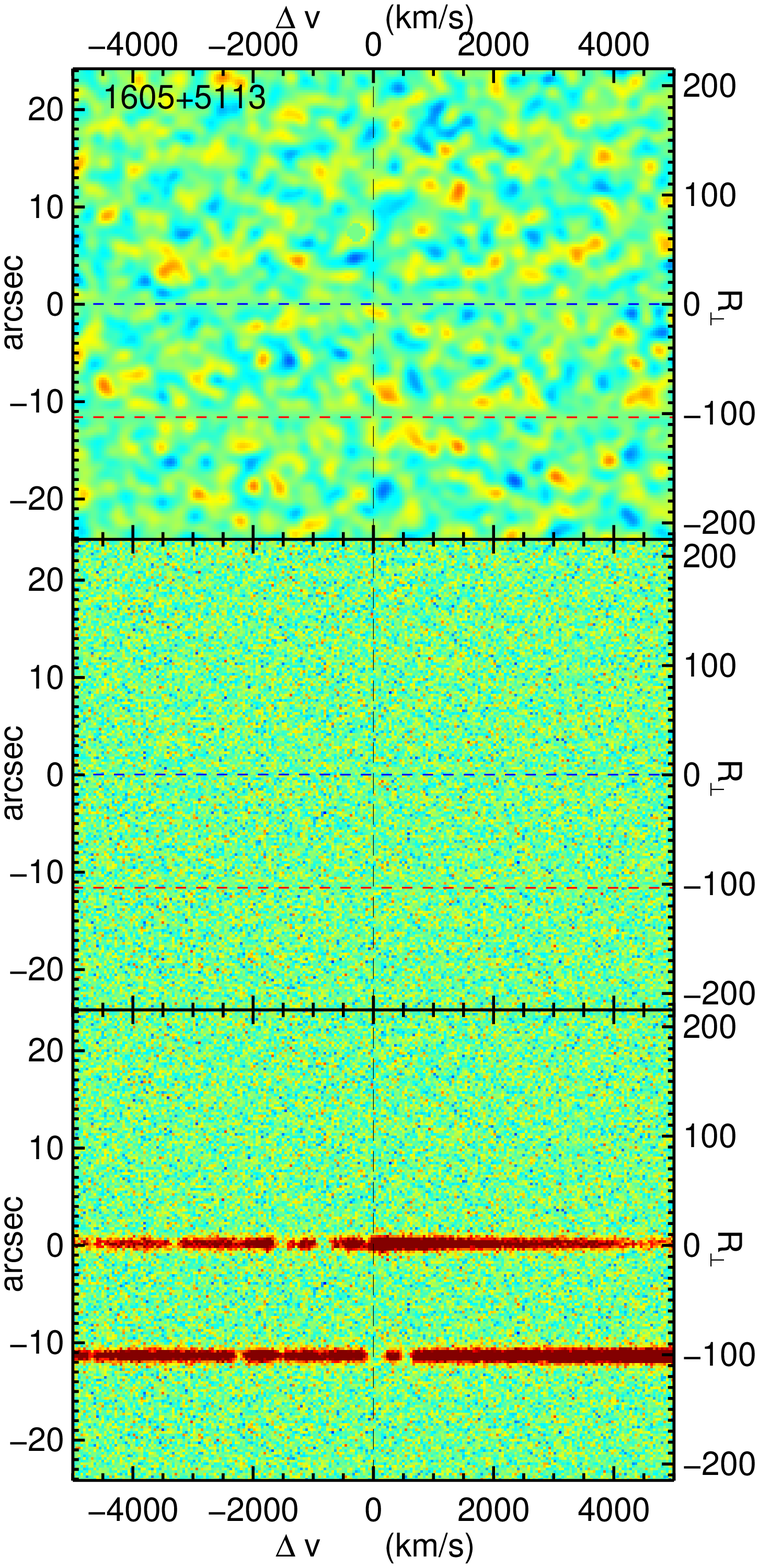,bb=10 0 540 860,width=0.45\textwidth,clip=}}
  \caption{Continued\label{fig:maps8}.}
\end{figure}

\clearpage

\appendix

\section{The Attenuation of \mlya\ by Dust}

In most, if not all, astrophysical environments that contain cool gas
and metals, a fraction of the refractory elements (e.g.\  Fe, Ni, Si)
have condensed to form dust grains.   
For the treatment of \mlya\ emission considered in this paper, dust is
an important factor. Because we detect absorption from gas which is optically
thick in the \mlya\ line, \mlya\ photons will be resonantly
trapped, and can only escape the system through diffusion, both
in frequency and spatially. If the resonant photons must traverse a much larger
pathlength to escape, this could result in a significant
increase in the extinction of resonant lines compared to the continuum
extinction at the same wavelength. Because the focus of this paper is
\mlya\ emission from the cold gas detected in absorption, we calculate the
magnitude of the attenuation of \mlya\ by dust grains given our
expectations for the physical conditions of the gas in absorbing
clouds.  Our estimates closely follow the discussion in \S15.7 of
\citet{draine11}. 

Consider an ionized spherical cloud with total hydrogen column density
$N_{\rm H}$, neutral column density $N_{\rm HI}$, neutral fraction
$x_{\rm HI} \equiv N_{\rm HI}\slash N_{\rm H}$, volume density
$n_{\rm H}$, and Doppler broadening parameter $b$.  We define the
radius of the cloud to be $r_{\rm c} \equiv 3N_{\rm H}\slash 4 n_{\rm
  H}$, where we have used the fact that the average absorption
pathlength $\langle l_{\rm c} \rangle \equiv N_{\rm H}\slash n_{\rm
  H}$ through a sphere of radius $r_{\rm c}$ equals $\langle l_{\rm
  c}\rangle = 4\slash 3 r_{\rm c}$. In what follows we
assume, that the cloud has a uniform ionization structure, whereas in
reality such clouds are optically thick to ionizing photons and will
be self-shielding. Uniform ionization structure is likely to be a good
approximation in a cooling radiation scenario where \mlya\ photons are
generated throughout the cloud, as the bulk of the \mlya\ photons will
be generated in the regions of the cloud which have properties close
to the mean. For photoionization of self-shielding clouds, this
approximation is likely to over-estimate the dust attenuation, because
most recombinations are generated in a thin self-shielding layer at
the surface of the cloud where the neutral fraction (and hence
\mlya\ optical depth) is much lower than the average value in the
cloud.

Following \citet{draine11}, we can write the mean square displacement 
of a \mlya\ photon from the point of emission as
\be
\langle r^2\rangle \approx \frac{2}{3}L_0^2 N_{\rm s}^3 \label{eqn:r2} 
\ee
where 
\be
L_0\equiv \frac{1}{n_{\rm HI}\sigma_{w}}, 
\ee
\be
\sigma_{w} =6.48\sci{-18}\,{\rm cm}^{-2}\left(\frac{b}{20\,\kms}\right)^{-2}
\ee
and $N_{\rm s}$ is the total number of resonant scatterings and $\sigma_{w}$ is the
\ion{H}{1} absorption cross-section. These
expressions are valid in the Lorentzian wings of the Voigt profile where
the bulk of the spatial diffusion takes place.  A reasonable criterion 
for escape is that $\langle r^2 \rangle = r_{\rm c}^2$, from which we 
can determine the total number of scatterings via eqn.~(\ref{eqn:r2})
\be
N_{\rm s} \approx 15\left(\frac{N_{\rm HI}}{10^{19}\,{\rm cm}^{-2}}\right)^{2\slash 3}\left(\frac{b}{20\,\kms}\right)^{-4\slash 3}. 
\ee
This modest number of scatterings is sufficient to frequency-shift
a typical photon to 
$\Delta \nu\slash \nu \approx N_{\rm s}^{1\slash 2} b\slash c$, such that our 
assumption of being in the damping wings of the Voigt profile is valid. 

To determine the amount of attenuation by dust, we must first estimate the 
total pathlength traveled by an escaping photon, which can be written as
$s \approx 1\slash 2 L_0 N_{\rm s}^2$ \citep{draine11}, or 
\be
s\approx 1.8\left(\frac{N_{\rm HI}}{10^{19}\,{\rm cm}^{-2}}\right)^{1\slash 3}\left(\frac{b}{20\,\kms}\right)^{-2\slash 3}\langle l_{\rm c}\rangle \label{eqn:s}. 
\ee
Thus given the cloud properties we have assumed, the pathlength and hence total 
optical depth through the cloud for  resonantly scattered photons is a 
factor of $\sim 2$ larger than it is for continuum photons. 

The amount of extinction corresponding to this pathlength depends on
the amount of dust associated with the gas, which is highly
uncertain. In quasar absorption line research, there is strong
evidence that refractory elements in DLAs are depleted onto dust
grains albeit at modest levels compared to the Galactic ISM
\citep{pettini94,pw01,vladilo11}.  The effects appear to be strongest
for the systems with highest metallicity \citep[e.g.][]{kph+10}.  For
LLSs where the gas is predominantly ionized, there has been
essentially no empirical assessment of dust.  One expects that the
physical conditions -- higher UV radiation field, higher temperature,
lower density -- lead to lower dust-to-gas ratios.  Nevertheless, dust
is apparent in highly ionized regions of the local universe
\citep[e.g.][]{draine11} and also the diffuse environment of galactic
halos \citep{msf+10}. The most extreme assumption is to assume that
all of the gas contains dust, and that the dust-to-gas ratio
corresponds to that of the Milky Way, which is equivalent to assuming
a solar metallicity and a high depletion factor. This is conservative,
because DLAs are observed to have lower depletions than the Milky Way
ISM and it is rather unlikely that all of gas in the CGM of quasars
has solar enrichment as we assume. For example, our detailed analysis
of \sdssj\ in \citep{QPQ3} showed metallicities in the range
$0.3-1.6\,Z_{\odot}$, and our expectation is that a more typical
metallicity of the quasar CGM $\sim 0.1\,Z_{\odot}$ \citep[][Prochaska
et al., in prep.]{QPQ5}.

The continuum extinction at $\lambda = 1216$\AA\ can be written
$A^{\rm cont}_{\rm 1216} = 0.18 (R_{\rm V}\slash 3.1)(N_{\rm H}\slash
10^{20}\,{\rm cm^{-2}})\,{\rm mag}$, corresponding to a visual
extinction of $A_{\rm V} = 0.05$, where we take $A^{\rm cont}_{\rm
  1216}\slash A_{\rm V}=3.3$ based on the extinction curves of
\citep{Fitzpatrick99} for $R_{\rm V}=3.1$.  In the absence of resonant
effects, 1216\AA\ photons would be attenuated according to the
continuum extinction $A^{\rm cont}_{\rm 1216}$.  Combining this extinction
with the increased resonant pathelngth in eqn.~(\ref{eqn:s}), we can write the total 
resonant extinction that \mlya\ photons suffer as they ecape the cloud as
\be
A^{\rm res}_{\rm 1216} = 0.32 
\left(\frac{x_{\rm HI}}{0.1}\right)\!
\left(\frac{N_{\rm HI}}{10^{19}\,{\rm cm}^{-2}}\right)^{4\slash 3}\!
\left(\frac{b}{20\,\kms}\right)^{-2\slash 3}. 
\ee
Once \mlya\ photons escape an individual cloud, they
propagate freely until they encounter other clouds, where they might
undergo additional resonant scattering. But the probability of this
occurring is small, since it requires both spatial and spectral
overlap. If the spatial covering factor of clouds is $f_{\rm C}$, and
if bulk motions of the clouds in the quasar halo follow a gaussian
with line-of-sight dispersion $\sigma$, then the probability of
additional resonant scattering is $\sim f_{\rm C} N_{\rm s}^{1\slash
  2} b\slash \sqrt{2}\sigma \sim 0.07$ for typical numbers ($N_{\rm s}
= 15$, $f_{\rm C} = 0.50$, $\sigma = 400\kms$, and $b=20\kms$), thus
additional scattering would amount to $\lesssim 10\%$ corrections to
the total extinction, which is already highly uncertain because of the
unknown dust-to-gas ratio. However, in traversing additional material
\mlya\ photons will still encounter continuum extinction, so we can
heuristically write the total extinction as $A_{1216} \approx A^{\rm
  res}_{1216} + f_{\rm C}A^{\rm cont}_{\rm 1216}$.  

Thus given the observed properties of the quasar CGM 
($N_{\rm HI}\sim 10^{19}\,{\rm cm}^{-2}$; $x_{\rm HI}\sim 0.1$; $b\sim 20\,\kms$), 
and  under the most conservative assumptions regarding the amount of
dust, namely a Milky Way dust-to-gas ratio (corresponding to solar
metallicity and high depletion), we conclude that the extinction of
\mlya\ photons would be in the range of $A_{1216} \approx 0.3-0.6$,
resulting in at most a factor of $\sim 2$ reduction in the total
\mlya\ surface brightness. This extinction scales linearly with the
metallicity, and as mentioned previously, it is unlikely that all gas
in the quasar CGM has solar enrichment as we have assumed.
Furthermore, our estimate of the extinction due to resonance effects
is appropriate for \mlya\ photons generated throughout the volume of
an emitting cloud, as expected in a cooling radiation scenario,
whereas it is likely overestimates the attenuation of flourescent
emission powered by photoionization.

Finally, it is worth asking whether the continuum extinction from CGM absorbers would
be detectable via the reddening of the b/g quasar spectra. Consider for
example the absorber in \sdssj\, which we analayzed in detail in \citep{QPQ3}. 
Our photoionization modeling gave a higher total gas column density of 
$N_{\rm H}=10^{20.5}\,{\rm cm^{-2}}$, which implies a continuum 
extinction of $A^{\rm cont}_{\rm
  1216}=0.57$, for solar metallicity and a Milky Way dust to gas
ratio.  This reddening would not be
distinguishable from either the SDSS photometry or our optical
spectra, since the relative reddening $A^{\rm cont}_{\rm 1216} - A^{\rm
  cont}_{\rm 2500}= 0.18$ between rest frame wavelengths $\lambda
=1216$\AA\ and 2500\AA\ (corresponding to the observed frame
$4000-8000$\AA\ covered by the optical spectra and photometry) is
small, and cannot be distinguished from intrinsic variations in the
quasar continuum slopes.

\end{document}